\documentclass[prd,aps,nofootinbib,preprintnumbers,showpacs,floatfix]{revtex4}
\usepackage{amsmath,amssymb}
\usepackage{epsfig}
\usepackage{graphicx}
\usepackage{subfigure}

\begin{document}
\title{Gravitational Waves from Abelian Gauge Fields and Cosmic Strings at Preheating}
\date{June 1, 2010}
\author{Jean-Fran\c{c}ois Dufaux$^{1, 2}$, Daniel G.~Figueroa$^{2,3}$, Juan Garc\'ia-Bellido$^{2,4}$}
\affiliation{$^{1}$APC, UMR 7164 (CNRS - Universit\'e Paris 7), 10 rue Alice Domon et L\'eonie Duquet, 
75205 Paris Cedex 13, France\\
$^{2}$Instituto de F\'isica Te\'orica UAM/CSIC, Universidad Aut\'onoma de Madrid, 28049 Madrid, Spain\\
$^{3}$Department of Physics, CERN - Theory Division, CH-1211 Geneva 23, Switzerland\\
$^{4}$D\'epartement de Physique Th\'eorique, Universit\'e de Gen\`eve, CH-1211 Gen\`eve 4, Switzerland}
\preprint{IFT-UAM/CSIC-10-38, CERN-PH-TH/2010-121}

\def\O{\mathcal{O}}
\def\lesssim{\mathrel{\hbox{\rlap{\hbox{\lower4pt\hbox{$\sim$}}}\hbox{$<$}}}}
\def\gtrsim{\mathrel{\hbox{\rlap{\hbox{\lower4pt\hbox{$\sim$}}}\hbox{$>$}}}}
\def\Mp{M_{\mathrm{Pl}}}
\def\be{\begin{equation}}
\def\ee{\end{equation}}
\def\bea{\begin{eqnarray}}
\def\eea{\end{eqnarray}}
\newcommand{\picdir}[1]{../Figs/#1}
\def\O{\mathcal{O}}
\def\lesssim{\mathrel{\hbox{\rlap{\hbox{\lower4pt\hbox{$\sim$}}}\hbox{$<$}}}}
\def\gtrsim{\mathrel{\hbox{\rlap{\hbox{\lower4pt\hbox{$\sim$}}}\hbox{$>$}}}}
\def\Mp{M_{\mathrm{Pl}}}
\def\be{\begin{equation}}
\def\ee{\end{equation}}
\def\bea{\begin{eqnarray}}
\def\eea{\end{eqnarray}}
\newcommand{\ba}{\begin{array}}
\newcommand{\ea}{\end{array}}
\newcommand{\nn}{\nonumber \\}
\newcommand{\lag}{{\mathcal L}}
\newcommand{\mn}{{\mu\nu}}
\newcommand{\ab}{{\alpha\beta}}
\newcommand{\hI}{\hspace{1cm}}
\newcommand{\hVII}{\hspace{.7cm}}
\newcommand{\hV}{\hspace{.5cm}}
\newcommand{\hu}{{\mathcal H}}
\newcommand{\bq}{{\bf q}}
\newcommand{\bk}{{\bf k}}
\newcommand{\bx}{{\bf x}}

\begin{abstract}
Primordial gravitational waves provide a very important stochastic background that could be
detected soon with interferometric gravitational wave antennas or indirectly via the induced
patterns in the polarization anisotropies of the cosmic microwave background. The detection 
of these waves will open a new window into the early Universe, and therefore it is important to 
characterize in detail all possible sources of primordial gravitational waves. In this paper we 
develop theoretical and numerical methods to study the production of gravitational waves from 
out-of-equilibrium gauge fields at preheating. We then consider models of preheating after 
hybrid inflation, where the symmetry breaking field is charged under a local $U(1)$ symmetry. 
We analyze in detail the dynamics of the system in both momentum and configuration space. 
We show that gauge fields leave specific imprints in the resulting
gravitational wave spectra, mainly through the appearence of new peaks at characteristic
frequencies that are related to the mass scales in the problem. We also show how these new
features in the spectra correlate with string-like spatial configurations in both the Higgs and
gauge fields that arise due to the appearance of topological winding numbers of the Higgs
around Nielsen-Olesen strings. We study in detail the time evolution of the spectrum of gauge
fields and gravitational waves as these strings evolve and decay before entering a turbulent
regime where the gravitational wave energy density saturates.
\end{abstract}

\maketitle

\begin{center}
{\it This paper is dedicated to the memory of Lev Kofman}
\end{center}


\section{Introduction}

Gravitational waves (GW) are a robust prediction of General Relativity~\cite{Maggiore}. They correspond to ripples in space-time that travel at the speed of light, and are typically produced whenever an astronomically large body of mass moves at relativistic speeds like in astrophysical binary systems, or whenever large density contrast waves collide against each other, like in early universe phase transitions. The change in the orbital period of a binary pulsar known as PSR 1913+16 was used by Hulse and Taylor~\cite{HulseTaylor} to obtain indirect evidence of their existence. Although gravitational radiation has not been directly detected yet, it is expected that the present universe should be permeated by a diffuse background of GW of either astrophysical or cosmological origin~\cite{Maggiore}. Astrophysical sources, like the gravitational collapse of supernovae or the neutron star and black hole binaries' coalescence, also produce a stochastic gravitational wave background (GWB) which comes from unresolved point sources. On the other hand, among the backgrounds of cosmological origin, we find the approximately scale-invariant background produced during inflation~\cite{Starobinsky}, or the GWB generated at hypothetical early universe thermal phase transitions~\cite{PhaseTransitions,Apreda,Nicolis,Caprini}, from relativistic motions of turbulent plasmas~\cite{Turbulence} or from the decay of cosmic strings~\cite{Maggiore}. Fortunately, these backgrounds have very different spectral shapes and amplitudes that might, in the future, allow gravitational wave observatories like the Laser Interferometer Gravitational Wave Observatory (LIGO)~\cite{LIGO}, the Laser Interferometer Space Antenna (LISA)~\cite{LISA}, the Big Bang Observer (BBO)~\cite{BBO} or the Decihertz Interferometer Gravitational Wave Observatory (DECIGO)~\cite{DECIGO}, to disentangle their origin. Unfortunately, due to the weakness of gravity, this task will be extremely difficult, requiring a very high accuracy in order to distinguish one background from another. It is thus important to characterize the many different sources of GW as best as possible. Recently there has been a significant improvement in the sensitivity of laser interferometers to a cosmological background of GW and there are now limits on the amplitude of this background that are just below the BBN~\cite{BBN} and CMB bounds~\cite{Elena} as recently reported by the joint LIGO and VIRGO collaborations~\cite{LigoVirgo}.

One cosmological GW background that is very well motivated by other observations is the approximately scale-invariant spectrum of GW produced from quantum fluctuations during inflation \cite{Starobinsky}. This spectrum extends over a very wide frequency range and its amplitude is directly related to the energy density during inflation. These GW may be detected indirectly by forthcoming CMB experiments through their effect on the B-mode polarization of CMB anisotropies~\cite{Planck}, or even directly in the longer term by interferometric experiments, if inflation occurs at sufficiently high energy scales. However, if inflation occurs at lower energies, as is the case in many models motivated by high-energy physics, GW from inflation would have an amplitude that is too low to be observable. On the other hand, in case that a global phase transition took place after the end of inflation, then a scale-invariant GW background is also generated~\cite{Krauss:91,JonesSmith:2007,Fenu:2009}. The detailed calculations show that the amplitude of this new GW background is two orders of magnitude greater than that expected from inflation for the same energy scale~\cite{Fenu:2009}, and thus might be detected directly by the future GW detectors or in the B-mode polarization of the CMB~\cite{GB:2010}. Finally, another source of GW that may be relevant for interferometric experiments and whose study will be our main target here, is provided by the violent period following the end of inflation. Indeed, in many models, the inflaton decays in an explosive and highly inhomogeneous way, in the process of preheating \cite{preheating}. The particular mechanism responsible for preheating is model-dependent, but it is generally dominated by a non-perturbative production of Bose fields with very high occupations numbers, far from thermal equilibrium. This leads to a second, longer stage characterized by turbulent-like interactions between classical waves, before the system eventually reaches a thermal state~\cite{MichaTkachev}. The large, time-dependent fields' inhomogeneities produced by preheating source a stochastic GW background~\cite{KTGW,JuanGW,EL1,EL2,FKGW,GF,GFS,DBFKU,DFKN}. When redshifted until today, this background may fall in the frequency range accessible by interferometric experiments if inflation and preheating occur at sufficiently low energy scales, providing an alternative to test inflation with GW. In addition, GW from preheating carry crucial information about the mostly unkown post-inflationary dynamics and, because the details of preheating depend very much on the model, they could be used in the future to discriminate between different inflationary models.

Gravitational waves from preheating have been intensively studied by different groups~\cite{KTGW,JuanGW,EL1,EL2,FKGW,GF,GFS,DBFKU,DFKN}. Different numerical methods have been used in early works, see \cite{DBFKU} for a critical comparison, but the later results obtained by different groups agree well with each others~\cite{EL2,GFS,DBFKU,DFKN}. Two main classes of models have been studied so far: preheating after chaotic inflation and preheating after hybrid inflation. In the first case, preheating proceeds via parametric resonance~\cite{preheating}. Because inflation and preheating occur at high energy scales in these models, the resulting GW have a typical frequency today in the range $10^6$ - $10^9$ Hz, which is too high for the signal to be observable by currently available experiments. On the other hand, preheating after hybrid inflation may occur at much lower energy scales and the resulting GW may fall in a frequency range below $10^3$ Hz, which is accessible by high-sensitivity interferometric experiments. GW from preheating after hybrid inflation~\cite{hybrid,CLLSW} were first studied in Ref.~\cite{JuanGW}, in the framework of parametric resonance studied in~\cite{KTGW}. It was later understood that hybrid inflation models preheat in an even more violent way, due to the tachyonic amplification of fluctuations of the symmetry breaking field when the fields roll towards the true minumum of the potential, a process called tachyonic preheating~\cite{tachyonic}. Gravitational waves from tachyonic preheating after hybrid inflation were first studied in Ref.~\cite{GF,GFS,DBFKU} and then fully explored in~\cite{DFKN}, where the regions of the parameter space that may lead to an observable signal were determined. Finally, it is also worth noting that the methods developed to study GW production from preheating may be applied to other out-of-equilibrium sources in the early universe. One example that is particularly interesting from the perspective of ground based interferometers is the GW background~\cite{flatdirGW} produced by the non-perturbative decay of flat direction condensates in supersymmetric theories~\cite{flatdir}~\footnote{GW may also be produced from Q-ball fragmentation~\cite{Qball}, see however~\cite{Qball2}.}. Another example is the evolution of unstable domain walls~\cite{domainwalls}.

One common point in these works has been to focus on models involving only scalar fields~\footnote{GW production from a decaying tachyon field coupled to a gauge field has been studied numerically in Ref.~\cite{mazumdar}, although lattice gauge techniques do not seem to be used in this work and the transverse-traceless part of GW does not seem to be correctly extracted.}. However, in realistic models, gauge (vector) fields may also be copiously produced during preheating~\cite{chern,CEWB,magnetic}\footnote{Similarly, flat direction condensates in the MSSM are charged under the gauge symmetries and gauge fields are produced by their non-perturbative decay~\cite{flatdir}.}. These lead to new terms in the anisotropic stress sourcing GW, in addition to the gradients of the scalar fields, and they may play an important role in GW production. Indeed, the numerical 
simulations~\cite{EL1,GFS,DBFKU,DFKN} indicate that scalar fields do not lead to a significant production of GW during 
the turbulent evolution towards thermal equilibrium after preheating. This was demonstrated in \cite{DBFKU} where it was shown that massless gauge fields may change this result. Out-of-equilibrium gauge fields are of course also ubiquitous in other sources of GW, such as thermal phase transitions and local topological defects~\cite{Maggiore}. Moreover, tachyonic preheating~\cite{tachyonic,Copeland2002,symbreak} could be responsible for copious production of dark matter particles~\cite{ester}, lepto and baryogenesis~\cite{CEWB,CLRT,chern,TS,Tranberg2009}, topological defects~\cite{tachyonic,Copeland2002}, primordial magnetic fields~\cite{magnetic}, etc, whose observational consequences could help put more stringent constraints on the period immediately after inflation responsible for the reheating of the Universe.

The main purpose of this paper is to study the GW background produced by physical models with scalar and gauge fields at preheating. We will develop numerical methods and derive theoretical results which can be applied to out-of-equilibrium gauge fields in general. We will then consider models of preheating after hybrid inflation where the symmetry breaking field is complex and coupled to a $U(1)$ gauge field. As we will see, a major consequence of the gauge field is to introduce new characteristic scales which are inherited by the GW spectra. This prompted us to study in detail the dynamics of the system in both Fourier and position spaces. We found in particular that a crucial role is played by the dynamics of cosmic string configurations of the gauge and scalar fields. In order to probe these different scales in the simulations, we had also to develop a lattice calculation of GW with gauge fields that is accurate up to second order in the lattice spacing and timestep.

The paper is divided as follows. In section \ref{SecModel}, we specify the model of hybrid preheating embeded in a gauge framework that we will study in this paper, briefly reviewing the fields' dynamics after inflation. Section \ref{SecTheory} is dedicated to theoretical perspectives on the effects of gauge fields on GW production from preheating. In section \ref{SecNumerics}, we present our numerical method to compute the GW spectra produced by scalar and gauge fields on the lattice. In section \ref{SecSpectra}, we study in detail the dynamics of an abelian-higgs model of tachyonic preheating in Fourier space and the resulting GW spectra. Section \ref{SecPosition} is dedicated to the sudy of spatial configurations in position space of both the sources and the GW themselves. We conclude in section \ref{SecConclu} with a summary of our results and directions for future works. In appendix \ref{AppNogo}, we derive a no-go result for abelian models discussed in the main text. The details of our lattice calculation are given in appendix \ref{AppLattice}.


\section{Abelian-Higgs Preheating After Hybrid Inflation}
\label{SecModel}

Hybrid inflation~\cite{hybrid} is a class of  inflationary models based on particle physics and in particular on spontaneous symmetry breaking by a scalar field. This field is coupled to a singlet scalar, the inflaton, and triggers the end of inflation when its mass-squared goes from positive to negative due to its 
coupling to the rolling inflaton. Let us consider a generic hybrid inflation model, as described by the potential
\begin{equation}\label{eq:HybridPotential}
 V(\varphi,\chi) =  \frac{\lambda}{4}\left(|\varphi|^2 - 
v^2\right)^2 + \frac{1}{2}g^2\chi^2|\varphi|^2
+ V_{\mathrm{infl}}(\chi)\,,
\end{equation}
where $g^2$ is the strenght of the coupling of the inflaton $\chi$ to a symmetry breaking field $\varphi$, with self-coupling $\lambda$ and vacuum expectation value (VEV) $v$. A relevant feature of these models is that a small $\lambda$ is not a prerequisite in order to generate the observed CMB anisotropies, while the scale of inflation can be choosen to range from GUT scales ($\sim 10^{16}$ GeV) all the way down to GeV scales. Depending on the particular model considered, the purely inflaton part of the potential $V_{\mathrm{infl}}(\chi)$ can take different forms, see e.g. Ref.~\cite{lythriotto}, and an apropiate choice of the parameters makes these models completely compatible with CMB constraints.\footnote{For instance, the original
model~\cite{hybrid}, with $V_{\mathrm{infl}}(\chi) = \mu^2 \chi^2$ and $\mu=1.4\times10^{14}$ GeV, $v=3.6\times10^{16}$ GeV, $\lambda=0.17$ and $g=0.001$, produces a spectrum with scalar tilt $n_s=0.98$ and tensor-to-scalar ratio $r=0.1$, in perfect agreement at 95\% c.l. with the WMAP-7yr data~\cite{Komatsu2010}. Other models based on logarithmic loop corrections to the flatness of the potential also give a negative tilt during inflation compatible
with WMAP data.}

We want to embed this setup into a gauge-invariant framework, such that the symmetry breaking field $\varphi$ is
coupled to the corresponding gauge fields. In general, the group of gauge symmetry could
be abelian or non-abelian (or a product of both) and $\varphi$ could even be the Higgs field of the Standard Model with gauge symmetry $SU(3) \times SU(2) \times U(1)$. In the very early universe the gauge group could for instance contain multiple $U(1)$'s, as described in some compactifications of string theory. In the numerical simulations, however, we will restrict ourselves to a complex field $\varphi = (\varphi_1+i\varphi_2)$ coupled to a single gauge field $A_\mu$, with a $U(1)$ symmetry. Nevertheless we will also consider other gauge groups in section \ref{SecTheory}, where we derive theoretical expectations about the production of GW from gauge fields. 

For convenience, independently of the physical origin of $\varphi$, we will often refer to this field simply as the ``Higgs" of the model. In the case of a $U(1)$ symmetry, we will consider the Abelian-Higgs model coupled to an inflaton, as described by
\begin{eqnarray}\label{eq:lagrangianGauge}
S = -\int d^4x\sqrt{-g}\left\lbrace {1\over4}F_\mn F^\mn + (D_\mu\varphi)^{\dag}(D^\mu\varphi) + \frac{1}{2}(\partial_\mu\chi)^2 + 
V(\varphi,\chi)\right\rbrace,
\end{eqnarray}
with $g$ the background metric determinant, $F_{\mu \nu} = \partial_\mu A_\nu - \partial_\nu A_\mu$ is the (anti-symmetric) field strength of the gauge field $A_\mu$, and $D_\mu = \partial_\mu-ieA_\mu$ the usual covariant derivative with $e$ the gauge coupling. After inflation, the metric is that of a (flat) Friedman-Robertson-Walker (FRW) background, with $g_\mn = {\rm diag}(-1,a^2(t),a^2(t),a^2(t))$ and $a(t)$ the scale factor. The (classical) equations of motion can then be derived varying the action~(\ref{eq:lagrangianGauge}) as
\begin{eqnarray}\label{eq:eom}
-\partial_\mu \partial^\mu \chi + 3H\dot\chi + V_{,\chi} &=& 0\\
-D_\mu D^\mu \varphi +3H\dot\varphi + V_{,\varphi^*} &=& 0\\
\dot{E}_i + H\,E_i - \epsilon_{ijk}\,\partial_j B_k - 2e\,{\rm Im}\left[(D_i\varphi)^*\varphi\right] &=& 0\,,
\end{eqnarray}
where a dot denotes the time derivative, $H=\dot a/a$ is the Hubble rate, $E^i = F^{0i}$ and 
$B_i = \frac{1}{2}\epsilon_{ijk}F^{jk}$ are the gauge invariant electric and magnetic fields and $\epsilon_{ijk}$ is 
the completely anti-symmetric 3-tensor. Here and in what follows, latin indices $i, j, k, ...$ run over the three spatial coordinates and repeated indices are summed unless stated otherwise. 

The equation associated to the time component of the gauge field,
\begin{equation}\label{eq:GaussConstraint}
 \partial_kE_k = -2e\,{\rm Im}\left[(D_0\varphi)^*\varphi\right]\,,
\end{equation} 
is not a dynamical equation, but rather a constraint equation equivalent to the Gauss Law of classical electromagnetism, 
$\vec\nabla\vec E = \rho$, with $\rho = -2e\,{\rm Im}\left[(D_0\varphi)^*\varphi\right]$.

In this model the characteristic time scale evolution of the fields after inflation is set by the inverse of $m \equiv \sqrt{\lambda}v$. It follows then that the expansion of the universe is negligible during preheating, since $H \sim (v/M_p)m \ll m$. For studying preheating it is thus sufficient to consider a flat background $\eta_{\mu \nu} = \mathrm{diag}(-1,1,1,1)$. Moreover, for convenience we will evolve the system in the temporal gauge, i.e. fixing $A_0 = 0$. The Higgs and gauge fields equations then look as
\begin{eqnarray}
\label{Xcont}
\ddot{\varphi} - D_i D_i \varphi +  V_{,\varphi^*} &=& 0\\
\label{Acont}
\ddot{A}_i - \partial_j \partial_j A_i + \partial_i \partial_j A_j &=& 
2e\,\mathrm{Im}\left[\varphi^*\,D_i \varphi\right]\\
\label{Gcont}
\partial_i \dot{A}_i &=& 2e\,\mathrm{Im}\left[\varphi^*\,\dot{\varphi}\right] \ .
\end{eqnarray}
The condition $A_0 = 0$ does not fix the gauge completely as we can still perform a gauge transformation 
$\varphi \rightarrow e^{i\alpha}\varphi$, $A_i \rightarrow A_i + \frac{1}{e}\partial_i\alpha$, with a time-independent function $\alpha = \alpha(\vec{x})$. Numerically, we will evolve the system in the temporal gauge and use the remaining gauge degree of freedom to set $A_i = 0$ initially. The initial conditions for the electric field are then read from satifying initially the Gauss constraint (\ref{Gcont}), see below.

The abelian Higgs model is known to give rise, upon symmetry breaking, to cosmic strings connecting Nielsen-Olesen vortices, where the Higgs winds $n$-times. The type of string (I and II) depends on the ratio of Higgs to gauge field mass, $\beta = 2\lambda/e^2 = m_\varphi^2/m_A^2$. Magnetic flux lines repel each other, while the scalar field produces an attractive force, and their range is controlled by the Compton wavelength of the mediating boson. Depending on which one 
is more massive we may have type I or type II superconductors. For $\beta < 1$ ($\lambda < e^2/2$) the penetration depth 
of the magnetic field inside the string is smaller than the coherence length of the string, and local vortices are 
stable for arbitrary winding $n$, as in type I superconductors, while for $\beta > 1$ ($\lambda > e^2/2$) we have the opposite situation, and the magnetic flux lines can live inside the string (vortices with $n > 1$ are unstable in that case). 
Some of the properties that we will encounter in the time evolution of the abelian Higgs model after symmetry breaking in fact depends on whether the cosmic strings that are formed are type I or II. In both cases we find cosmic strings with windings of the Higgs around them, and their energy density will be a strong source for gravitational wave production. We leave for Sections V and VI the detailed description of the rather complicated dynamics.

\subsection{Hybrid Preheating}
\label{subsec:Field Dynamics}

The general qualitative dynamics of the system is as follows. For $\chi > \chi_c$, where $\chi_c \equiv m/g$ is called the critical point, the Higgs field have positive mass squared and the potential has a valley at $\varphi = 0$. During inflation the expansion is driven by the false vacuum energy, $V_0 \simeq \lambda v^4 / 4$, while the inflaton decreases slowly along the valley due to the uplifting term $V_{inf}(\chi)$ in~(\ref{eq:HybridPotential}). Inflation ends either when $\chi$ reaches the critical point or when the slow-roll conditions are violated, whichever occurs first. In both cases, when $\chi < \chi_c$, the effective square Higgs mass becomes negative and this triggers the symmetry breaking process from $\langle |\varphi|^2 \rangle = 0$ to the true vacuum $\langle |\varphi|^2 \rangle = v^2$. The time at which the inflaton reaches $\chi_c$ is called the critical time $t_c$. Around that time, the inflaton's time evolution can be linearly approximated by $\chi (t)= \chi_c (1  - V_c m(t-t_c))$, where and $V_c$ is the (dimensionless) inflaton's velocity at $t_c$,
\begin{equation}\label{defVc}
V_c \equiv \frac{1}{\chi_c}\left|\frac{d\chi(t)}{d(mt)}\right|_{t_c} \, = \, \frac{g |\dot{\chi}_c|}{\sqrt{\lambda} v} \, .
\end{equation}
Through the Higgs-inflaton coupling, the evolution of $\chi(t)$ around $\chi_c$ induces a time dependence in the Higgs effective mass, $m_\varphi^2 = -2 V_c m^3 (t-t_c)$, which changes from positive at $t < t_c$ ($\chi > \chi_c$) to negative at $t > t_c$ ($\chi < \chi_c$). As described in Ref.~\cite{symbreak}, the quantum evolution of the system around $t_c$ can indeed be solved exactly, as long as the non-linearities in the Higgs field (i.e.~the self-interactions of $\varphi$), the interactions with the inflaton fluctuations and with the gauge fields are neglected. In such a case the Higgs behaves as a free scalar field with time dependent mass $m_\varphi(t)$, whose quantum evolution can be solved in terms of Airy functions~\cite{symbreak}. When $t \gtrsim t_c$, the low momentum $k < m$ Higgs modes grow exponentially fast due to the negative square mass, in a process known as ``tachyonic preheating''~\cite{tachyonic,Copeland2002,symbreak,chern}. This in turn sources the production of the gauge field through its interactions with the Higgs.

The low momentum Higgs modes acquire large occupation numbers during tachyonic preheating and therefore become quasi-classical very fast. The transition from quantum to classical of the Higgs' long wavelength modes, takes place very rapidly indeed as compared with the time scale of the symmetry breaking process itself. This way, soon after the end of inflation but much earlier than the Higgs reaches the true vaccum, the modes within the tachyonic band have become fully classical, while all other modes remain in their quantum ground state. Due to the non-linear $\lambda\phi^4$ interaction term, the non-tachyonic modes will eventually get populated thanks to the highly occupied low-momentum modes. Moreover, since the Higgs and the gauge fields are directly coupled, the gauge fields are also excited during and after tachyonic preheating. 

Our strategy here, following~\cite{symbreak,chern}, will be therefore to introduce the system into a lattice in the precise moment in which the Higgs modes within the tachyonic band have became classical (slightly after the inflaton crossed the critical point) but the non-linearities and other interactions are yet negligible. In order to solve for the evolution of the system we will then use the lattice approach, consisting of replacing the quantum evolution by the classical one. This way, the quantum nature of the problem remains in the stochastic character of the initial conditions. This approach is, of course, only valid in the particular scenario under study because the quantum dynamics of the low momentum modes of $\varphi$ drives the system into a regime of classical field behaviour, which ultimately justifies the use of the lattice. It is the subsequent non-linear classical behaviour of the symmetry breaking field that induces the growth of the inflaton and gauge fields, which then develop a non-trivial anisotropic stress-tensor which will source gravitational waves. This approximation has the clear advantage that it fully captures the non-linear and non-perturbative nature of the problem, and allow for the use of gauge fields in a relatively simple way. The full non-linear evolution of the system can then be studied using lattice techniques, discretizing both in space and time the classical equations of all fields~(\ref{Xcont}), (\ref{Acont}) and (\ref{Gcont}), see subsection~\ref{subsecLatt} below. Note that other authors have also used the lattice formulation to study hybrid preheating in the presence of gauge fields, studying some of its phenomenological consequences like baryogenesis~\cite{CEWB,chern,TS,Tranberg2009} or the generation of the primordial seed of cosmological magnetic fields~\cite{magnetic}. Here, for the first time, we study the production of GW associated to the presence of such gauge fields. 

Finally, note that in general the model-dependent inflaton part of the potential $V_{inf}(\chi)$ does not affect significantly the dynamics of preheating, except by setting the initial velocity with which the inflaton reaches the critical point. We will thus ignore $V_{\mathrm{infl}}(\chi)$ and treat the initial velocity $V_c$ as a free parameter. 

\begin{figure}[htb]
\begin{tabular}{cc}
\begin{minipage}[t]{8.1cm}
\begin{center}
\includegraphics[width=8cm]{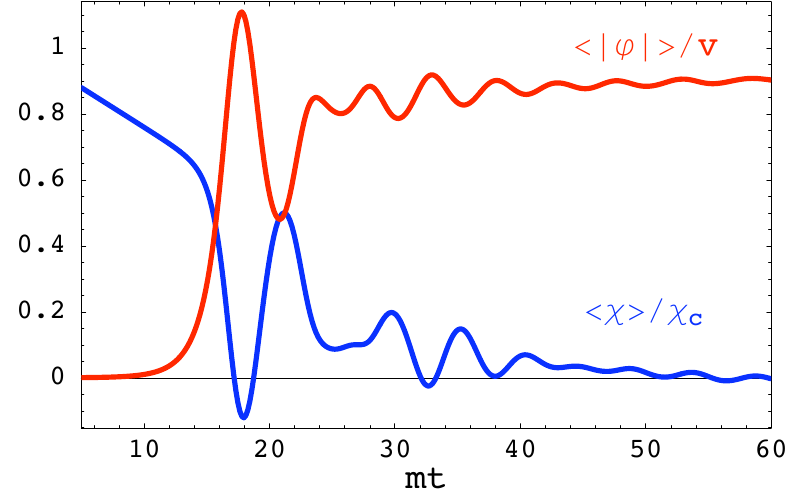}
\caption{Time evolution of the inflaton's mean $\langle \chi \rangle / \chi_c$ (blue) and of the Higgs' root mean squared 
$\langle |\varphi| \rangle / v$ (red) in a hybrid model with $\lambda = g^2 / 2 = 0.125$, $e = 0.3$ and $V_c = 0.024$. One can clearly see the growth of the Higgs towards the true vacuum, while the inflaton rolls down to the bottom of the potential. Once the Higgs is sufficiently close to the true vacuum, 
its self-interactions compensate the tachyonic mass and the field oscillates close to the VEV, while the inflaton oscillates around $\chi = 0$.}
\label{meanrms}
\end{center}
\end{minipage}&
\hspace*{1.5cm}
\begin{minipage}[t]{8cm}
\begin{center}
\includegraphics[width=8cm]{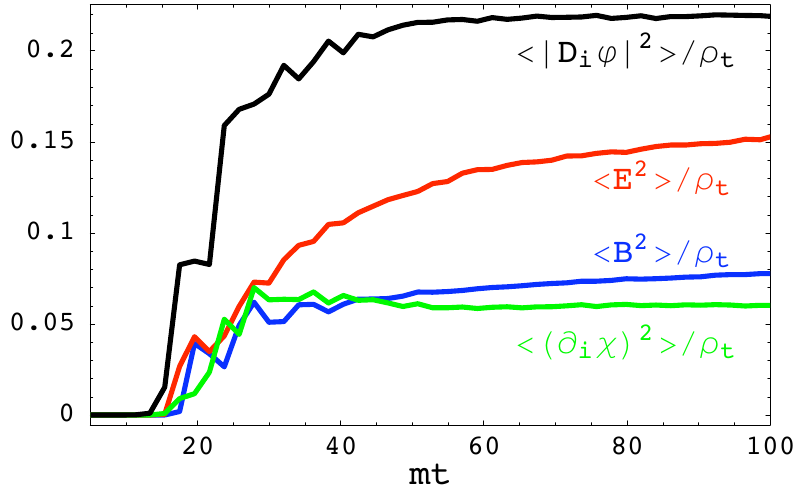}
\caption{Time evolution of the covariant energy density of the Higgs $\langle D_i \varphi (D_i \varphi)^* \rangle$ (black), the electric $\langle E_i E_i \rangle$ (red) and magnetic $\langle B_i B_i \rangle$ (blue) energy densities, and the inflaton's gradient energy density $\langle \partial_i \chi \partial_i \chi \rangle$ (green), all normalized to the total energy of the system. The parameters are the same as in Fig.~\ref{meanrms}. 
A fast growth of the energy components is experienced during tachyonic preheating, first in the Higgs field and then in the rest of the fields due to the their interactions with the Higgs. 
}
\label{rhocomp}
\end{center}
\end{minipage}
\end{tabular}
\end{figure}

\subsection{Initial Conditions}
\label{subsec:InitialConditions}

The initial conditions of the fields follow the prescription from Refs.~\cite{symbreak,chern}. The Higgs modes $\varphi_k$ are solutions of the evolution equations obtained from~(\ref{eq:lagrangianGauge}), which can be rewritten for each component, as oscillators with time-dependent frequencies $\ddot\varphi_k+\left[k^2-2m^3V_c(t-t_c)\right]\varphi_k=0$. If $V_c$ is not extremelly small and the couplings $\lambda$, $e$ and $g$ are not very big, see Ref.~\cite{symbreak}, there is always a time $t_i$ greater than the critial time $t_c$, but much shorter than the time scale in which the non-linearities are important, for which the Higgs modes within the tachyonic band have become classical, whereas those out of the tachyonic band can be set classically to zero. The amplitude of the tachyonic modes can then be found at the time $t_i$, as distributed according to a Gaussian random field of zero mean, which translates into a Rayleigh distribution
\begin{equation}\label{eq:InitialConditions}
\indent 
P(|\varphi_k|)d|\varphi_k|d\theta_k = \frac{|\varphi_k|}{\pi\sigma_k^2}\exp{\left\lbrace-{|\varphi_k|^2/\sigma_k^2}
\right\rbrace}\,d|\varphi_k|d\theta_k\,,
\end{equation}
for the modulus, with a uniform random phase $\theta_k\in[0,2\pi]$. The dispersion $\sigma_k^2$ is given by $\sigma_k^2 \equiv P(k,t_i)/k^3$, where $P(k,t_i) = |f_k|^2$ is the power spectrum of the initial Higgs quantum fluctuations in the background of the homogeneous inflaton~\cite{symbreak}. In the classical limit, the conjugate momentum $\dot\varphi_k(\tau)$ is uniquely determined through $\dot\varphi_k(t_i) = F(k,t_i)\varphi_k(t_i)$, where $F(k,\tau) = {\rm Im}(f_k^*(\tau)g_k(\tau))$, with $f_k$ and $g_k$ functions defined in Eqs.~(43) and (44) of~\cite{symbreak}. 
The rest of the fields (the gauge fields, the non-zero modes of the inflaton and the gravitational waves) are supposed to start from the vacuum, and therefore we will semi-classically set them to zero initially in the simulations. Their coupling to the Higgs modes will drive their evolution, giving rise to a rapid growth of the gauge and inflaton modes and GW subsequently. Their non-linear evolution will then be well described by the lattice simulations.

One could think that, initially at time $t_i$, since the vector fields are in vacuum (they have not been excited yet), one could put their amplitude and conjugate momenta to zero, once the classical regime in the Higgs sector has been established. However, the \textit{gauss constraint} in Eq.~(\ref{eq:GaussConstraint}) imposes a relation between the gauge fields and the Higgs' components, which must be fulfilled at any time during the evolution. Therefore, given our initial conditions for the Higgs described above, we initialize the gauge field in such a way that Gauss constraint is satisfied initially, following \cite{chern}. In the temporal gauge $A_0 = 0$, the Gauss constraint~(\ref{Gcont}) in Fourier space reads
\begin{eqnarray}
\label{GaussFourier}
+ik_i{\dot A}_i({\bf k},t_i) = j_0({\bf k},t_i)\,,
\end{eqnarray}
with $j_0 = -2e^2\,{\rm Im}\left[\dot\varphi^*\varphi\right]$. Note that our initial conditions for the Higgs ensures 
that the initial charge vanishes, $\int d^3x \,j_0(t_i) = 0$. This is a necessary condition for Eq.~(\ref{Gcont}) to be satisfied initially. We then set the initial conditions for the electric field in Fourier space, according to
\begin{eqnarray}\label{coulombGauge}
&&\dot A_i({\bf k}, t_i) = -i\frac{k_i}{k^2}j_0({\bf k},t_i) \\
&&\dot A_i(\vec k = \vec 0, t_i) = 0 
\end{eqnarray}
which is a particular solution of (\ref{GaussFourier}). Thus our initial conditions for the gauge field are such that initially the electric field is purely longitudinal, $\dot{A}_i(t_i) \propto k_i$, while the magnetic field vanishes, 
$A_i(t_i) = 0$.


\section{Theoretical Perspectives on the GW Production in Scalar Gauge Theories}
\label{SecTheory}

Our main purpose in this paper is to study the details of the production of a stochastic GWB during the reheating of the universe after hybrid inflation. Such GWB has indeed been extensively studied recently in global models of hybrid inflation, in the absence of gauge fields~\cite{GF,GFS,DBFKU,DFKN}. There the dynamics of the interacting fields generate a non-trivial anisotropic stress-tensor in the scalar fields, which source a significant background of GW. Here our aim is to embed the hybrid model described in Eq.~(\ref{eq:HybridPotential}) into the gauge invariant framework of Eq.~(\ref{eq:lagrangianGauge}). The symmetry breaking field $\varphi$ will then be coupled to the corresponding gauge fields associated to the gauge symmetry, and the dynamics of all the fields will be described by the set of differential coupled equations~(\ref{Xcont}), (\ref{Acont}) and (\ref{Gcont}). We will then study in detail numerically (see sections~\ref{SecNumerics},~\ref{SecSpectra} and ~\ref{SecPosition}) the production of GW as coming form the new sources due to the presence of the gauge fields. Before going into the details of this model, however, we will first highlight some theoretical and model-independent results about the production of GW from gauge fields at preheating.

In this section, therefore, we start laying down the basic formalism to study GW production in scalar gauge theories. Following Ref.~\cite{DBFKU}, we then apply this formalism to very simple, but relevant, wave-like sources which allow for an easy analytical treatment. We will review the argument why massless vector fields might enhance GW production during the turbulent evolution towards thermal equilibrium after preheating and we will extend this argument to gauge-invariant theories. We will then show that massless gauge fields are not produced at preheating in abelian scalar gauge theories.

Gravitational waves on top of a (spatially flat) FLRW background correspond to (gauge invariant) linear perturbations of the metric that are symmetric, transverse and traceless (TT), $\delta g_{ij} = a^2(t)\,h_{ij}$ with 
$\partial_i h_{ij} = h_{ii} = 0$~\footnote{In the following, we will raise or lower indices of the metric perturbations with the spatial metric, which reduces at first order in $h$ to the Kronecker $\delta_{ij}$, so 
$h_{ij} = h^{i}_{j} = h^{ij}$ and so on.}. The perturbed Einstein equations describe the evolution of these tensor perturbations~\cite{Mukhanov} as
\begin{equation}\label{eq:GWeq}
\ddot h_{ij} + 3H\dot h_{ij} - \frac{1}{a^2}\nabla^2h_{ij} =
16\pi G\,\Pi_{ij}^{\rm TT}\, .
\end{equation}
The source term is the TT part - verifying $\partial_i\Pi_{ij}^{\rm TT} = \Pi_{ii}^{\rm TT} = 0$ - of the anisotropic stress, $a^2\,\Pi_{ij} = T_{ij} - p\,g_{ij}$ where $p$ is the background pressure. In the present case, both the scalar fields (Higgs and inflaton) and the vector fields (gauge bosons) contribute to the GW source. 

Defining $\bar{h}_{ij} = a\,h_{ij}$ and going to Fourier space, the GW equations~(\ref{eq:GWeq}) read
\be
\label{GWF}
\bar{h}''_{ij} + \left(k^2 - \frac{a''}{a}\right)\,\bar{h}_{ij} = 16\pi\,G\,a^3\,\Pi_{ij}^{\rm TT}
\ee
where $k^2 = \vec{k}^2$ is the square of the comoving wave-number and primes denote derivatives with respect to conformal time $d\tau = dt/a$. The TT part of a symmetric tensor in Fourier space is obtained by the projection
\begin{eqnarray}
\label{TTpart}
\Pi^{\mathrm{TT}}_{ij}(\vec{k}) = \mathcal{O}_{ij,lm}(\hat{k}) \, \Pi_{lm}(\vec{k})\\
\mathcal{O}_{ij,lm}(\hat{k}) = P_{il}(\hat{k})\,P_{jm}(\hat{k}) - \frac{1}{2}\,P_{ij}(\hat{k})\,P_{lm}(\hat{k})\,,
\end{eqnarray}
where $P_{ij}(\hat{k}) = \delta_{ij} - \hat{k}_i\,\hat{k}_j$ and $\hat{k} = \vec{k} / k$ is the unit vector in the direction of $\vec{k}$.

For causal processes like preheating, most of the GW are produced well inside the Hubble radius, where the term in $a''/a$ can be neglected in (\ref{GWF}). The solution of (\ref{GWF}) is then expressed in terms of a simple Green 
function~\footnote{This is also the exact solution for all the modes in a radiation-dominated universe where $a'' = 0$. This is a good approximation in a wide class of preheating models where the equation of state quickly jump towards 
$w = 1/3$ at the beginning of preheating~\cite{eos}. In a matter-dominated universe, the Green function for wavelengths 
of the order of, or larger than the Hubble radius is given by other Bessel functions.}
\be
\label{GWGreen}
\bar{h}_{ij}(\vec{k},\tau) = \frac{16\pi G}{k}\,\int_{\tau_i}^{\tau} d\tilde{\tau}\,
\mathrm{sin}\left[k (\tau - \tilde{\tau})\right]\,a^3(\tilde{\tau})\,\Pi^{\mathrm{TT}}_{ij}(\vec{k},\tilde{\tau})
\ee
for initial conditions $h_{ij} = h'_{ij} = 0$ at $\tau = \tau_i$.

The expansion of the universe will not play an important role for the discussion in this section and it is usually negligible during tachyonic preheating after hybrid inflation that we will study in the next sections. Therefore, from now on, we will work with a Minkowski background, $a = 1$. In that case, the TT part of the anisotropic stress for the 
abelian model (\ref{eq:lagrangianGauge}) reads
\be
\label{Tcont}
\Pi_{ij}^{\mathrm{TT}} = 
\left[\partial_i \chi \partial_j \chi + 2\,\mathrm{Re}\left[D_i \varphi \left(D_j \varphi\right)^*\right] 
- B_i B_j - E_i E_j\right]^{\mathrm{TT}}
\ee
where pure-trace terms have been removed by the TT projection. Non-abelian gauge fields would lead to similar contributions, with the corresponding modifications of the covariant derivative and the electric and magnetic fields.


\subsection{GW from wave-like sources}

In general, the calculation of GW production through (\ref{GWGreen}) requires the knowledge of the time evolution of the source. It is interesting to study the very simple case where the source corresponds to a superposition of waves with wave-like dispersion relation and adiabatically evolving frequencies~\cite{DBFKU}. This case covers to a good approximation different situations that occur in the context of preheating, in particular the stage of turbulent evolution towards thermal equilibrium\footnote{Of course, there are also several situations in the context of preheating where the time-evolution of the fields is {\it not} wave-like, such as the exponential amplification of fluctuations or the collision of non-linear bubble configurations.}. Specifically, consider the following time evolution for the Fourier modes of some generic scalar field $\phi$,  
\be
\label{wavelike}
\phi(\vec{k},\tau) = \alpha^{+}_k(\tau)\, e^{i \, \omega_\phi(k) \, \tau} + \alpha^{-}_k(\tau)\, e^{-i \, \omega_\phi(k) \, \tau} \;\;\;\; 
\mbox{with} \;\;\;\; \omega^2_\phi(k) = k^2 + m_\phi^2 
\ee
and similarly for the other fields. In the context of preheating, we deal generally with interacting waves where the interactions contribute to the effective mass $m_\phi$. The cases in which the frequency $\omega_\phi$ is not constant but is nevertheless evolving {\em adiabatically} with time ($\dot\omega\ll \omega^2$), can be treated in the same way, since there the WKB approximation gives $\phi(\vec{k},\tau) \propto \alpha_k^\pm\,\exp\left(\pm i \, \int^\tau \omega_\phi(k,\tilde{\tau}) \, d\tilde{\tau}\right)$. 

\begin{figure}[htb]
\begin{center}
\begin{tabular}{ccc}
\includegraphics[width=4cm]{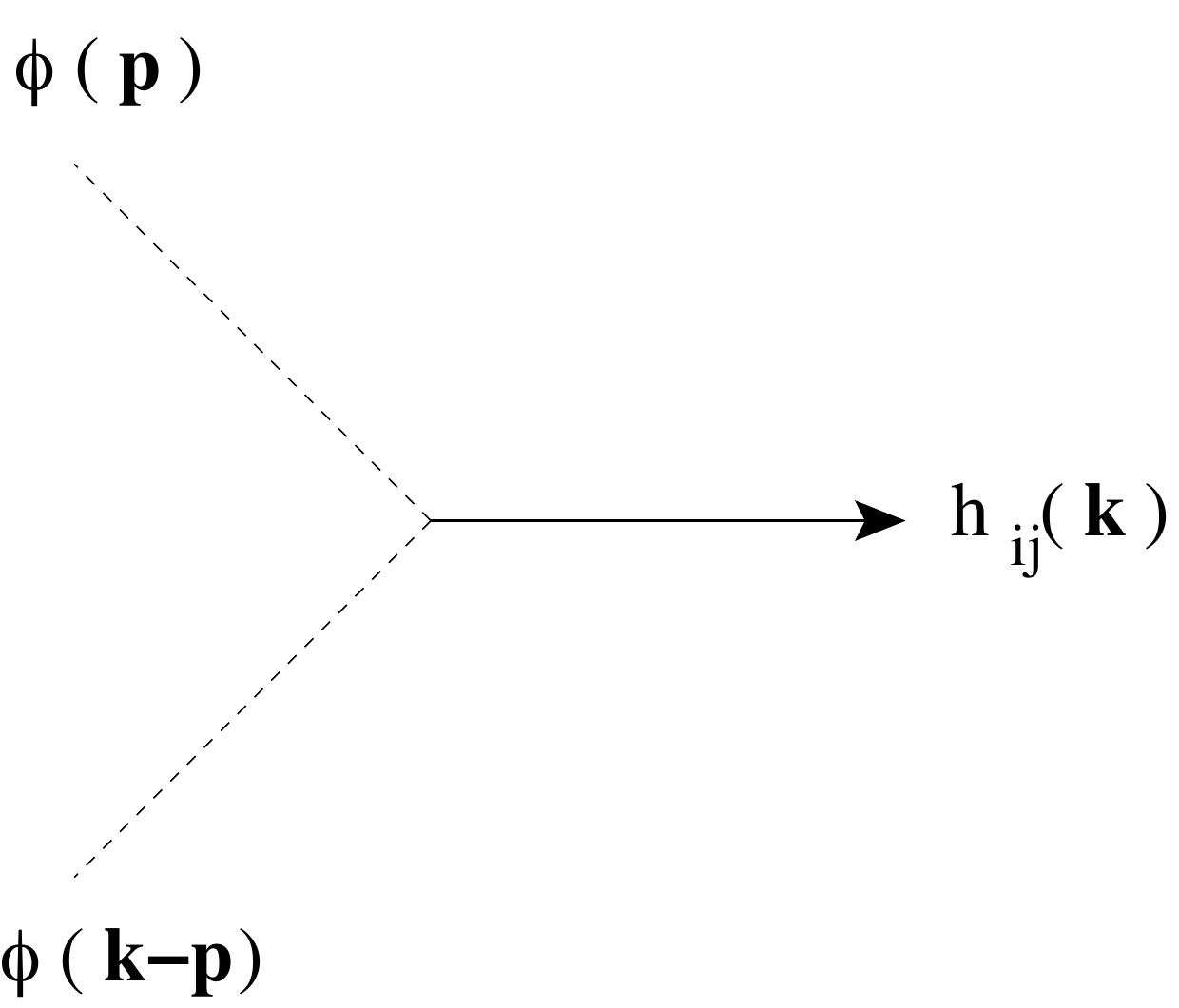} \hspace*{1.5cm}
\includegraphics[width=4cm]{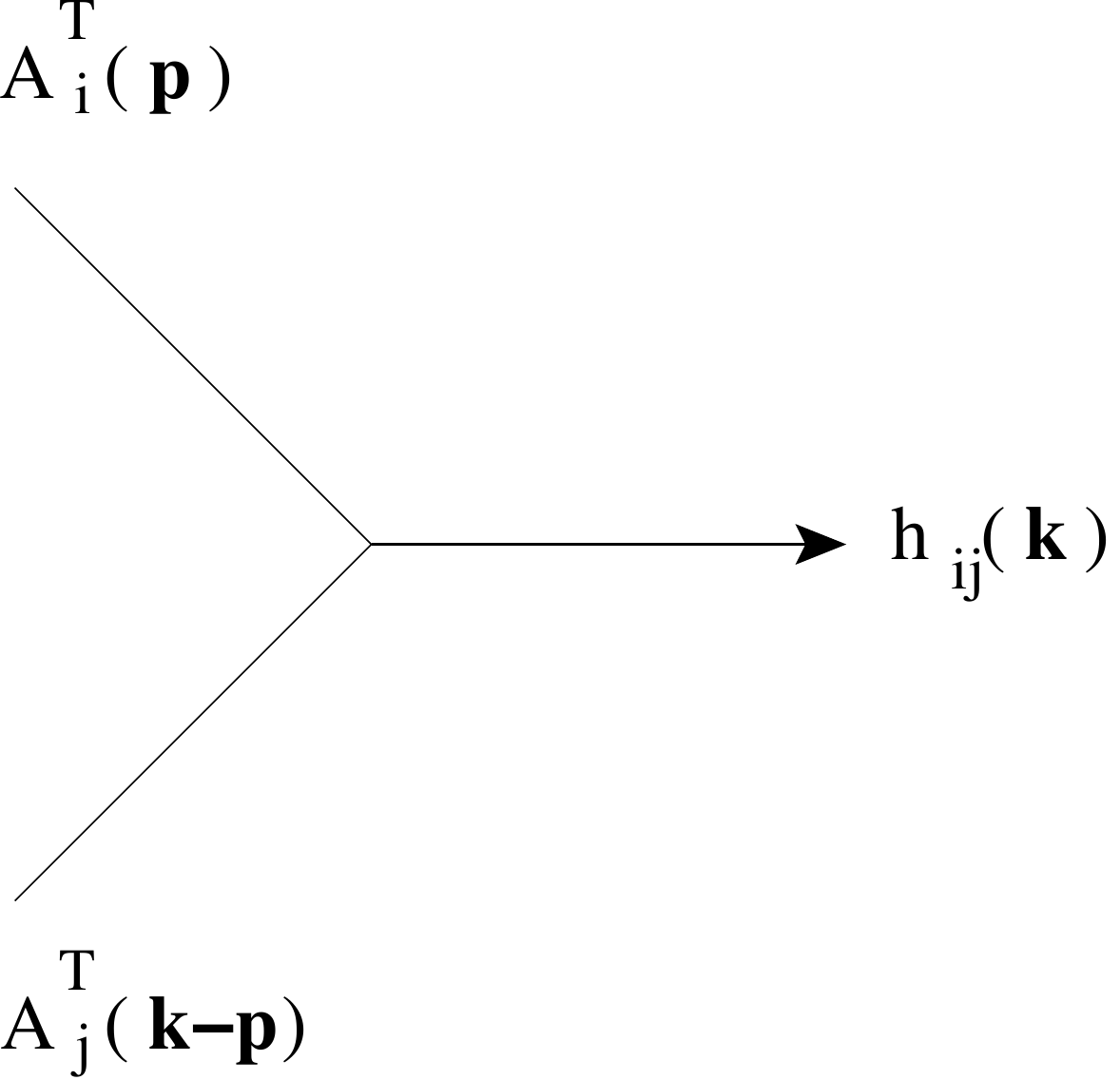} \hspace*{1.5cm}
\includegraphics[width=4cm]{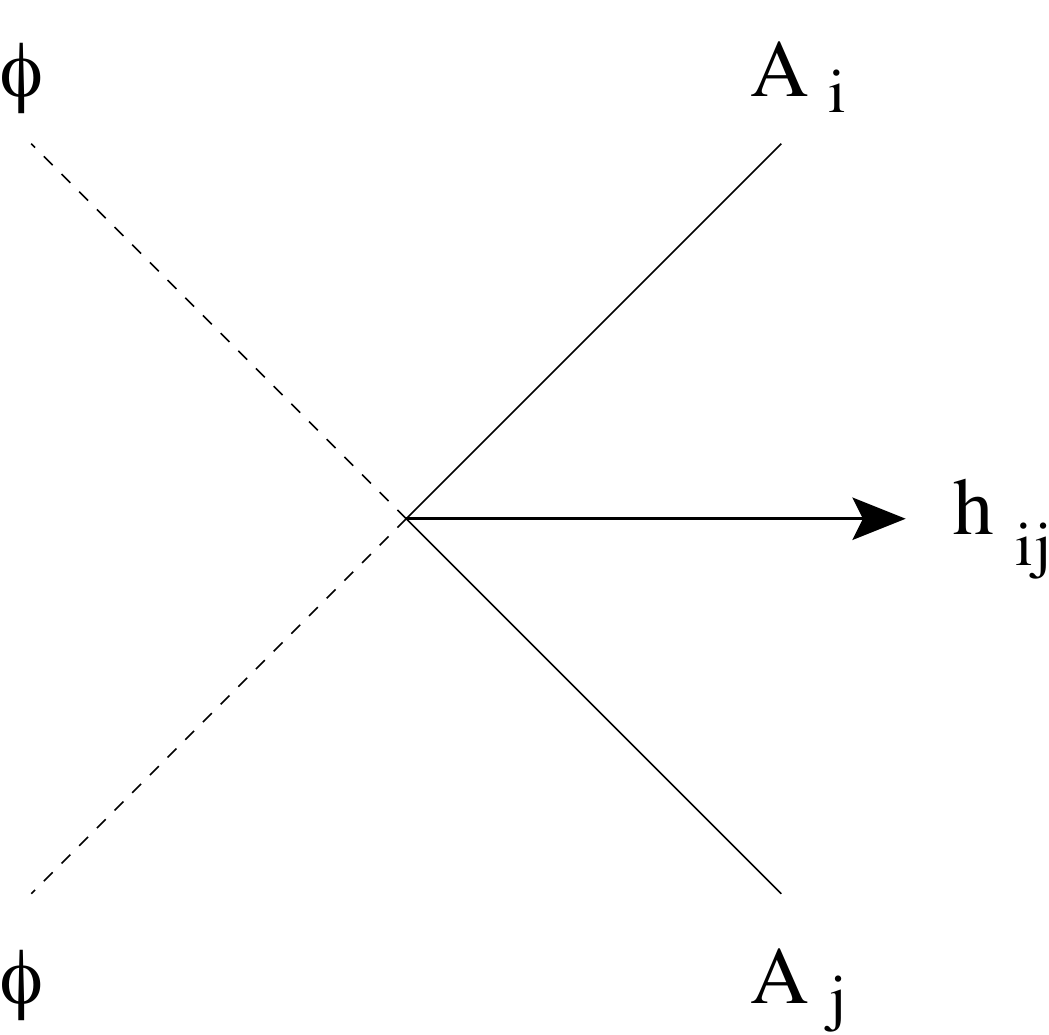}
\end{tabular}
\caption{Contribution of different source terms to GW production from wave-like fields. Left pannel: contribution associated to the source term in $\partial_i \phi \partial_j \phi$ for scalar fields, corresponding to the interaction between two scalar waves and a graviton (forbidden by helicity conservation). Middle pannel: contribution associated to the terms in (\ref{AAh}), corresponding to the interaction between two vector waves and a graviton (allowed if the vector field is massless). Right pannel: contribution associated to the second term in (\ref{AAXXh}), corresponding to the interaction between several scalar and vector waves and a graviton.}
\label{diags}
\end{center}
\end{figure}

In theories involving only scalar fields, the source terms for GW have the same form as the first term in the RHS of 
(\ref{Tcont}). In Fourier space, the product of the spatial derivatives of the scalar field leads to the convolution
\be
\label{Tcont2}
\Pi^{\mathrm{TT}}_{ij}(\vec{k},\tau) \propto \mathcal{O}_{ij , lm}(\hat{k})\,\int d^3\vec{p}\ p_l\,p_m\,
\phi(\vec{p},\tau)\,\phi(\vec{k} - \vec{p},\tau) 
\ee
where we have used $\mathcal{O}_{ij , lm}\,k_m = 0$. Inserting~(\ref{Tcont2}) into~(\ref{GWGreen}), and using the time dependence (\ref{wavelike}), we get, whenever the coefficients $\alpha^\pm_k(\tau)$ evolve adiabatically with time,  
\be
\label{hijWavelike}
h_{ij}(\vec{k},\tau) \propto G\,e^{\pm i\,k\,\tau}\,\mathcal{O}_{ij , lm}(\hat{k})\,\int d^3\vec{p}\ p_l\,p_m\,
\alpha^\pm_p\,\alpha^\pm_{k-p}\,\int_{\tau_i}^{\tau} d\tilde{\tau}\,\exp\left[i\,\tilde\tau\,\left(\pm \omega_\phi(p) \pm \omega_\phi(|\vec{k}-\vec{p}|) 
\mp k\right)\right]
\ee
where we have decomposed the sine in (\ref{GWGreen}) into imaginary exponentials. In the limit of large time $\tau$ 
with respect to the frequencies, the time integrals above reduces to Dirac delta distributions, enforcing energy conservation for trilinear processes involving two ``particles" of the field $\phi$ and one graviton. For instance, 
we have
\be
\label{Epcons}
\omega_\phi(p) + \omega_\phi(|\vec{k}-\vec{p}|) = k \ .
\ee
for two $\phi$-particles emitting a graviton (left panel of Fig.~\ref{diags}). Not only the conservation of energy, but also the conservation of momentum is taken into account in (\ref{Epcons}). Other signs in the phase inside the time integral in (\ref{hijWavelike}) correspond to other trilinear processes, such as the decay of a $\phi$-particle into another $\phi$-particle and a graviton. Energy and momentum conservations are possible only for massless $\phi$-particles~\footnote{That Eq.~(\ref{Epcons}) implies $m_\phi = 0$ simply reflects the well-kown fact that a massless particle (here a graviton) cannot be emitted by two massive particles (here the $\phi$-particles), since in the frame of the center of mass of the two incident particles the massless particle should be emitted with zero momentum.} and for $\vec{k} \parallel \vec{p}$. However, for $\vec{k} \parallel \vec{p}$, the TT projection brings the GW amplitude to zero, $\mathcal{O}_{ij , lm}(\hat{k})\,p_l\,p_m = 0$ in (\ref{hijWavelike}). The reason for this is clearly helicity conservation, which forbids interactions between scalar waves and a graviton at leading order in the gravitational coupling constant $G$. Interactions involving several gravitons are possible beyond the linear order in $h_{ij}$, but they are highly suppressed by extra powers of the Newton constant $G$. 

On the other hand, the presence of massless vector fields (photons) may change this result, since interactions between two vector waves with helicity $1$ and a graviton (middle pannel of Fig.~\ref{diags}) are not forbidden by helicity conservation. Indeed, consider the two terms $B_{i} B_{j}$ and $E_i E_j$ in the source (\ref{Tcont}) of GW, and decompose the vector field $A_i$ into a longitudinal part $A_i^L = \partial_i L$ and a transverse part $A_i^{\mathrm{T}}$ with $\partial_i A_i^{\mathrm{T}} = 0$. Proceding as above leads to energy-momentum conservation like in (\ref{Epcons}), which again implies the momenta of the two vector particles to be parallel to the momentum of the graviton. As before, each term involving a spatial derivative $\partial_i$ or $\partial_j$ then leads to a factor of $k_i$ or $k_j$ in Fourier space, which vanishes when contracted with $\mathcal{O}_{ij , lm}(\hat{k})$. However, the terms $B_{i}B_{j}$ and $E_i E_j$ in~(\ref{Tcont}), include contributions as
\be
\label{AAh}
\partial_k A_i^{\mathrm{T}} \, \partial_k A_j^{\mathrm{T}} \;\; 
\mbox{ and } \;\; \dot{A}_i^{\mathrm{T}} \, \dot{A}_j^{\mathrm{T}}
\ee 
which do not involve $\partial_i$ and $\partial_j$. These terms can therefore lead to a non-zero GW amplitude from a 
wave-like source, through processes like the one illustrated in the middle pannel of Fig.~\ref{diags}. Note however that energy-momentum conservation for these processes still requires the vector field to be effectively massless. 

In addition, extra channels of GW production from wave-like fields arise in scalar gauge theories from the covariant derivative of the scalars. Indeed, consider the second term in the source (\ref{Tcont}) of GW. It involves contributions of the form
\be
\label{AAXXh}
e\,\partial_i \varphi \, A_j \, \varphi^* \;\; \mbox{ and } \;\; e^2\,A_i\,A_j\,|\varphi|^2 \ .
\ee
Since these source terms are cubic and quartic in the fields, GW production from these terms for wave-like fields correspond to interactions between three and four particles and a graviton, see e.g. the right pannel of Fig.~\ref{diags}. Clearly, these higher-order interactions are less restrictive as for instance they are now possible even for massive scalar and gauge fields.  

These qualitative arguments, although valid only for a very simple time-dependence of the source, illustrate that gauge fields can lead to new channels of GW production. In particular, they suggest that gauge fields can keep GW production active well after preheating, during the stage of turbulent evolution towards thermal equilibrium. Since this stage can last for a long period of time, this could significantly increase the amplitude of the resulting GW spectrum. As we will see, this does not happen for the hybrid model described by (\ref{eq:HybridPotential}) and (\ref{eq:lagrangianGauge}). In the context of the hybrid abelian-Higgs models, the gauge field will have specific effects on GW production during preheating but the GW amplitude will eventually saturate during the later stages of the dynamics. Indeed, in such abelian model the gauge field acquires a mass through the Higgs mechanism during the dynamical spontaneous symmetry breaking. Thefore, as discussed above, processes like the one in the middel pannel of Fig.~\ref{diags} are forbidden~\footnote{On the other hand, processes like the one in the right pannel of Fig.~\ref{diags} are in principle possible, although they are further suppressed by the gauge coupling constant.}. It is possible that massless gauge fields produced at preheating could still enhance GW production during the subsequent evolution towards thermal equilibrium. However, this does not happen in abelian scalar gauge theories, as we discuss next.


\subsection{No massless gauge fields from abelian scalar fields during preheating}

In the rest of this section, we argue that massless gauge fields are not produced during preheating in abelian scalar gauge theories, at least with canonical kinetic terms. More precisely, we show that in this case a would-be massless gauge field is either decoupled from the other scalar and gauge fields and therefore not produced~\footnote{Here we do not consider fermions, which of course can also couple to gauge fields. Since the production of fermions is limited by Pauli blocking, they are not expected to lead to an abundant production of gauge fields at preheating.}, or it receives an effective mass due to its interactions during preheating and the subsequent turbulent evolution~\footnote{When the system equilibrates and cools down, such an effective mass will eventually become negligible, but this occurs at much later stages in the dynamics, much beyond our simulations' range.}. Here and in the following, the term ``mass" will be used in a very ``loose" sense and should be understood as a contribution to the dispersion relation of the fields in the adiabatic regime (21), which is relevant for our discussion of GW production from wave-like sources in section III.A. We will consider a generic form of the scalar potential and an arbitrary number of U(1) symmetries and of scalar and gauge fields charged under these symmetries. However, the argument does not apply to non-abelian theories, like $SU(2) \times U(1)$, where massless photons may be produced at preheating~\cite{magnetic}.

Before considering the general case, consider again the system (\ref{eq:lagrangianGauge}) of a single abelian gauge field 
$A_\mu$ coupled to a single scalar field $\varphi$. A first possibility to obtain a massless gauge field would be 
that the VEV of $\varphi$ vanishes, since otherwise the gauge field acquires a mass through the Higgs mechanism. 
However, in that case, $A_\mu$ acquires an effective mass through its interaction with the fluctuations 
$\delta \varphi$ of $\varphi$
\be
\partial^\nu F_{\mu\nu} + 2 e^2\, |\delta \varphi|^2 \,A_\mu = 2e\,\mathrm{Im}\left[\delta \varphi^*\,\partial_\mu 
\delta \varphi\right] \ .
\ee
In the second term of the LHS, the large and classical fluctuations of the scalar field behave as a classical VEV 
providing a mass to the gauge field by the Higgs mechanism.  
The fluctuations $\delta \varphi$ are produced whenever $A_\mu$ is produced, since both of them act as a source for the production of the other. Either the fluctuations of $\varphi$ are responsible for the production of $A_\mu$ or the gauge field is produced by another mechanism and leads in turn to the production of $\delta \varphi$ fluctuations through its interaction with $\varphi$. Thus it does not seem possible to produce a massless gauge field at preheating in this model.

Another possibility is to consider several gauge fields so that a combination of them can remain massless like in 
the $SU(2) \times U(1)$ Standard Model. Consider therefore an abelian model with two gauge fields $A_\mu^{(1)}$ 
and $A_\mu^{(2)}$, coupled to a single scalar field $\varphi$ with coupling constants 
$e_1$ and $e_2$, $D_\mu \varphi = \left(\partial_\mu - i e_1 A_\mu^{(1)} - i e_2 A_\mu^{(2)}\right)\,\varphi$. The 
system is invariant under the gauge transformation
\be
\varphi \rightarrow e^{i\alpha(x) + i\beta(x)}\,\varphi \hspace*{0.3cm} , \hspace*{0.3cm}
A^{(1)}_\mu(x) \rightarrow A^{(1)}_\mu(x) + \frac{1}{e_1}\,\partial_\mu \alpha(x) \hspace*{0.3cm} , \hspace*{0.3cm}
A^{(2)}_\mu(x) \rightarrow A^{(2)}_\mu(x) + \frac{1}{e_2}\,\partial_\mu \beta(x) \ .
\ee
The equations of motion are
\be
\partial^\nu F^{(k)}_{\mu \nu} = 2e_k\,\mathrm{Im}\left[\varphi^*\,D_\mu \varphi\right]
\ee
for $k = 1, 2$ and where $F^{(k)}_{\mu \nu} = \partial_\mu A^{(k)}_\nu - \partial_\nu A^{(k)}_\mu$. 
It is easy to see that the combination $\tilde{A}_\mu = e_2\,A^{(1)}_\mu - e_1\,A^{(2)}_\mu$ is massless. However, 
both the mass term and the source term vanish for $\tilde{A}_\mu$, so it is a free field
\be
\partial^\nu \tilde{F}_{\mu \nu} = 0
\ee
where $\tilde{F}_{\mu \nu} = \partial_\mu \tilde{A}_\nu - \partial_\nu \tilde{A}_\mu$ is the gauge field strength 
of the massless combination $\tilde{A}_\mu$. Thus the massless gauge field is decoupled from 
the system and therefore not produced. This is similar to the photon of the $SU(2) \times U(1)$ Standard Model, which does 
not couple to the Higgs at tree level. However, it couples to the $W$ and $Z$ bosons because of the non-linear nature of non-abelian theories. In that case, Higgs fluctuations amplified at preheating may source the production of $W$ and $Z$ bosons, which in turn source the production of photons~\cite{magnetic}. On the other hand, in the case of $U(1)$ symmetries, interactions between gauge fields would require non-canonical kinetic terms in the lagrangian, like $\left(F_{\mu\nu} F^{\mu\nu}\right)^2$, or higher order terms like $\theta F_{\mu\nu}\tilde F^{\mu\nu}$ or $\phi^\dagger\phi F_{\mu\nu} F^{\mu\nu}$, in order to avoid the tree-level decoupling of the photons. 

It is straightforward to extend the arguments above to an arbitrary number of scalar and gauge fields, see Appendix~\ref{AppNogo}. We conclude thus that massless gauge fields are not significantly produced during preheating in abelian scalar gauge theories. As discussed in the previous subsection, we therefore do not expect a significant production of GW during the stage of turbulent interactions after preheating in such theories. The situation could be 
different, however, for non-abelian theories.


\section{Numerical Calculation of GW with Gauge Fields}
\label{SecNumerics}

In this section, we present our numerical method to calculate GW production in scalar gauge theories on the lattice. The basic methods developed in \cite{GFS,DBFKU,DFKN} for scalar fields can be directly generalized to such theories. We briefly review these methods in the first subsection. On the other hand, in the presence of gauge fields, special care must be paid in the lattice calculation of GW in order to reproduce the continuum theory up to $\O(dx^2)$ and $\O(dt^2)$ accuracy in the lattice spacing $dx$ and timestep $dt$. This is discussed in subsection \ref{subsecLatt}. 

The numerical results presented in this paper were obtained with lattices of $128^3$ and $256^3$ points. We performed a number of checks to verify that the results are physical and not affected by lattice artefacts like insufficient IR or UV coverage or too large timestep.

\subsection{Numerical Method}

Since the expansion of the universe is negligible during preheating after hybrid inflation, we will work in a Minkowski background to simplify the notations. The extension to an expanding universe is straightforward~\cite{GFS,DBFKU,DFKN}.
In a Minkowski background, the equation (\ref{eq:GWeq}) describing the evolution in time of the $TT$ tensor perturbations representing GW reduces to
\be
\label{hcont}
\ddot{h}_{ij} - \partial_k \partial_k h_{ij} = 16\pi G\,\Pi_{ij}^{\mathrm{TT}}
\ee 
where the source term $\Pi_{ij}^{\mathrm{TT}}$ was given in (\ref{Tcont}). In the lattice simulations, however, 
we can solve an alternative equation in position space,
\be
\label{unonTT}
\ddot{u}_{ij} - \partial_k \partial_k u_{ij} = 16\pi G\,\Pi_{ij}
\ee
where the source term~\footnote{Here we neglect again pure-trace terms in $\Pi_{ij}$ because they will be removed 
by the TT projection.} 
\be
\label{TnonTT}
\Pi_{ij} = \partial_i \chi \partial_j \chi + 2\,\mathrm{Re}\left[D_i \varphi \left(D_j \varphi\right)^*\right] - B_i B_j - E_i E_j
\ee 
is \textit{not} Transverse-Traceless. We can thus solve eq.~(\ref{unonTT}) together with the evolution equations of the scalar and gauge fields sourcing GW. The reason is the following: the TT part of the source is most easily calculated in Fourier space through the projection (\ref{TTpart}), which is non-local in position space. One could of course Fourier transform the anisotropic stress tensor at each timestep in order to calculate its TT part and then evolve the GW 
equation, but this would be highly time-consuming. However, since the equation (\ref{hcont}), the TT projection 
(\ref{TTpart}) and the Fourier transform are all linear in $h_{ij}$, we can just solve for Eq.~(\ref{unonTT}) with a 
non-TT source term, and apply the TT projection (\ref{TTpart}) on $u_{ij}$ in Fourier space only at those moments of time when we want to output the GW spectra. This is certainly a faster procedure since it does not require to take Fourier transforms at each timestep. This method was originally proposed in~\cite{GFS}. A modified version was developed 
in~\cite{DFKN} based on the formalism of \cite{DBFKU}. Another alternative method is to use the formal solution 
of~(\ref{GWF}) in terms of its Green function~(\ref{GWGreen}) to directly calculate the GW spectra from the source 
$\Pi^{\mathrm{TT}}_{ij}$~\cite{DBFKU}. The results presented in the next section were obtained using the methods of both Refs.~\cite{GFS} and \cite{DFKN} and we checked that the method of \cite{DBFKU} give the same results. We refer the reader to those references for more details.

The GW propagate freely after their production and their energy density can be calculated by the spatial average
\be
\label{defrhogw}
\rho_{\mathrm{gw}} = \frac{\langle \dot{h}_{ij} \, \dot{h}_{ij} \rangle}{32\pi G} \, .
\ee  
Performing an extra time average over a full period in order to eliminate the fast oscillations of the waves with time, 
the energy density is given by the sum of the kinetic and gradient terms
\be
\rho_{\mathrm{gw}} = \frac{1}{64\pi G\,V}\,\int d^3\vec{k} \,
\left(\dot{h}_{ij}(\vec{k})\,\dot{h}_{ij}(\vec{k})^* + k^2\,h_{ij}(\vec{k})\,h_{ij}(\vec{k})^*\right)
\ee
where $V$ is the volume of the lattice box and $h_{ij} = \mathcal{O}_{ij , lm} u_{lm}$ is the TT component calculated
in Fourier space. The spectrum of energy density in GW per logarithmic frequency interval 
at the time of production is then given by
\be
\left(\frac{d\rho_{\mathrm{gw}}}{d\mathrm{ln} k}\right)_p = \frac{k^3}{16 G\,V}\,
\left(\dot{h}_{ij}(\vec{k})\,\dot{h}_{ij}(\vec{k})^* + k^2\,h_{ij}(\vec{k})\,h_{ij}(\vec{k})^*\right) 
\ee 
where we have used the fact that the spectrum is statistically isotropic. 

Finally, the quantity of interest is the present-day spectrum of energy density in GW per logarithmic frequency interval divided by the critical density
\be
h^2 \Omega_{\mathrm{gw}}(f) = \left(\frac{1}{\rho_c}\,\frac{d \rho_{\mathrm{gw}}}{d\mathrm{ln} f}\right)_0 
\ee
where the subscript ``$0$" refers to today. Assuming a ``standard" thermal history after reheating, the frequency and amplitude of the GW today are obtained from the spectrum at the time of production according to
\bea
\label{f0}
f &=& 4 \times 10^{10} \, \mathrm{Hz}\ \frac{k}{a_p\,\rho_p^{1/4}} \, \left(\frac{a_p}{a_*}\right)^{\frac{1}{4}(1-3\bar{w})}  \\
\label{h0}
h^2\,\Omega_{\mathrm{gw}} &=& 9.3 \times 10^{-6}\,  
\left(\frac{1}{\rho}\,\frac{d\rho_{\mathrm{gw}}}{d\mathrm{ln} k}\right)_p \, \left(\frac{a_p}{a_*}\right)^{1 - 3 \bar{w}}
\eea
where $a_p$ and $\rho_p = \lambda v^4 / 4$ are the scale factor and total energy density at the time of GW production, 
$a_*$ is the scale factor when the universe becomes radiation-dominated and $\bar{w}$ is the mean equation of state 
between these two moments of time, see e.g. \cite{DBFKU} for details. Since tachyonic preheating is a very fast process, we assume that the universe becomes radiation-dominated in less than a Hubble time, so that the dependence on the scale factor in Eqs.~(\ref{f0},\ref{h0}) is negligible.

\subsection{Lattice Formulation}
\label{subsecLatt}

Our lattice formulation for the evolution of the scalar and gauge fields is standard and described in full details in Appendix~\ref{AppLattice}. The scalar fields are defined at the lattice points and the gauge field in the segments between lattice points. We start from a discretized version of the continuum action of the abelian-Higgs model~(\ref{eq:HybridPotential})
\be
\label{Sdis}
S = - dt d^3\vec{x}\,\sum_{\{x^\lambda\}}\,\left[\frac{1}{2}\,\partial^+_\mu\tilde{\chi}\,\partial^{+ \mu}\tilde{\chi} + 
D^+_\mu \tilde{\varphi}\,\left(D^{+ \mu} \tilde{\varphi}\right)^* 
+ \frac{1}{4}\,\tilde{F}_{\mu \nu}\,\tilde{F}^{\mu \nu} + V(\tilde{\chi}, |\tilde{\varphi}|)\right]
\ee
where the sum is over all the spacetime lattice points. We denote the lattice fields with a tilde to distinguish them from their continuum analog. The lattice expressions for the forward partial derivative $\partial^+_\mu$, the forward gauge covariant derivative $D^+_\mu$ and the gauge field strength $\tilde{F}_{\mu \nu}$ are given in Eqs.~(\ref{forpar}), 
(\ref{forcov}) and (\ref{Ftilde}) respectively. The action (\ref{Sdis}) is invariant under the lattice gauge 
transformation
\be
\label{gaugedis}
\tilde{\varphi}(x) \rightarrow e^{i\,\tilde{\alpha}(x)}\,\tilde{\varphi}(x) \hspace*{0.3cm} , \hspace*{0.3cm} 
\tilde{A}_\mu\left(x + \frac{\hat{\mu}}{2}\right) \rightarrow \tilde{A}_\mu\left(x + \frac{\hat{\mu}}{2}\right) 
+ \frac{1}{e}\,\partial^+_\mu \tilde{\alpha} 
\ee
which is a discretized version of the gauge transformation in the continuum. This ensures in particular that the discretized equations of motion derived form (\ref{Sdis}) lead to the conservation of the discretized version of Gauss constraint~\cite{chern}. These equations are derived in Appendix~\ref{AppLattice}, where we show that they reproduce the continuum equations up to $\mathcal{O}(dx^2)$ and $\mathcal{O}(dt^2)$ accuracy in the lattice spacing $dx$ and timestep $dt$. 

On the other hand, this property would be lost by a naive discretization of the equations of motion 
(\ref{hcont}, \ref{unonTT}) for GW. The non-TT tensor perturbations $\tilde{u}_{ij}(x)$ are defined at the lattice points (as the scalar fields) and their discretized equation of motion is
\be
\label{hdis}
\partial_{\mu}^{+} \partial^{+ \mu} \tilde{u}_{ij} = 16\pi G\,\tilde{\Pi}_{ij} \ .
\ee 
Contrary to the equations of motion of the scalar and gauge fields, the lattice expressions $\partial^+_\mu$, $D^+_\mu$ 
and $\tilde{F}_{\mu \nu}$ reproduce their continuum analog up to $\mathcal{O}(dx,dt)$ only as $dx, dt \rightarrow 0$. Therefore, the same would 
be true for the RHS of Eq.~(\ref{hdis}) if we use these lattice expressions to calculate the different source terms 
(\ref{TnonTT}) of 
$\tilde{\Pi}_{ij}$. Instead of this, we have to construct new lattice expressions that reproduce their continuum analog 
up to $\mathcal{O}(dx^2,dt^2)$ accuracy and which lead to a stress-energy tensor that is still invariant under the gauge transformation (\ref{gaugedis}). The details of this procedure are derived in Appendix~\ref{AppLattice} and the final expression for $\Pi_{ij}$ is given in Eq.~(\ref{Tdis}). 

\begin{figure}[htb]
\begin{tabular}{cc}
\begin{minipage}[t]{8cm}
\begin{center}
\includegraphics[width=8cm]{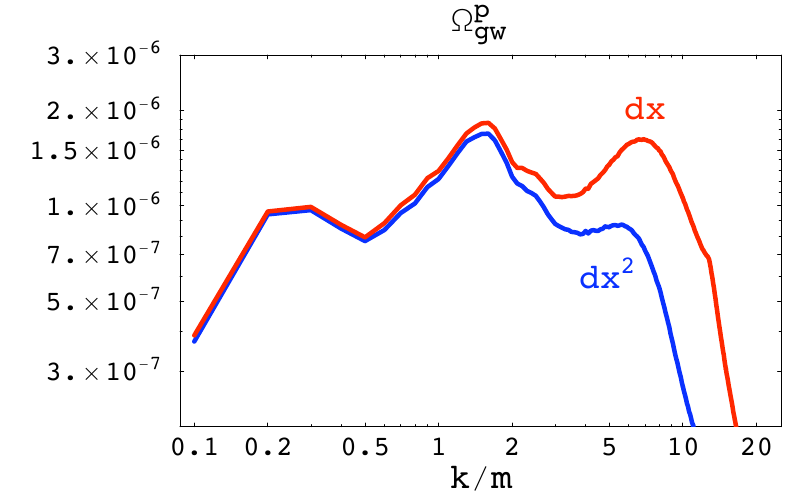}
\caption{Example of GW spectra calculated with $\mathcal{O}(dx)$ (red) and $\mathcal{O}(dx^2)$ (blue) accuracy 
(see Appendix~\ref{AppLattice} for details), for the model (\ref{eq:HybridPotential},\ref{eq:lagrangianGauge}). The two spectra were obtained 
with the same model and lattice parameters and they were output at the same moment of time ($mt = 100$).}
\label{dxVsdx2}
\end{center}
\end{minipage}&
\hspace*{1.5cm}
\begin{minipage}[t]{8cm}
\begin{center}
\includegraphics[width=8cm]{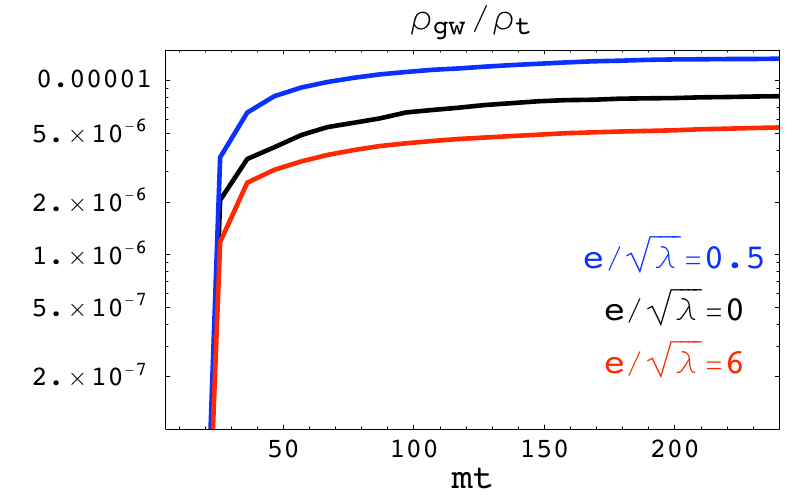}
\caption{Evolution with time of the fraction of energy density in GW during preheating, 
$\rho_{\mathrm{gw}}/\rho_{\mathrm{tot}}$, for the model (\ref{eq:HybridPotential},\ref{eq:lagrangianGauge}) with 
$e/\sqrt{\lambda} = 6$ (red), $0.5$ (blue) and $0$ (black). The last case simply corresponds to the 
model without gauge field. The other parameters were $v = 10^{-3}\,\Mp$, $\lambda = g^2 / 2 = 10^{-4}$ 
and $V_c = 0.024$.}
\label{rhogw}
\end{center}
\end{minipage}
\end{tabular}
\end{figure}

Note that the calculation of GW production with $\mathcal{O}(dx^2)$ accuracy ensures a much better control on the UV part of the GW spectra. This is illustrated in Fig.~\ref{dxVsdx2}, where we compare the GW spectra obtained by computing the GW source term with $\mathcal{O}(dx)$ and $\mathcal{O}(dx^2)$ accuracy, for the same lattice and model parameters. The two spectra agree well in the IR, but the $\mathcal{O}(dx)$ spectrum (in red) displays a larger (unphysical) growth in the UV. 

Finally, we conclude this section by showing in Fig.~\ref{rhogw} how the total energy density in GW is accumulated with time during preheating in the model (\ref{eq:HybridPotential},\ref{eq:lagrangianGauge}) for typical values of the parameters. As anticipated in section \ref{SecTheory}, GW production eventually saturates at late times after symmetry breaking. Another observation that we can already make form Fig.~\ref{rhogw} is that the total energy density in GW varies in a non-monotonic way with the ratio of the gauge coupling and the Higgs' self-coupling $e/\sqrt{\lambda}$. 
For $e/\sqrt{\lambda} \sim 1$, the gauge field leads to higher GW energy density than in the case with only scalar fields. This does not result from the mere addition of an extra field, but rather from an increase of the terms sourcing GW due to the dynamics of the coupled system of scalar and gauge fields. As we will see in section \ref{SecPosition}, a crucial role in the dynamics of preheating in the model (\ref{eq:HybridPotential},\ref{eq:lagrangianGauge}) is played by cosmic string configurations of the fields in position space. When $e/\sqrt{\lambda} \sim 1$, strings of the Higgs field and strings of the gauge field have the same width and lie on top of each other. In that case, the different source terms in 
(\ref{Tcont}) add to each others at the position of the strings. As we will now discuss, not only the amplitude of the GW signal is affected by the gauge field, but also its spectral properties.


\section{Scalar, Vector and Gravity Waves Spectra}
\label{SecSpectra}

In this section we present our results for the GW spectra produced by the system of scalar and 
gauge fields at preheating in the model (\ref{eq:HybridPotential}, \ref{eq:lagrangianGauge}). In order to highlight the 
consequences of the presence of gauge fields, let us first quickly review the gross 
features of the GW spectra produced from preheating in models invovling only scalar fields. In 
that case, the spectra of the scalar fields amplified by preheating are usually strongly peaked 
around some typical momentum $k_*$, which depends on the particular model considered and can usually 
be calculated analytically as a function of the parameters. The final GW spectrum depends essentially 
on this typical scale $k_*$, with a peak frequency and amplitude today that can be estimated according 
to 
\be
\label{fstar}
f_* \,\approx\, \frac{k_*}{\rho_p^{1/4}}\,4 \times 10^{10}\, {\rm Hz} \hspace*{0.5cm} , \hspace*{0.5cm}
h^2 \Omega^*_{\rm gw} \,\approx\, 10^{-6}\, \left(\frac{H_p}{k_*}\right)^2 \ ,
\ee
where $H_p$ and $\rho_p$ are the Hubble parameter and the total energy density at preheating
when gravity waves are produced. The factor $10^{-6}$ arises  from  the redshift of the GW radiation. 
In configuration space, $R_* \sim 1/k_*$ corresponds to the typical size of the ``bubble-shaped" 
fluctuations of the scalar fields amplified by preheating. Not surprisingly, similar estimates 
hold for the GW produced by bubble collisions in first order phase transitions, where in that case 
$R_*$ is the typical size of the bubbles when they collide. 

As we will see, a major consequence of the presence of gauge fields is to introduce, in addition to 
$k_*$, new characteristic scales in the problem, which will be inherited by the final GW spectra. We 
will therefore first study the appearance of these scales in the spectra of the scalar and gauge fields 
themselves, before considering their consequences for the GW spectra. In the next section, we will 
see how these new scales arise from the dynamics of string-like spatial configurations of the scalar and 
gauge fields. Of course, at the practical level, the presence of different scales in the problem makes 
numerical simulations more difficult, since each scale has to be resolved efficiently in a single 
simulation. One can tune the parameters in such a way that the different scales coincide with each 
other, but important consequences of the gauge field can then be missed. It was therefore important 
for us to develop a lattice formulation accurate up to second order in the lattice spacing $dx$ 
(see Appendix \ref{AppLattice}), as it allowed us to obtain reliable results for the UV behavior of 
the GW spectra while keeping a higher IR resolution as compared to a calculation accurate up to 
$\O(dx)$ only. Nevertheless, we will naturally be able to simulate only a restricted region of the 
parameter space.  We will therefore study in some detail how our results vary with the model parameters, 
in order to extrapolate them to other regions of the parameter space.

As discussed in section \ref{SecModel}, the model involves five independent parameters: the Higgs  
VEV $v$, its self-coupling $\lambda$, its coupling to the inflaton $g$, the gauge coupling constant 
$e$ and the initial velocity of the inflaton condensate at the critical point $V_c$ (\ref{defVc}). 
The GW produced in this model without gauge field were first studied in \cite{GF,GFS,DBFKU} and more 
in detail, exploring the parameters space, in~\cite{DFKN}. Without gauge field ($e=0$), the GW spectra 
are already very sensitive to the remaining parameters. In general, however, very small values of the 
coupling constants $\lambda$ and/or $g^2$ are required for these GW to have a sufficiently small frequency 
today to be observable. Neglecting the expansion of the universe (which is a good approximation for preheating 
after hybrid inflation, unless $v$ is very high), the VEV $v$ can be scaled out of the field equations and of 
the initial conditions by suitable redefinition of the fields and variables, so the dependence of physical 
quantities on $v$ is known exactly. For the GW spectrum redshifted into present-day variables, one finds 
that the GW frequencies do not depend on $v$ at all~\footnote{Despite the fact that different values of $v$ 
lead to different values of the energy density $\rho = \lambda v^4/4$ during inflation and preheating.}, while 
the GW energy density scales as $h^2 \Omega_{gw} \propto Gv^2$. These scalings with $v$ are preserved in the presence 
of gauge fields and we will simply take $v = 10^{-3}\,\Mp$ throughout this section. Depending on the remaining 
parameters $\lambda$, $g$ and $V_c$, three different dynamical regimes of GW production from preheating after 
hybrid inflation were identified in Ref.~\cite{DFKN}. In each regime, the scale $k_*$ and the resulting 
GW spectra vary in a very different way with the parameters $\lambda$, $g$ and $V_c$. As we will see, the 
effects of the gauge field may also depend on which regime is considered. Our main interest here is on 
the consequences of the gauge field for GW production and, as could be easily expected, 
an important parameter in this respect is now the ratio $e^2/\lambda$. 
Note that, in a regime where a very small $\lambda$ is required for the GW to fall in an observable frequency 
range, the ratio $e^2/\lambda$ may be huge.  

\begin{figure}[htb]
\begin{tabular}{cc}
\begin{minipage}[t]{8cm}
\begin{center}
\includegraphics[width=8cm]{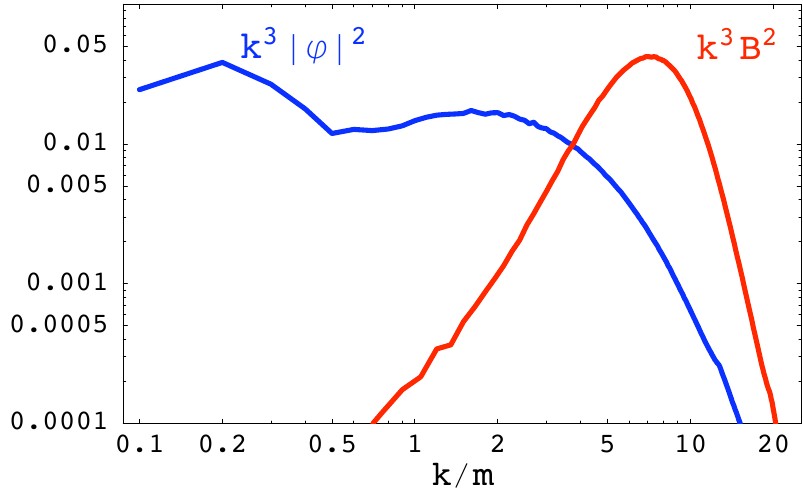}
\caption{Spectra $k^3\,|\varphi|^2$ and $k^3\,|B|^2$ of the Higgs and magnetic fields at $mt = 250$ for 
$\lambda = g^2 / 2 = 10^{-4}$, $V_c = 0.024$ and $e/\sqrt{\lambda} = 8$. Here the normalization of the 
amplitude of the two spectra is arbitrary and has been chosen only for convenience. The results were checked 
with $N=256$ simulations with $k_{IR} / m = 0.075$, $0.1$ and $0.15$.}
\label{k3X2B2}
\end{center}
\end{minipage}&
\hspace*{1.5cm}
\begin{minipage}[t]{8cm}
\begin{center}
\includegraphics[width=8cm]{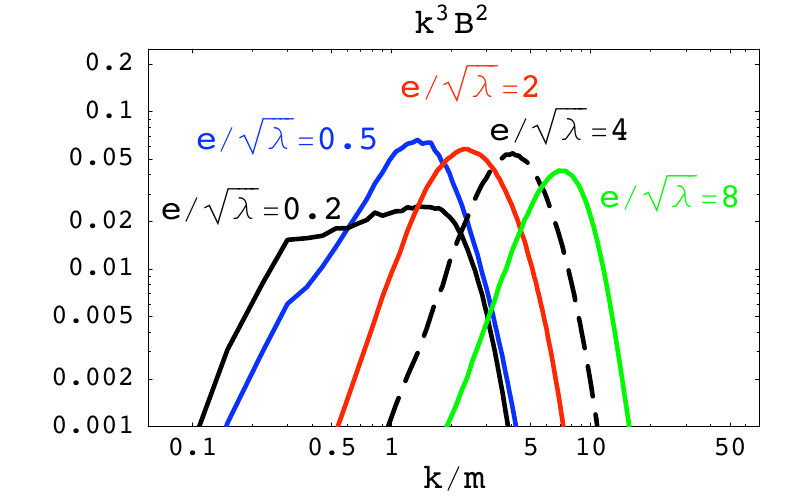}
\caption{Spectrum of magnetic energy density per logarithmic frequency interval divided by the total energy 
density, $\frac{1}{\rho_t}\,\frac{d\rho_{\mathrm{mag}}}{d\mathrm{ln}k} \, \propto k^3\,|B|^2$, at $mt = 250$
for different values of $e/\sqrt{\lambda}$. From left to right, $e/\sqrt{\lambda} = 0.2$ (black), $0.5$ (blue), 
$2$ (red), $4$ (black, dashed) and $8$ (green). The other parameters are the same as in Fig.~\ref{k3X2B2}.}
\label{k3B2eL}
\end{center}
\end{minipage}
\end{tabular}
\end{figure}

Let us first study the consequences of the gauge field for $\lambda \sim g^2$ and a significant initial 
velocity $V_c$. As far as the scalar sector is concerned, this is the easiest case to simulate as 
different dynamical scales are of the same order of magnitude. In that case, the scalar fields are 
amplified with a typical momentum $k_* \sim V_c^{1/3}\,m$, where $m = \sqrt{\lambda}\,v = m_\varphi / \sqrt{2}$ 
is the mass of the Higgs' fluctuations (divided by $\sqrt{2}$) around the minimum of the potential. In 
Fig.~\ref{k3X2B2}, we show the spectra $k^3\,|\varphi|^2$ and $k^3\,|B|^2$ of the Higgs and magnetic fields 
at late time, $mt \sim 250$ (when the field distributions have saturated and evolve very slowly with time), 
for $\lambda = g^2 / 2 = 10^{-4}$, $V_c = 0.024$ and $e/\sqrt{\lambda} = 8$. The main observation is 
that the two spectra are peaked around well distinct scales, $k \sim V_c^{1/3}\,m$ for the Higgs and 
the vector mass $k \sim e\,v$ for the magnetic field. This is a rather unusual situation in the context 
of preheating with only scalar fields, where mode to mode interactions tend to smooth the differences in 
the spectra of the different fields. In the present case, the vector mass is the typical scale for the 
width and interactions of string-like spatial structures of the gauge field, see the next section. We also 
note a ``UV bump" in the Higgs' spectrum, which is absent without gauge fields. As we will see, these 
features will be imprinted in the resulting GW spectrum. We found that the spectra of the gauge field 
are always peaked around its mass as long as it is more massive than the Higgs. This is illustrated in 
Fig.~\ref{k3B2eL}, where we plot the spectra of the magnetic energy density for different ratios 
of the vector and Higgs' masses, $m_A / m_\varphi = e / \sqrt{2\lambda}$. The spectra 
of the electric energy density behave in a similar way. This ratio plays a crucial role in the theory of 
cosmic strings (in the Abelian Higgs model), distinguishing between Type I ($m_A > m_\varphi$) and Type II 
($m_A < m_\varphi$) strings. For instance, the dynamics of multivortex solutions is governed by the fact that 
the interactions between vortices is attractive for $m_A > m_\varphi$ and repulsive for $m_A < m_\varphi$. Here 
we note that the spectrum of the gauge field tends to be peaked around the greatest of these scales, i.e. 
$k \sim \sqrt{\lambda} v$ if $m_\varphi > m_A$ and $k \sim e v$ if $m_\varphi < m_A$. Note however the 
``IR features" of the spectra for $m_\varphi > m_A$. 

\begin{figure}[htb]
\begin{tabular}{cc}
\begin{minipage}[t]{8cm}
\begin{center}
\includegraphics[width=8cm]{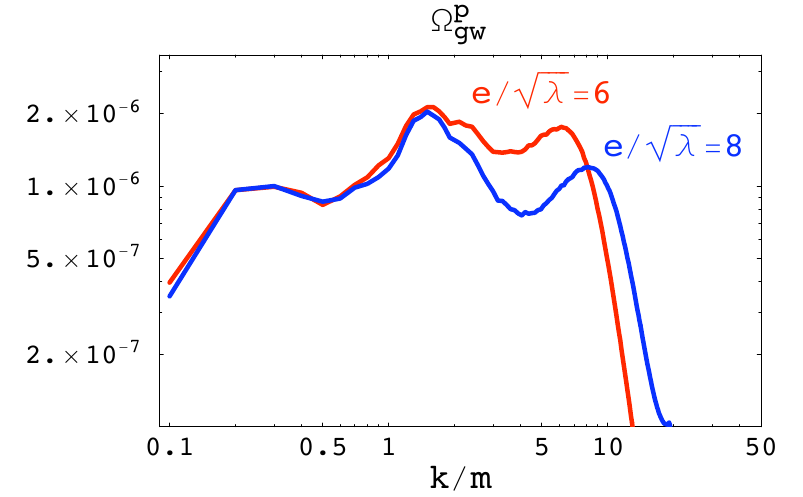}
\caption{GW spectra (\ref{SpecP}) for $e/\sqrt{\lambda} = 6$ (red) and $8$ (blue) at $mt = 250$. 
The other parameters are the same as in Fig.~\ref{k3X2B2}. The results were checked 
with $N=256$ simulations with $k_{IR} / m = 0.075$, $0.1$ and $0.15$.}
\label{GWeL68}
\end{center}
\end{minipage}&
\hspace*{1.5cm}
\begin{minipage}[t]{8cm}
\begin{center}
\includegraphics[width=8cm]{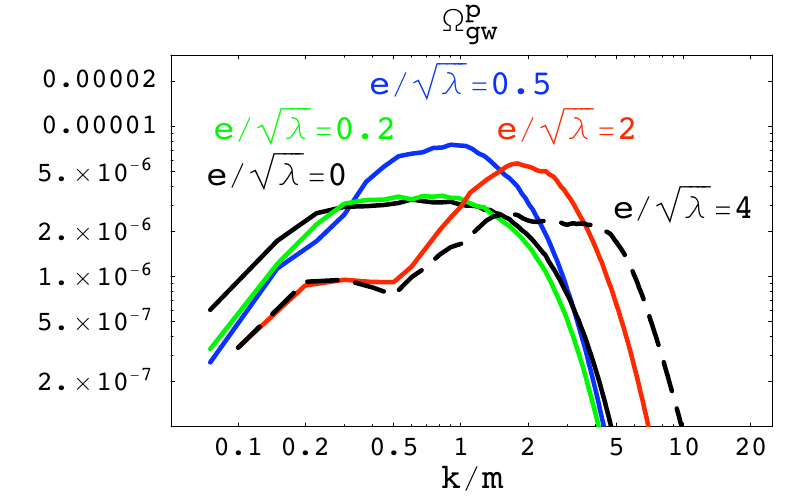}
\caption{Same as Fig.~\ref{GWeL68} for $e/\sqrt{\lambda} = 0$ (black), $0.2$ (green), $0.5$ (blue), $2$ (red) and $4$ (black, dashed).}
\label{GWeLsmall}
\end{center}
\end{minipage}
\end{tabular}
\end{figure}

The presence of well-distinct characteristic scales for the scalar and gauge fields leads to specific signatures 
in the resulting GW spectra, as illustrated in Fig.~\ref{GWeL68}. In order to compare the GW and matter fields' 
spectra, we will first consider the spectrum of energy density in GW per logarithmic frequency interval
\be
\label{SpecP}
\Omega_{\mathrm{gw}}^p(k) = \left(\frac{1}{\rho_t}\,\frac{d\rho_{\mathrm{gw}}}{d\mathrm{ln} k}\right)_p
\ee
as a function of the wave-number $k$, both quantities being evaluated at the time of GW production 
(i.e. during preheating). Present-day redshifted spectra will be considered later on. We see in 
Fig.~\ref{GWeL68} the presence of three distinct peaks in the GW spectrum for $e \gg \sqrt{\lambda}$: 
an IR peak around $k \sim 0.25\,m$, a middle peak located around the Higgs mass, $k \sim m$, and a UV peak 
located around the vector mass $k \sim e\,v$ (i.e. $k / m \sim e/\sqrt{\lambda}$). We never encountered such 
features in models without gauge field, where the GW spectra are usually peaked around a single frequency 
(see e.g. the black spectrum in Fig.~\ref{GWeLsmall}) even when different scales are present in the model. 
Contrary to the UV peak, the position and amplitude of the IR and middle peaks in Fig.~\ref{GWeL68} are 
independent of $e/\sqrt{\lambda}$, as long as $e \gg \sqrt{\lambda}$. We will see how they vary with the other 
parameters later on. The frequency of the IR peak tends to be smaller than in the case without gauge field, 
see Figs.~\ref{GWeL68} and \ref{GWeLsmall}. When $e \sim \sqrt{\lambda}$, the middle and UV peaks are superimposed. 
The resulting GW amplitude is greater than in the cases $e \gg \sqrt{\lambda}$ and $e \ll \sqrt{\lambda}$, 
see Fig.~\ref{GWeLsmall}. This is already the case for $e/\sqrt{\lambda} = 6$ in Fig.~\ref{GWeL68} (red spectrum), 
where the UV peak has a higher amplitude than for $e/\sqrt{\lambda} = 8$ (blue spectrum). Finally, for 
$e \ll \sqrt{\lambda}$ the GW spectrum becomes indistinguishable from the case without gauge field, see e.g. the 
black and green spectra in Fig.~\ref{GWeLsmall} for $e/\sqrt{\lambda} = 0$ and $0.2$ respectively.

Up to now, we have considered the spectra of the scalar, vector and gravity waves at late times, when the distributions
have saturated and evolve very slowly with time. It is interesting to see how they build up with time, as for instance 
the different peaks in the GW spectra of Fig.~\ref{GWeL68} are actually produced at different moments of time. 
Furthermore, although the spectra of the magnetic or electric energy density are useful to single out the vector 
sector, the covariant gradient energy of the Higgs contributes to a greater fraction of the total energy density, see 
Fig.~\ref{rhocomp}, and similarly the source of GW is dominated by the covariant gradient terms in (\ref{Tcont}). 
By analogy with (\ref{SpecP}), we can consider the spectrum of the covariant gradient energy density of the Higgs,
per logarithmic frequency interval, divided by the total energy density
\be
\label{SpecCov}
\Omega_{\mathrm{cov}}^p(k) = \left(\frac{1}{\rho_{\mathrm{tot}}}\,\frac{d\rho_{\mathrm{cov}}}{d\mathrm{ln} k}\right)_p
\ee
where $\rho_{\mathrm{cov}} = \langle D_i \varphi\,\left(D_i \varphi\right)^* \rangle$ is the covariant gradient energy density of the Higgs. 

\begin{figure}[htb]
\begin{center}
\begin{tabular}{cc}
\includegraphics[width=5.5cm]{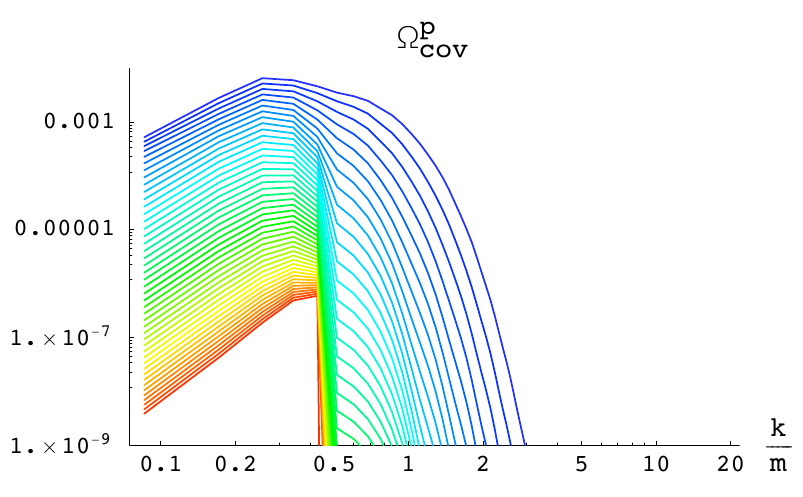} \hspace*{0.5cm}
\includegraphics[width=5.5cm]{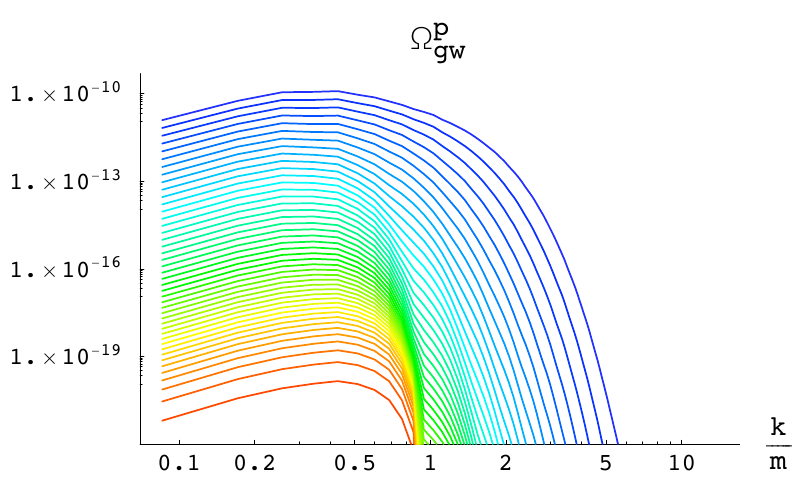}\\
\includegraphics[width=5.5cm]{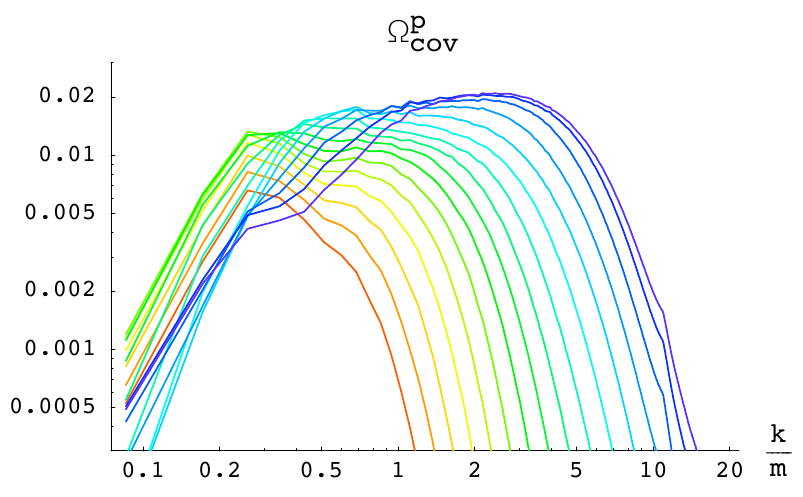} \hspace*{0.5cm}
\includegraphics[width=5.5cm]{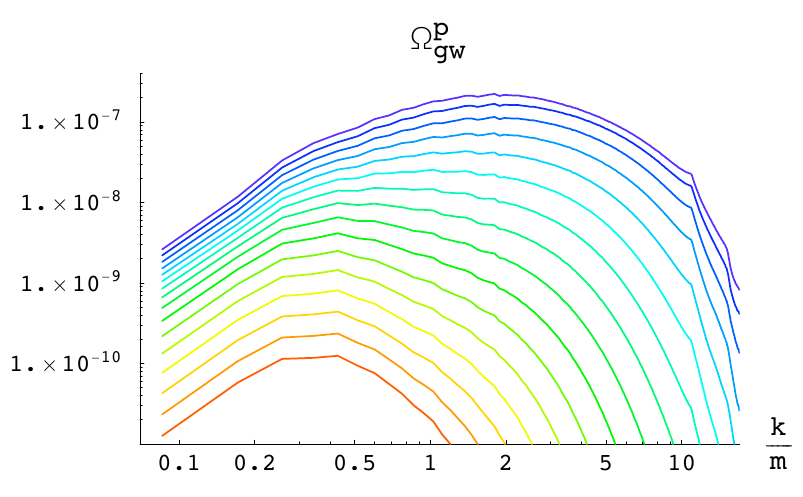}\\
\includegraphics[width=5.5cm]{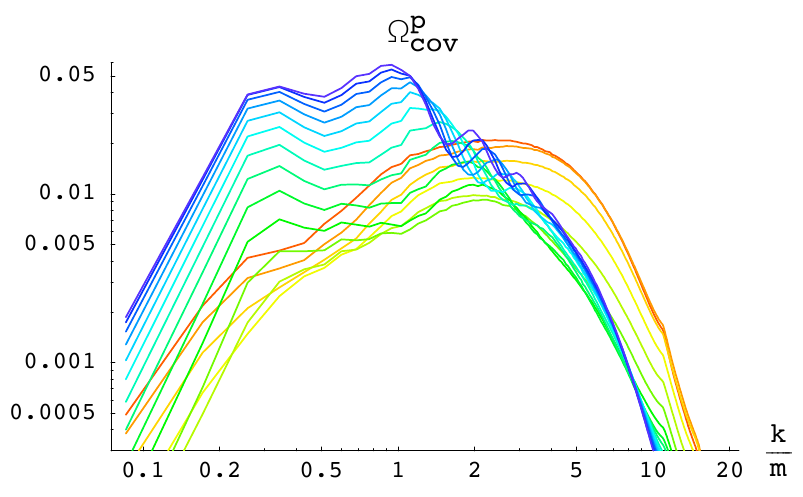} \hspace*{0.5cm}
\includegraphics[width=5.5cm]{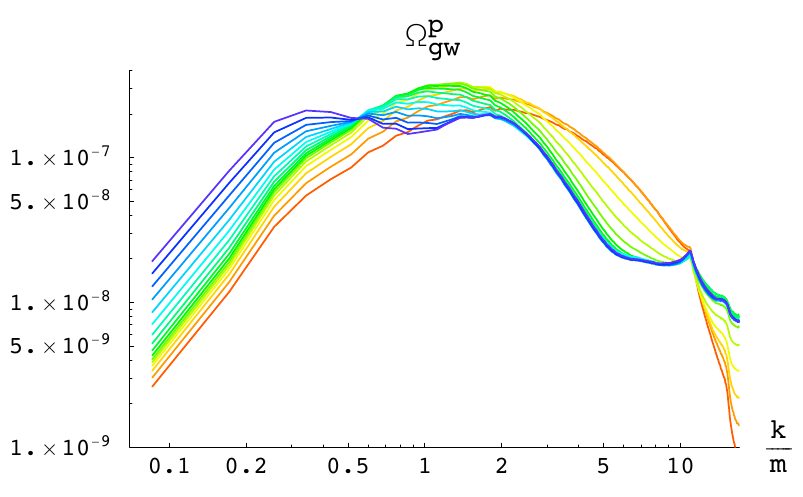}\\
\includegraphics[width=5.5cm]{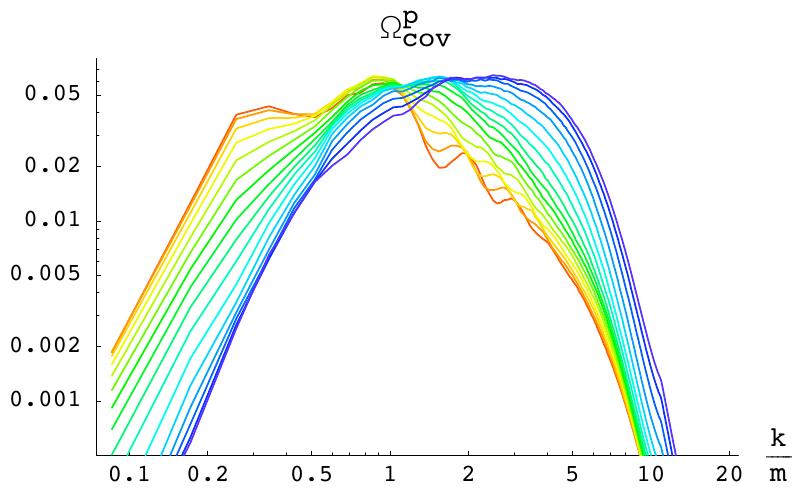} \hspace*{0.5cm}
\includegraphics[width=5.5cm]{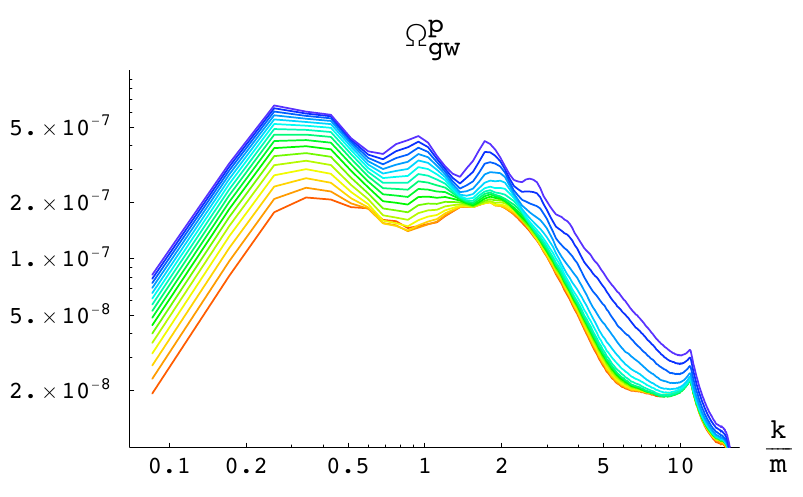}\\
\includegraphics[width=5.5cm]{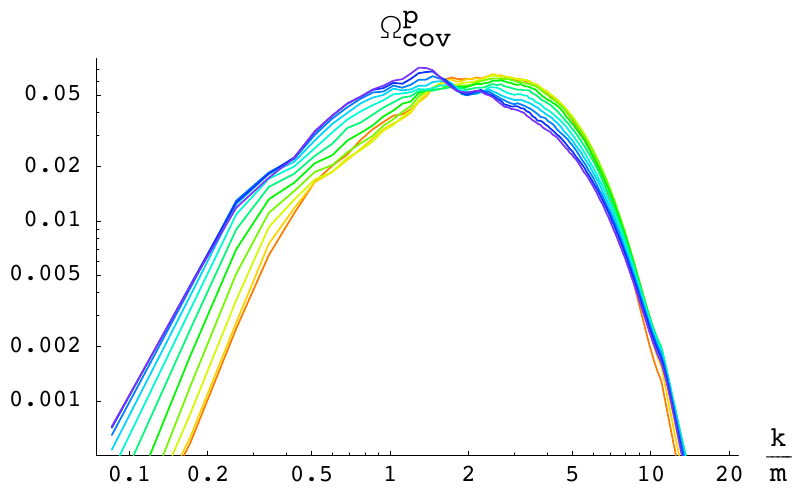} \hspace*{0.5cm}
\includegraphics[width=5.5cm]{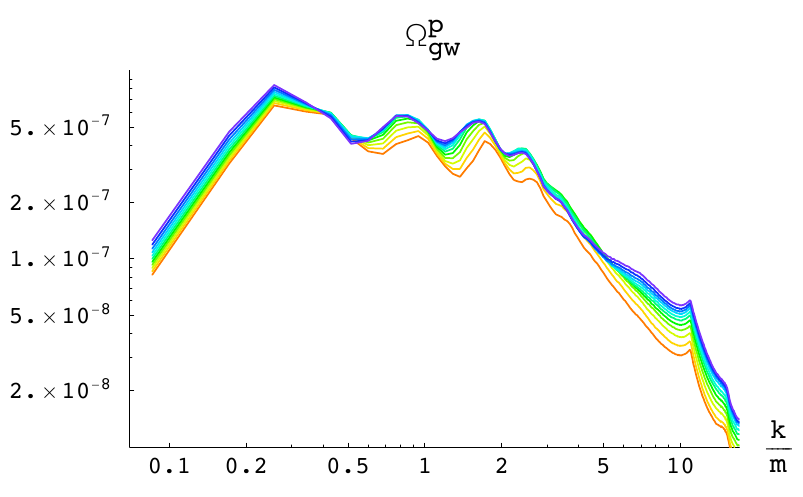}\\
\includegraphics[width=5.5cm]{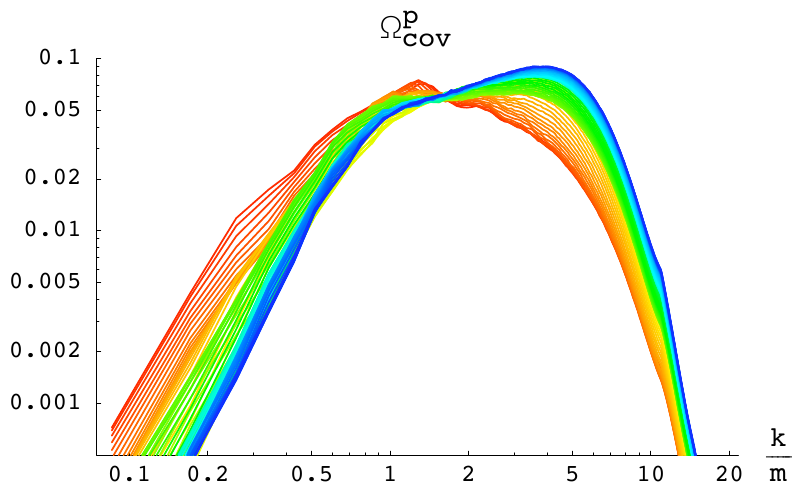} \hspace*{0.5cm}
\includegraphics[width=5.5cm]{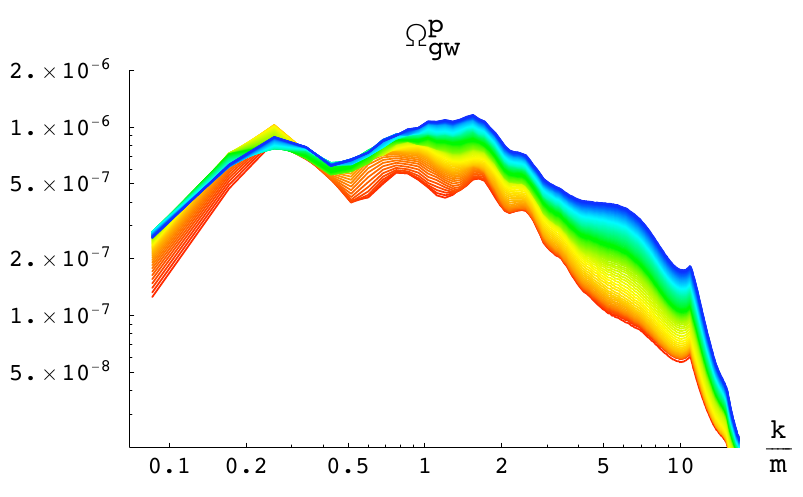}\\
\end{tabular}
\caption{Evolution with time of the spectrum of covariant gradient energy density of $\varphi$ (\ref{SpecCov}) (left pannels) 
and of the spectrum of energy density in GW (\ref{SpecP}) (right pannels) for $\lambda = g^2 / 2 = 10^{-4}$, $V_c = 0.024$ 
and $e/\sqrt{\lambda} = 6$. The spectra are shown between $mt = 5$ and $15.5$ on the first line, between $mt = 15.5$ and 
$19$ on the second line, between $mt = 19$ and $23$ on the third line, between $mt = 23$ and $26.5$ on the fourth line, between $mt = 26.5$ and $29$ on the fifth line and between $mt = 29$ and $50$ on the sixth line. In each case, the spectra are output every $mt = 0.24$ and red spectra correspond to earlier moments of time while blue spectra correspond to later moments of time (the colors go from red to yellow, green and blue as time evolves).}
\label{dteL6}
\end{center}
\end{figure}

The evolution with time of $\Omega_{\mathrm{cov}}^p(k)$ is shown in the left pannels of Fig.~\ref{dteL6}, together with 
the evolution of the spectrum of energy density in GW (\ref{SpecP}) in the right pannels. Each line in Fig.~\ref{dteL6} corresponds to a different interval of time. From $mt = 5$ to $15.5$ (first line), both spectra $\Omega_{\mathrm{cov}}^p$ and $\Omega_{\mathrm{gw}}^p$ are peaked around $k_* \sim V_c^{1/3}\,m$ and their amplitude increases exponentially with time as the tachyonic instability amplifies the Higgs' fluctuations. From 
$mt = 15.5$ to $19$ (second line), the peak of the spectra moves clearly towards UV, while their amplitude still increases. During this interval of time, the amplitude of the Higgs' fluctuations reach its vev in more and more regions of space and 
start to oscillate back in the potential. This leads to the collision of bubbles of the fluctuations and thiner and thiner regions of space where the Higgs is locally small and where it is energetically more favorable for the gauge field to be localized, as we will see in Section~\ref{SecPosition}. 

Next, from $mt = 19$ to $23$ (third line in Fig.~\ref{dteL6}), the spectrum $\Omega_{\mathrm{cov}}^p$ moves back towards the IR, where two peaks form around $k_* \sim 0.25\,m$ and $k_* \sim m$. During this interval of time, the fluctuations of the Higgs have oscillated back in the potential and become small in more and more regions of space, where the gauge field tends to be localized. Indeed, we will see in the next section that the string-like configurations of the gauge field get significantly fatter during this interval of time, the increase of their width corresponding to the shift of the spectrum towards IR that we observe here. Meanwhile, the IR part of the GW spectrum starts to increase during this interval of time. Then from $mt = 23$ to $26.5$ (fourth line in Fig.~\ref{dteL6}), the two IR peaks in $\Omega_{\mathrm{cov}}^p$ disappear and the spectrum oscillates back towards UV. During this time interval, the amplitude of the Higgs' fluctuations goes back to its vev in more and more regions of space. We will see in the next section that the string-like configurations of the gauge field tend to fragment into smaller structures during this interval of time. We see here that the IR part of the GW spectrum grow significantly during this period. In fact, the IR peak of $\Omega_{\mathrm{gw}}^p$ around $k_* \sim 0.25\,m$ reaches almost its final amplitude during this interval of time, while the UV part of the spectrum will still significantly increase. This IR peak is inherited from the peak of $\Omega_{\mathrm{cov}}^p$ at the same location that formed during the previous interval of time and which has now disappeared. 

Finally, $\Omega_{\mathrm{cov}}^p$ moves slightly back towards IR from $mt = 26.5$ to $29$ (fifth line in Fig.~\ref{dteL6}) with a new peak forming around $k_* \sim m$, before going back to the UV from $mt = 29$ to $50$ (sixth line in Fig.~\ref{dteL6}). During this interval of time, the GW spectrum increases significantly around $k_* \sim m$. From there on the spectrum of $\Omega_{\mathrm{cov}}^p$ has saturated and it is peaked around the vector mass, $k_* \sim 6\,m$. The GW spectrum then slowly increases around this momentum to eventually reach its final form displayed in Fig.~\ref{GWeL68} 
(red curve). 

To sum up, we see that the different peaks in the GW spectrum appear at different moments of time during the process of tachyonic preheating and symmetry breaking. They can be traced back to similar features in the spectrum of the covariant gradient energy density of the Higgs. However, whereas these features disappear in $\Omega_{\mathrm{cov}}^p$, which becomes eventually peaked around the vector mass, the IR peaks remain in the GW spectrum, since GW decouple as soon as they are produced and their spectral shape remains unchanged since then.

\begin{figure}[htb]
\begin{tabular}{cc}
\begin{minipage}[t]{8cm}
\begin{center}
\includegraphics[width=8cm]{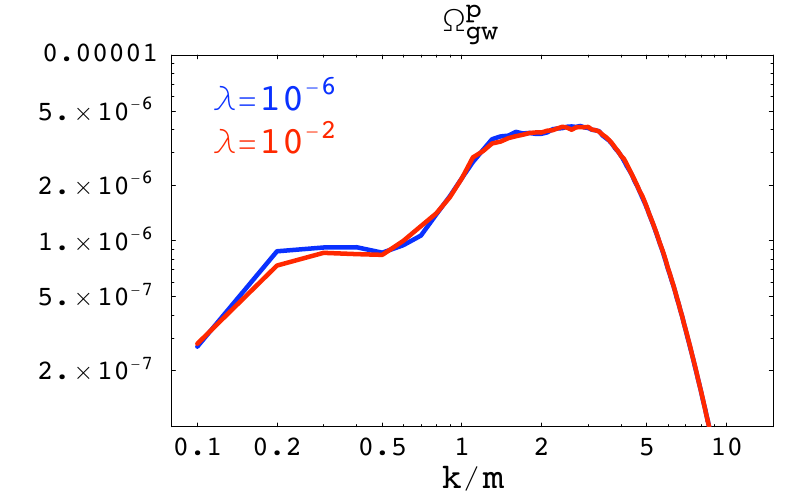}
\caption{GW spectra for $\lambda = 10^{-2}$ (red) and $\lambda = 10^{-6}$ (blue) at $mt = 300$. 
The other parameters are $V_c = 0.024$, $g^2 = 2 \lambda$ and $e = 3 \sqrt{\lambda}$.}
\label{GWlam}
\end{center}
\end{minipage}&
\hspace*{1.5cm}
\begin{minipage}[t]{8cm}
\begin{center}
\includegraphics[width=8cm]{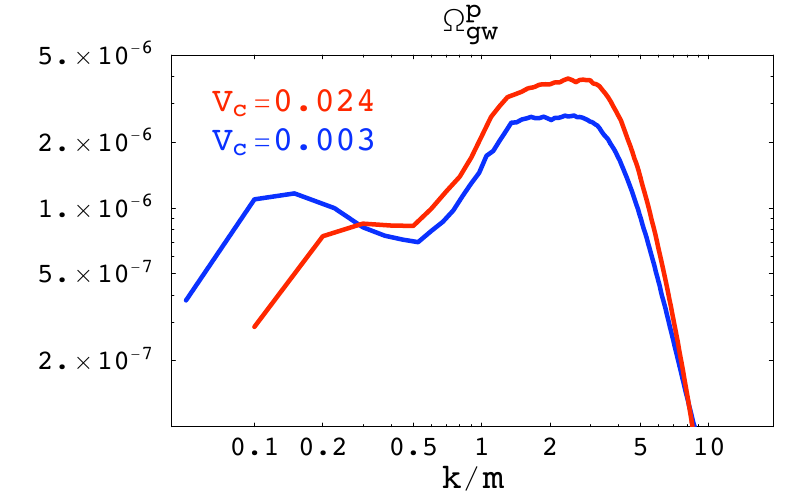}
\caption{GW spectra for $V_c = 0.024$ (red) and $V_c = 0.003$ (blue) at $mt = 250$. 
The other parameters are $\lambda = g^2 / 2 = 10^{-4}$ and $e/\sqrt{\lambda} = 3$. The results were checked 
with $N=256$ simulations with $k_{IR} / m = 0.05$, $0.075$ and $0.1$.}
\label{GWVc}
\end{center}
\end{minipage}
\end{tabular}
\end{figure}

Let us now study how the GW spectra varies with the other parameters. Without gauge field, and still in the regime $\lambda \lesssim g^2$ and significant initial velocity, the peak frequency of the GW spectrum at the time of production varies as $k \propto V_c^{1/3}\,m$ and the peak amplitude as $\Omega_{\mathrm{gw}}^p \propto V_c^{-2/3}\,v^2$, see Ref.~\cite{DFKN}. In Fig.~\ref{GWlam}, we show the GW spectra computed for $V_c = 0.024$, $g^2 = 2 \lambda$, $e = 3 \sqrt{\lambda}$ and two values of the Higgs' self-coupling: $\lambda = 10^{-2}$ and $10^{-6}$. The two spectra 
are almost on top of each others (in the units of $k/m$) so, as in the case without gauge field, the GW amplitude is independent of $\lambda$ while their frequency at the time of production varies as $k \propto m \propto \sqrt{\lambda}$, 
as far as the ratios $g^2/\lambda$ and $e^2/\lambda$ are kept constant. The dynamics and the shape of the final GW spectrum are very sensitive to these ratios of the coupling constants but not to the absolute value of $\lambda$~\footnote{Rescaling the coordinates as $x \rightarrow m\,x$ and the fields as $\phi \rightarrow \phi / v$, the coupling constant $\lambda$ drops out of the equations of motion and enters only in the initial conditions for the amplitude of Higgs' fluctuations. These have very little consequences on the shape of the final GW spectrum, as illustrated in Fig.~\ref{GWlam}.}.   

In Fig.~\ref{GWVc}, we show the GW spectra computed for $\lambda = g^2 / 2 = 10^{-4}$, $e/\sqrt{\lambda} = 3$
and two values of the initial velocity of the inflaton at the critical point: $V_c = 0.024$ and $V_c = 0.003$. The frequency of the IR peak of the GW spectrum varies exactly as $V_c^{1/3}$, as without gauge field. Its amplitude does 
not vary exactly as $V_c^{-2/3}$, which even without gauge field is only approximate anyway, but nevertheless increases as 
$V_c$ decreases. Thus we see that the IR peak of the GW spectrum, which is present when $e \gtrsim \sqrt{\lambda}$, 
varies with the parameters approximately in the same way as the GW spectrum produced without gauge field. For 
$e \gtrsim \sqrt{\lambda}$, however, extra peaks are still present around the Higgs mass $k \sim m$ and the vector mass 
$k \sim e\,v$. These are on top of each other in Figs.~\ref{GWlam} and \ref{GWVc}. We see in Fig.~\ref{GWVc} that the amplitude of at least one of these extra peaks seems to decrease when $V_c$ decreases, while the amplitude of the IR peak behaves in the opposite way. However, a quantitative estimate of the GW amplitude as a function of the intial velocity of the inflaton at the critical point would require to consider a larger range of values of $V_c$. Since the frequency of the IR peak decreases with $V_c$ while the frequency of the two other peaks remains constant for $e \gtrsim \sqrt{\lambda}$, 
it becomes soon impossible to catch accurately the different scales in a single simulation. On the other hand, for 
$e \ll \sqrt{\lambda}$ the gauge field becomes negligible and we recover the results of \cite{DFKN}.

As discussed in \cite{DFKN}, the regime with $\lambda \lesssim g^2$ and a significant intial velocity $V_c$ that we have considered up to now is the easiest one to simulate but not the most interesting one from an observational perspective. 
Without gauge field, another regime of GW production in the model (\ref{eq:HybridPotential}) occurs for 
$\lambda \lesssim g^2$ and a small initial velocity $V_c$, where the onset of preheating is driven by quantum diffusion of the fields around the critical point \cite{DFKN}. We will not study the consequences of the gauge field in this case here, but the main difference of this case compared to the previous one is in the onset of preheating, while the gauge field is amplified during the later stages. We thus expect the gauge field to have similar consequences as above.

\begin{figure}[htb]
\begin{tabular}{cc}
\begin{minipage}[t]{8cm}
\begin{center}
\includegraphics[width=8cm]{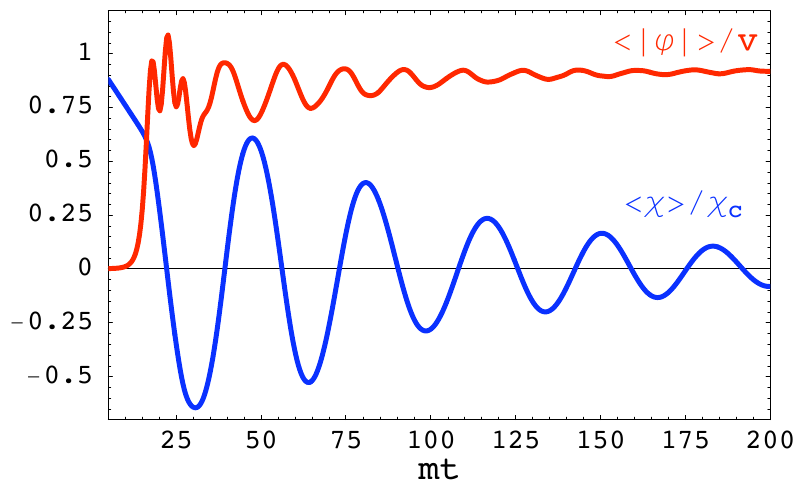}
\caption{Time evolution of the inflaton's mean $\langle \phi \rangle / \phi_c$ (blue) and of the Higgs' root mean squared 
$\langle |\varphi| \rangle / v$ (red) for $\lambda / g^2 = 16$, $e / \sqrt{\lambda} = 0.5$, $\lambda = 10^{-4}$ and 
$V_c = 0.024$. Notice the large oscilations of $\langle \phi \rangle$ as compared to Fig.~\ref{meanrms}.}
\label{meanrmslg}
\end{center}
\end{minipage}&
\hspace*{1.5cm}
\begin{minipage}[t]{8cm}
\begin{center}
\includegraphics[width=8cm]{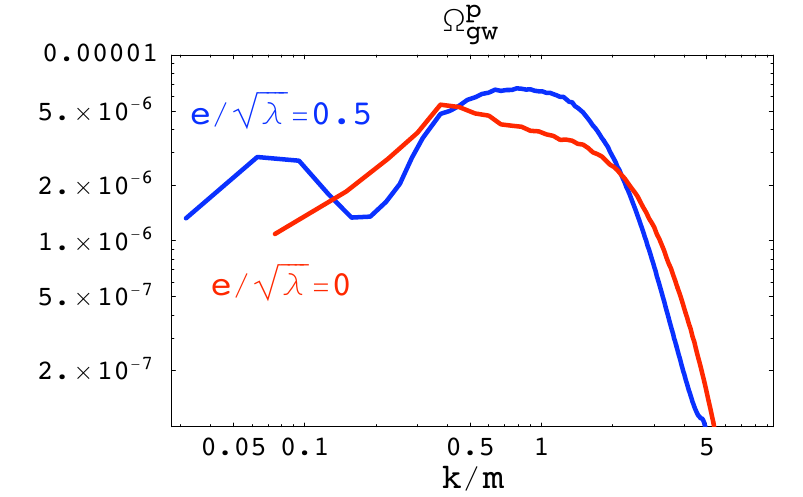}
\caption{GW spectra for $\lambda / g^2 = 16$ and $e / \sqrt{\lambda} = 0$ (red, no gauge field) and 
$e / \sqrt{\lambda} = 0.5$ (blue) at $mt = 400$. The other parameters are $\lambda = 10^{-4}$ and $V_c = 0.024$. 
The results were checked with $N=256$ simulations with $k_{IR} / m = 0.03$, $0.04$ and $0.075$.}
\label{GWlgScal}
\end{center}
\end{minipage}
\end{tabular}
\end{figure} 

\begin{figure}[htb]
\begin{tabular}{cc}
\begin{minipage}[t]{8cm}
\begin{center}
\includegraphics[width=8cm]{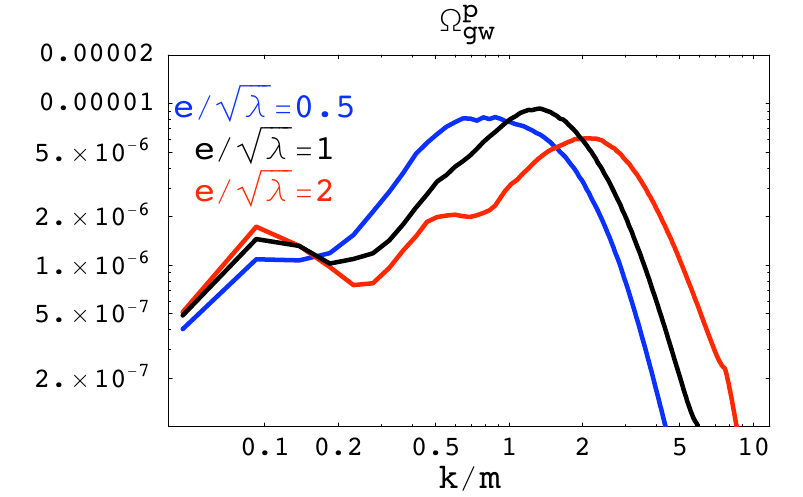}
\caption{GW spectra for $\lambda / g^2 = 8$ and $e / \sqrt{\lambda} = 0.5$ (blue), $1$ (black) 
and $2$ (red) at $mt = 350$. The other parameters are $V_c = 0.024$ and $\lambda = 10^{-4}$. The results 
were checked with $N=256$ simulations with $k_{IR} / m = 0.045$ and $0.06$.}
\label{GWlgel}
\end{center}
\end{minipage}&
\hspace*{1.5cm}
\begin{minipage}[t]{8cm}
\begin{center}
\includegraphics[width=8cm]{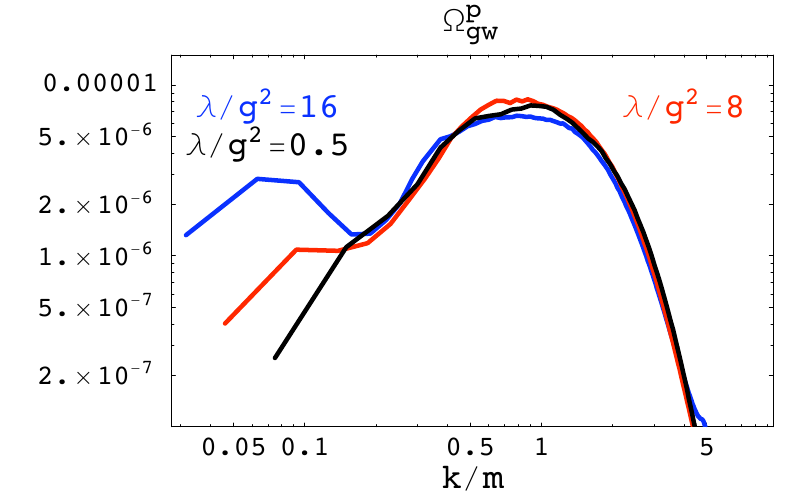}
\caption{GW spectra for $\lambda / g^2 = 0.5$ (black), $8$ (red) and $16$ (blue) at $mt = 400$.   
The other parameters are $V_c = 0.024$, $\lambda = 10^{-4}$ and $e / \sqrt{\lambda} = 0.5$. The results were 
checked with $N=256$ simulations with $k_{IR} / m = 0.03$, $0.04$ and $0.075$.}
\label{GWlg}
\end{center}
\end{minipage}
\end{tabular}
\end{figure}

Finally, the third regime of GW production identified in \cite{DFKN} corresponds to the case $g^2 \ll \lambda$. In that case, the hybrid potential is much flatter in the inflaton direction than in the Higgs direction. As a result, the inflaton condensate oscillates several times around the minimum of the potential with relatively large amplitude, see Fig.~\ref{meanrmslg}.  This leads to successive amplifications of the inflaton fluctuations by a combination of tachyonic instability and non-adiabatic resonance. In Fig.~\ref{GWlgScal}, we show the GW spectrum for $\lambda / g^2 = 16$ with (blue) and without (red) gauge field. As before, the gauge field leads to GW spectra with several peaks. We see here that the frequency of the IR peak can be significantly lower than the frequency of the peak without gauge field. This prevents us from simulating a significant range of parameters, since both the IR and UV scales have to be included in the same simulations. Nevertheless, the results are consistent with a UV peak located around the vector mass, see Fig.~\ref{GWlgel}, as was the case for $\lambda \sim g^2$. As before, we also expect the emergence of a middle peak located around the Higgs mass when $e \gg \sqrt{\lambda}$, although we could not simulate this case for $g^2 \ll \lambda$. 

In any case, the main feature seen in Fig.~\ref{GWlgScal} is the effect of the gauge field on the IR part of the GW spectrum, which can be of particular interest from the observational perspective. When $\lambda / g^2$ increases, the frequency of the IR peak decreases while its amplitude increases, see Fig.~\ref{GWlg}. This is similar to the behavior of the GW spectrum without gauge field, which could be studied more accurately for a much wider range of values of $\lambda / g^2$ in \cite{DFKN}. It is thus possible that the gauge field merely enhances this behavior but that the frequency and amplitude of the IR peak remain relatively well described by the predictions of \cite{DFKN} without gauge field. On the other hand, it is perhaps not surprising to observe specific effects of the gauge field for $g^2 \ll \lambda$. As shown in Fig.~\ref{meanrmslg}, the inflaton condensate oscillates with relatively large amplitude around the minimum of the potential in that case. When it is far away from the minimum, the minimum in the Higgs direction is at $|\varphi| < v$ and the Higgs amplitude is small in relatively large regions of space. As we will see in the next section, these regions play a crucial role as magnetic strings are produced there. We therefore expect a different dynamics of the strings when $g^2 \ll \lambda$. A detailed study of this case is certainly interesting, but it lies beyond the scope of this paper.

To conclude this section, let us summarize our results and relate them to the position space picture discussed in the next section. For $e > \sqrt{\lambda}$, the final GW spectrum can be understood as made of three different peaks, which are produced at different moments of time during the process of tachyonic preheating and symmetry breaking. A IR peak appears first, when bubbles of the Higgs start to collide and strings are formed in between the bubbles. The frequency of this IR peak tends to be smaller than the frequency of the peak of the GW spectrum produced without gauge field, but our results indicate that it varies in the same way with the model parameters. The frequency of the IR peak is set by the typical size of the bubbles of the Higgs field when they collide and by the correlation length of straight string segments at that time (these two scales are approximately equal). Next, a middle peak appears, whose frequency is set by the Higgs mass. This is the typical scale for the width and interactions of the Higgs' strings. Finally, a UV peak is formed when a significant fraction of energy has already been radiated away from the strings, see the next section. At that time, the spectrum of the gauge field is peaked around the vector mass, which is the typical scale for the width and interactions of the strings of the gauge field, and this scale sets the frequency of the UV peak of the GW spectrum.  

Once the GW are redshifted until today, their frequency is given by Eq.(\ref{f0}). We can then predict the 
present-day frequency of the three peaks as
\bea
\begin{array}{lcl}
\label{f123}
f_1 &\lesssim& f\left(g, \lambda, V_c\right) \hspace*{0.5cm} (\mbox{IR peak}) \\[2mm]
f_2 &\approx& \lambda^{1/4}\, 10^{11}\,\mathrm{Hz} \hspace*{0.5cm} (\mbox{Middle peak}) \\[2mm]
f_3 &\approx& \frac{e}{\sqrt{\lambda}}\,\lambda^{1/4}\, 10^{11}\,\mathrm{Hz} \hspace*{0.5cm} (\mbox{UV peak})
\end{array}
\eea
where the frequency $f_1$ of the IR peak depends on the parameters $g$, $\lambda$ and $V_c$ and is of the order 
of, or smaller than, the frequency of the peak of the GW spectrum produced without gauge field. We can then 
use the predictions of \cite{DFKN} 
\be
\label{IRpeak}
f\left(g, \lambda, V_c\right) \approx
\left\{
\begin{array}{l}
\lambda^{\frac{1}{4}}\,V_c^{\frac{1}{3}}\,10^{11}\,\mathrm{Hz}
\;\; \mbox{ for } \;\; g^2 \, \gtrsim \, \lambda \; \mbox{ and } \; V_c \, \gtrsim \, 500\,g^3 \\[2mm]
\lambda^{\frac{1}{4}}\,g\,10^{11}\,\mathrm{Hz}
\;\; \mbox{ for } \;\; g^2 \, \gtrsim \, \lambda \; \mbox{ and } \; V_c \, \lesssim \, 500\,g^3 \\[2mm]
\lambda^{\frac{1}{4}}\,\frac{g}{\sqrt{\lambda}}\,10^{10}\,\mathrm{Hz}
\;\; \mbox{ for } \;\; g^2 \, \ll \, \lambda
\end{array}
\right.
\ee
for the frequency of this peak. When $e \sim \sqrt{\lambda}$, the middle and UV peaks merge into a single one with 
higher amplitude. For $e \ll \sqrt{\lambda}$, the gauge field becomes negligible and the results reduce to the case 
with only scalar fields, characterized by a GW spectrum with a single peak around $f\left(g, \lambda, V_c\right)$. 
Note that, depending on the parameters, the frequencies $f_1$, $f_2$ and $f_3$ of the three peaks can differ by many 
orders of magnitude. It is of course not possible to simulate such cases on the lattice, but we expect the estimates 
(\ref{f123}) to remain valid in such cases since they result directly from the presence of well-defined characteristic scales in the problem. 

Finally, we come to the amplitude of the GW spectrum. Without gauge field, or for $e \ll \sqrt{\lambda}$, we can use the predictions of \cite{DFKN}
\be
\label{omegagw*}
h^2\,\Omega_{\mathrm{gw}}^* \sim
\left\{
\begin{array}{l}
10^{-6}\,V_c^{-2/3}\,\left(\frac{v}{M_{\mathrm{Pl}}}\right)^2 
\;\; \mbox{ for } \;\; g^2 \, \gtrsim \, \lambda \; \mbox{ and } \; V_c \, \gtrsim \, 500\,g^3 \\[2mm]
\frac{10^{-8}}{g^2}\,\left(\frac{v}{M_{\mathrm{Pl}}}\right)^2 
\;\; \mbox{ for } \;\; g^2 \, \gtrsim \, \lambda \; \mbox{ and } \; V_c \, \lesssim \, 500\,g^3 \\[2mm]
10^{-5}\,\frac{\lambda}{g^2}\,\left(\frac{v}{M_{\mathrm{Pl}}}\right)^2 
\;\; \mbox{ for } \;\; g^2 \, \ll \, \lambda
\end{array}
\right.
\ee
for the peak amplitude of the GW spectrum today. For $e \gtrsim \sqrt{\lambda}$, we could not vary the parameters 
over a sufficiently large range to study quantitatively how they affect the amplitude of each peak in the GW spectrum. However, 
our results indicate that the amplitude of the IR peak behaves roughly as in the case without gauge field and agrees within an order 
of magnitude with (\ref{omegagw*}). The amplitude of the UV peaks should then be also well-described by these estimates if a significant fraction of the total energy density is indeed efficiently converted into small-scale structures 
of the Higgs and gauge fields. When $e \sim \sqrt{\lambda}$, the middle and UV peaks are superimposed and their amplitude 
is slightly larger than in the case without gauge field. On the other hand, when $e \gg \sqrt{\lambda}$, the amplitude 
of each peak is slightly smaller than in the case without gauge field. In any case, the amplitude of the three peaks is mostly sensitive to the VEV of the Higgs and the corresponding behavior is known exactly, 
$\Omega_{\mathrm{gw}} \propto Gv^2$.


\begin{figure}[htb]
\begin{center}
\includegraphics[height=6.15cm]{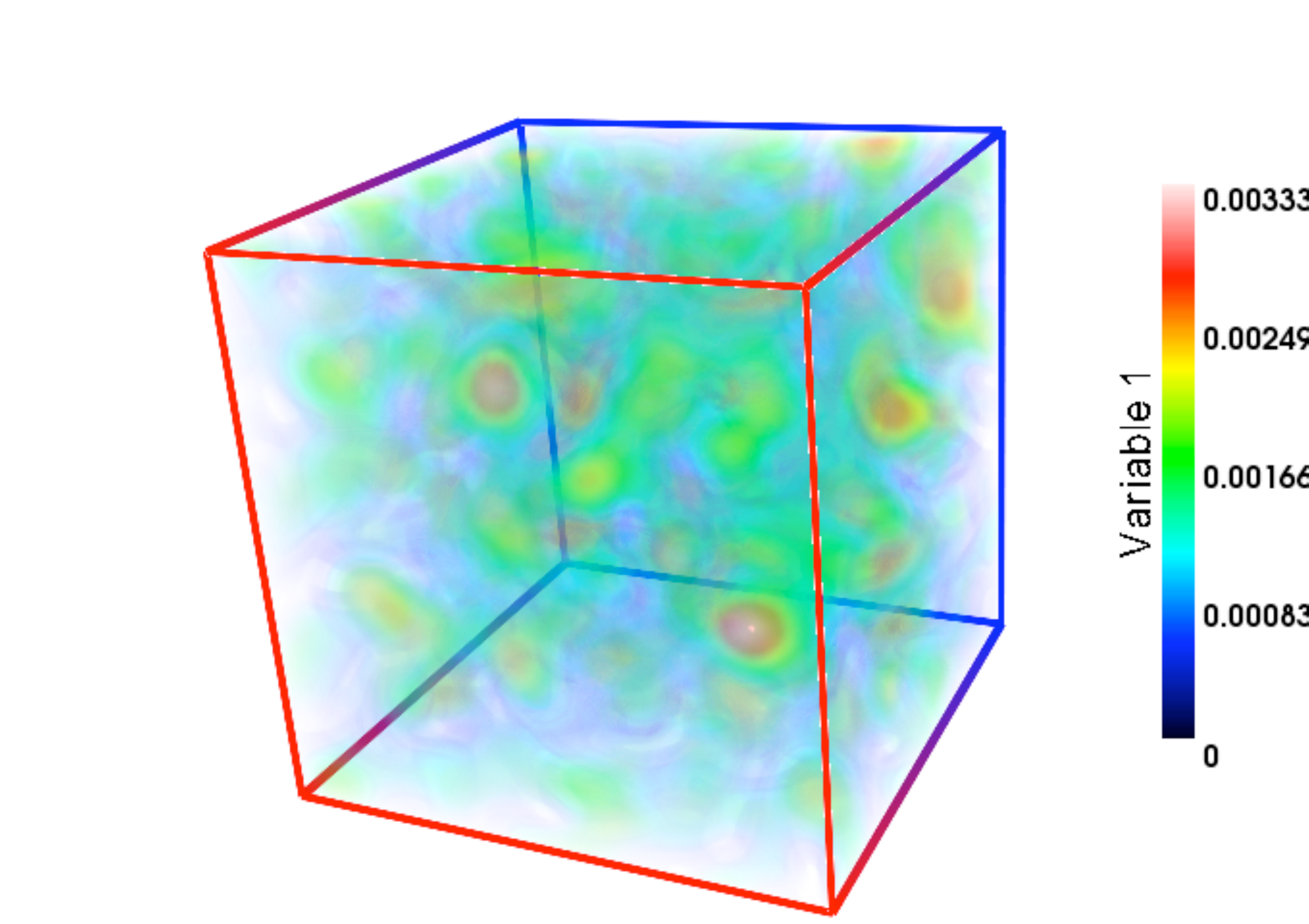}
\includegraphics[height=6.15cm]{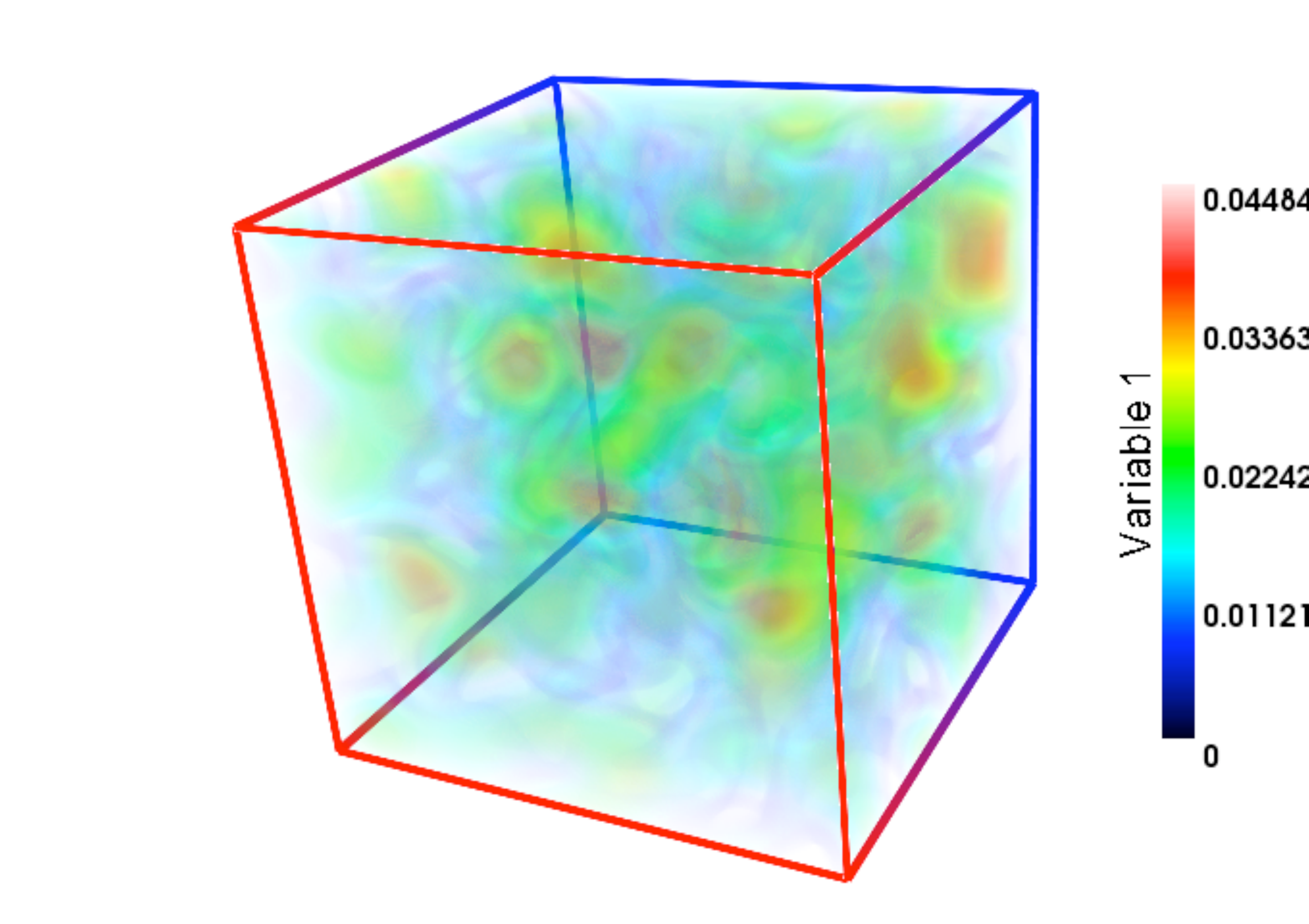}
\includegraphics[height=6.15cm]{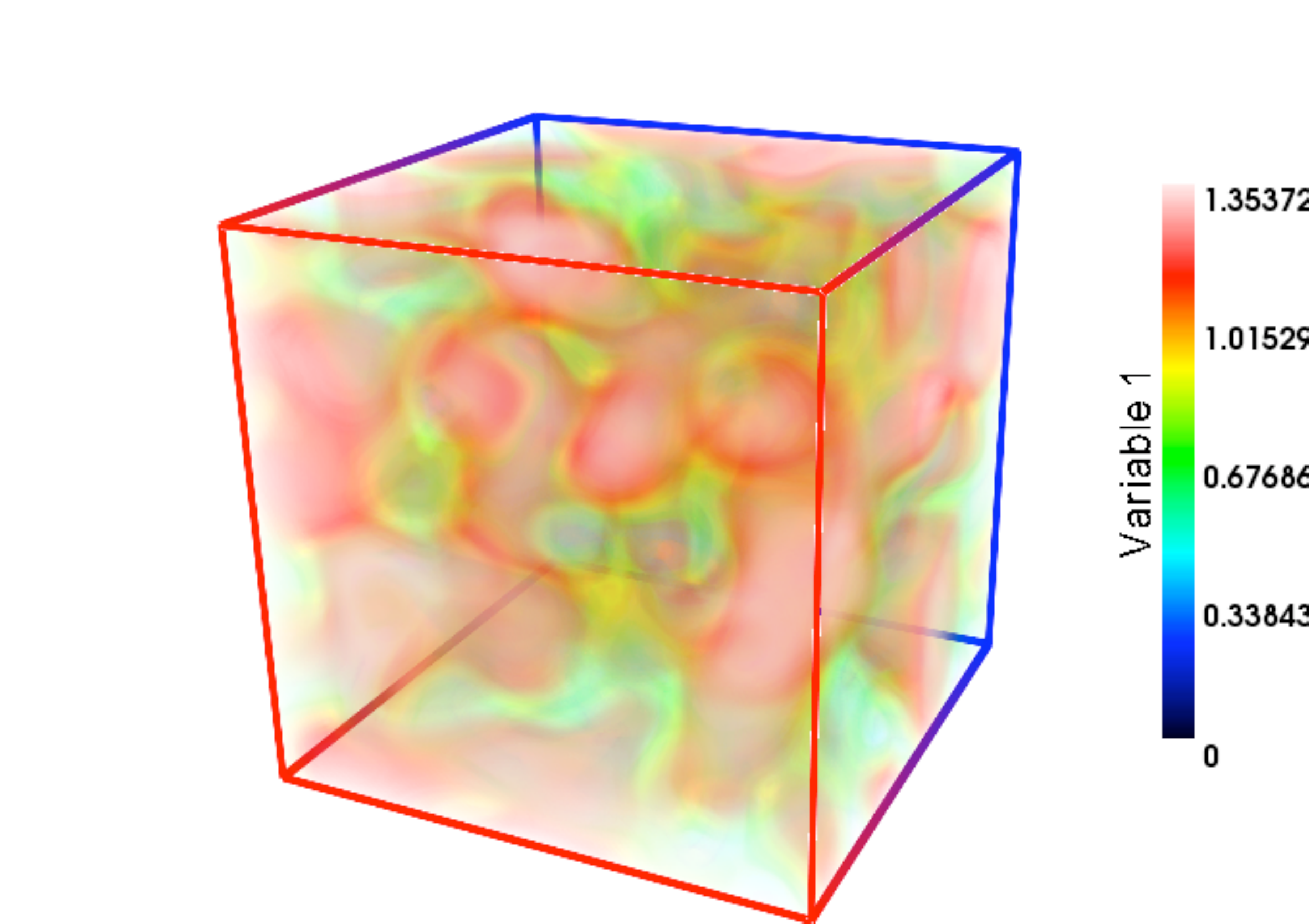}
\includegraphics[height=6.15cm]{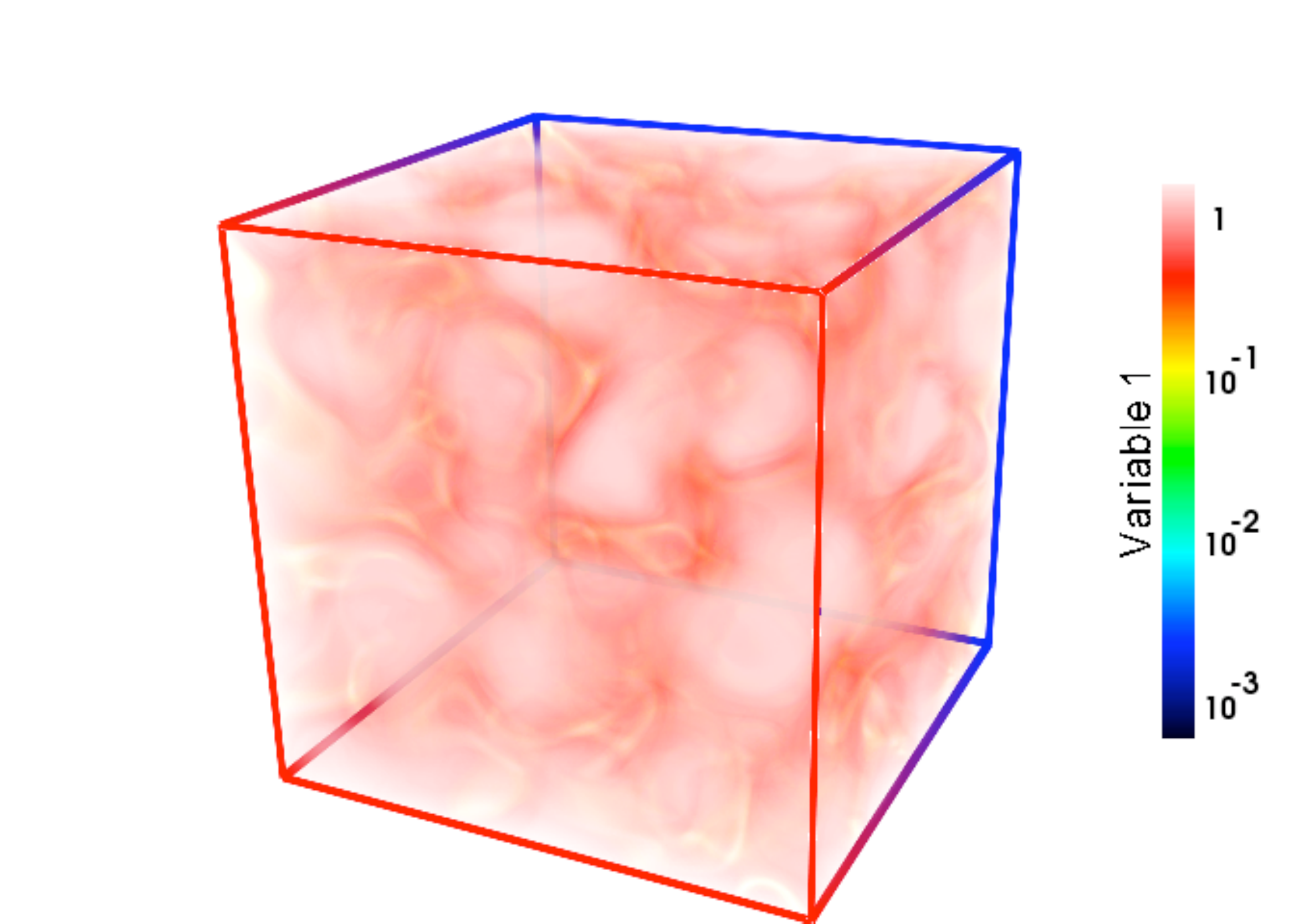}
\end{center}
\vspace*{-5mm}
\caption{Time evolution of the spatial distribution of the modulus of the Higgs field, $|\varphi|(\textbf{x},t) = \sqrt{\varphi_1^2 + \varphi_2^2}$, during the process of symmetry breaking. The images have been obtained with a N = 256 lattice simulation with an IR cut-off $k_{IR} = 0.15 $m, and parameters $g^2 = 2 \lambda = 0.25$, $V_c = 0.024$ and $e = 6\,\sqrt{\lambda}$. From left to right, top to bottom, the snapshots correspond to mt = 5.5, 11.0, 17.3 and 23.0.
}
\label{fig:figSpatialHiggs}
\vspace*{-3mm}
\end{figure}

\section{Spatial Configurations}
\label{SecPosition}

The spectra of the anisotropic stresses of the matter fields and GW power spectra give us only a partial information on the evolution of the fields and the origin of the peaks in the spectrum. In order to understand the detailed dynamics one has to use all the information available, and in particular, follow the spatial configurations in detail as a function of time, since then one can understand how specific features (like topological strings configurations) are formed and give rise to the observed peaks in the spectrum. Moreover, apart from both spatial images and power spectra, a very useful tool for this detailed understanding is the time evolution of histograms of both the Higgs and the magnetic fields' energy densities. These histograms allows us to identify the moments when the Higgs's oscillations make its vev reach zero and induce non-trivial windings at places where topological defects form.

For this purpose we turn to the discussion of the production of GW in 
configuration space, describing the spatial distributions and 
correlations between the energy density of the scalar and vector 
sources, and that of the GW.  We will see how the Higgs field forms bubbles that expand and 
collide and how the gauge field is excited during the symmetry breaking, forming elongated 
structures (tubes) in between the Higgs' bubbles. These string-like spatial configurations of 
the time-dependent Higgs and gauge fields follow from the dynamical equations of motion 
of the coupled system and will of course exhibit some differences with respect to the usual 
Nielsen-Olesen solution for static and infinite strings \cite{strings}. 
We will follow the formation and evolution of these strings during and after the symmetry breaking, 
but well before any scaling regime of string networks has been achieved and on length scales 
much smaller than the Hubble radius.

For Nielsen-Olesen strings, the magnetic flux decays away from the core of the string with a typical length scale that is given by the inverse of the gauge field mass. However, during the process of symmetry breaking, the Higgs field provides an effective mass for the gauge field which oscillates in time as the Higgs oscillates around its VEV, with relatively large amplitude, see e.g. Fig.~\ref{meanrms}. Furthermore, the magnetic flux tends to be confined in the regions of space where the amplitude of the Higgs is small, i.e. between the bubbles of large Higgs amplitude, and the width of these regions is also strongly time-dependent as the Higgs bubbles oscillate and collide. Therefore, we will see that the width of the string-like configurations of the gauge field oscillate with time, becoming thicker when the amplitude of the Higgs is a maximum. The same occurs for the Higgs field whose effective mass is also time-dependent via its coupling to the oscillating inflaton field. At later times, we will see how magnetic energy density is shedded away from the core of the strings. Similar effects have been observed in field theory simulations of cosmological networks of cosmic strings in the Abelian-Higgs model, see e.g.~\cite{hindmarsh} and references therein. Such effects are usually suppressed by the ratio of the width of the string over the length of straight string segments, which is not small for the system under study.

\begin{figure}[htb]
\begin{center}
\includegraphics[height=5.8cm]{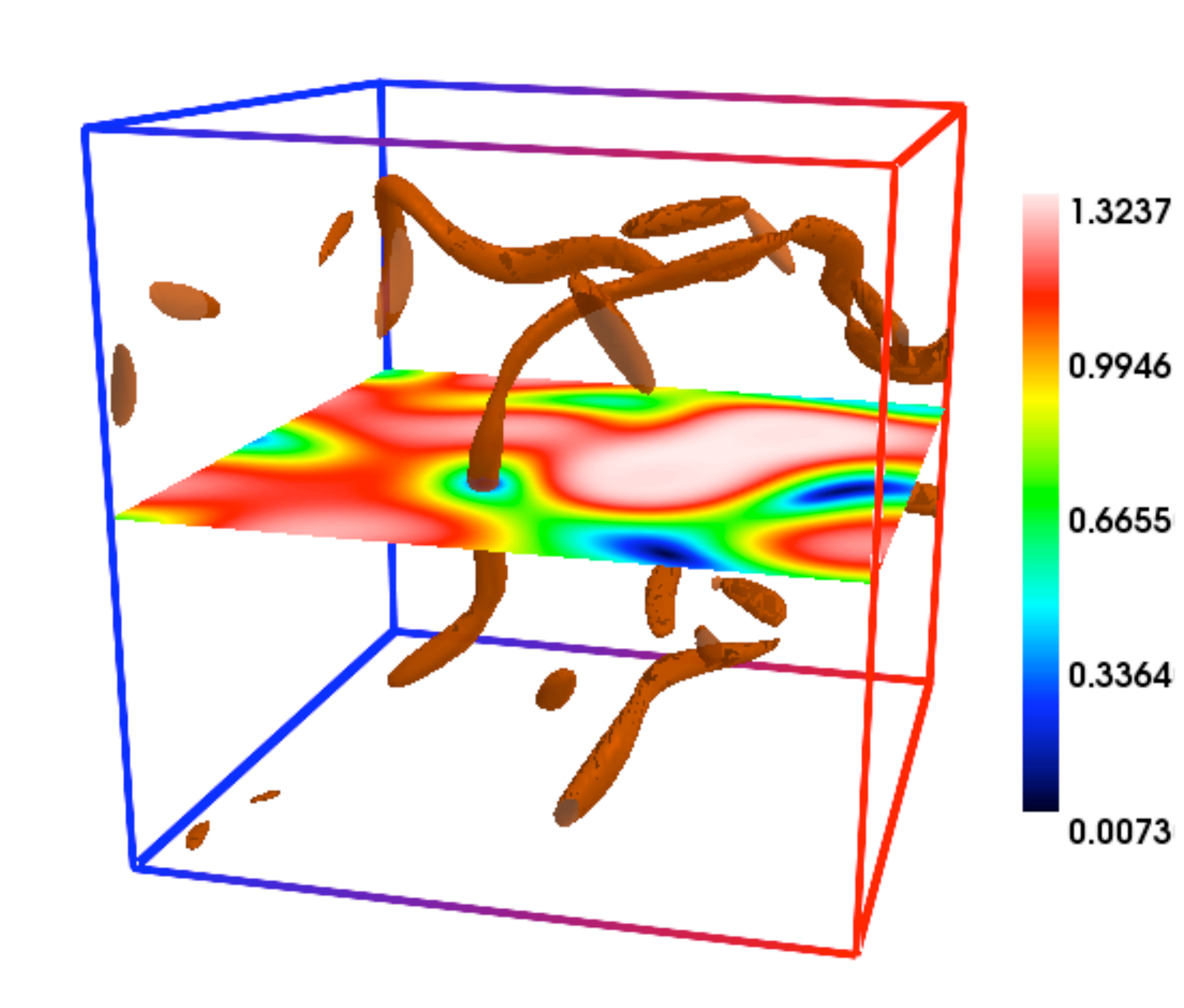}
\hspace{1cm}
\includegraphics[height=5.8cm]{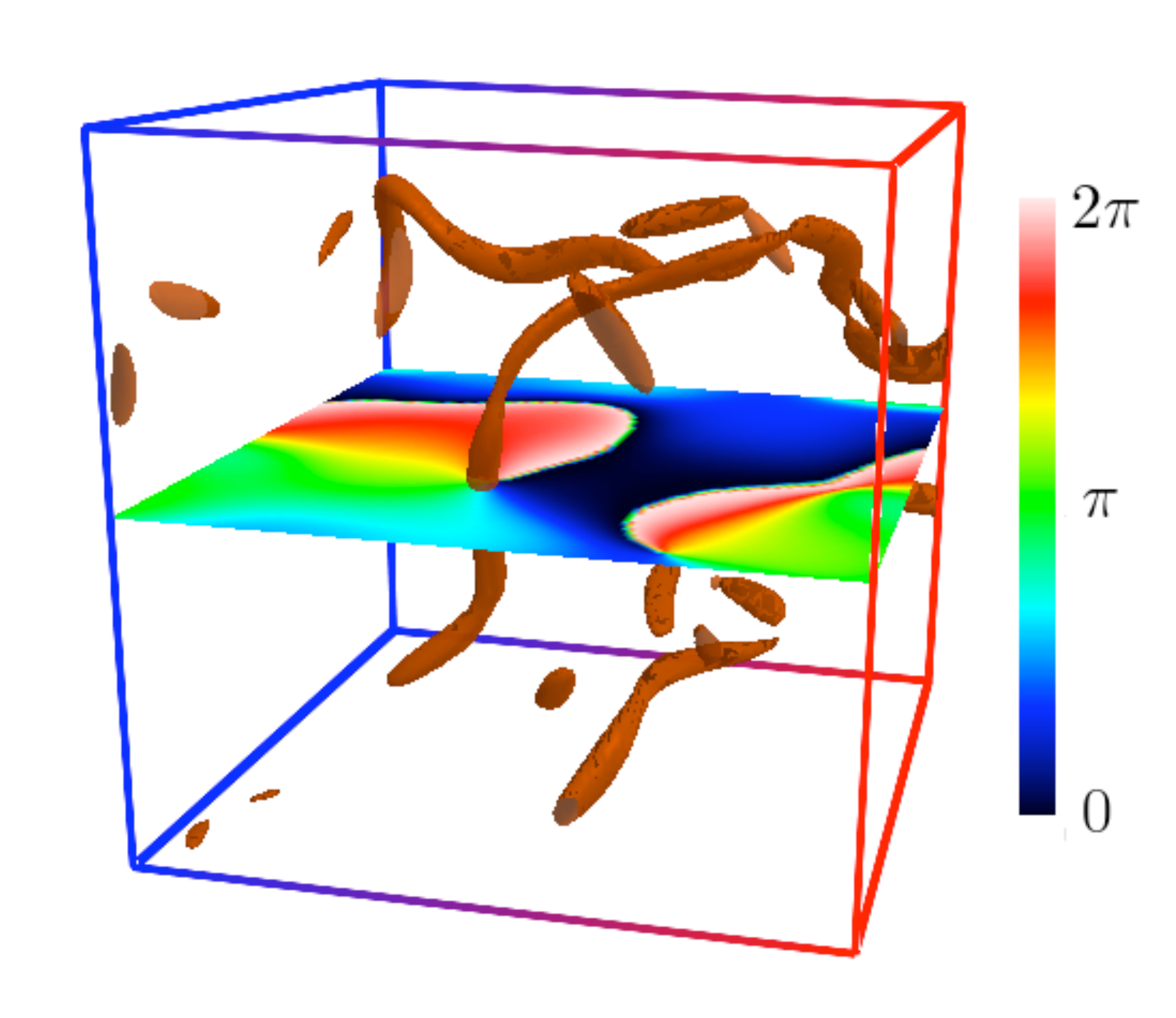}
\end{center}
\vspace*{-5mm}
\caption{Snapshots at time $mt = 17$ of the spatial distribution of the magnetic energy density $B^2$ (in units of $m^4$). The images have been obtained for a N = 256 lattice with an IR cut-off $k_{IR} = 0.1$m, and parameters $g^2 = 2 \lambda = 0.25$, $V_c = 0.024$ and $e = 6\,\sqrt{\lambda}$. In the left we see clearly how the string-like configuratinos of the magnetic fields are localized where the minima of the Higgs are, as described by the coloured transverse plane which plots the Higgs amplitude (normalized to the VEV) at that moment. On the right, the analagous figure where now the transverse coloured plane shows the phase of the Higgs, thus clearly demonstrating the existence of non-trivial winding in the Higgs around the positions of the the magnetic strings.} 
\label{fig:3Dmag2DhiggsWinding} \vspace*{-3mm}
\end{figure}

\begin{figure}[htb]
\begin{center}
\includegraphics[height=6.15cm]{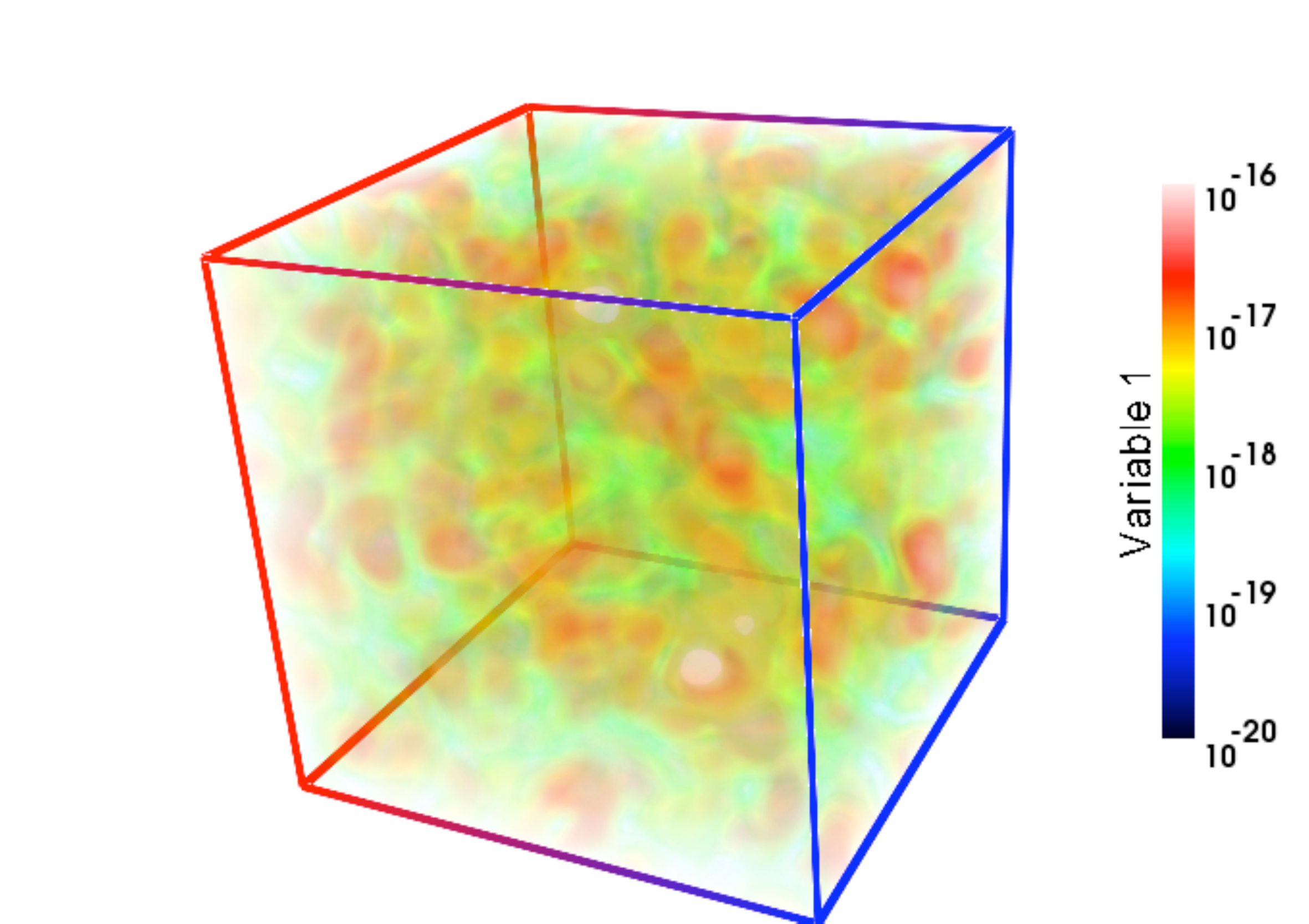}
\includegraphics[height=6.15cm]{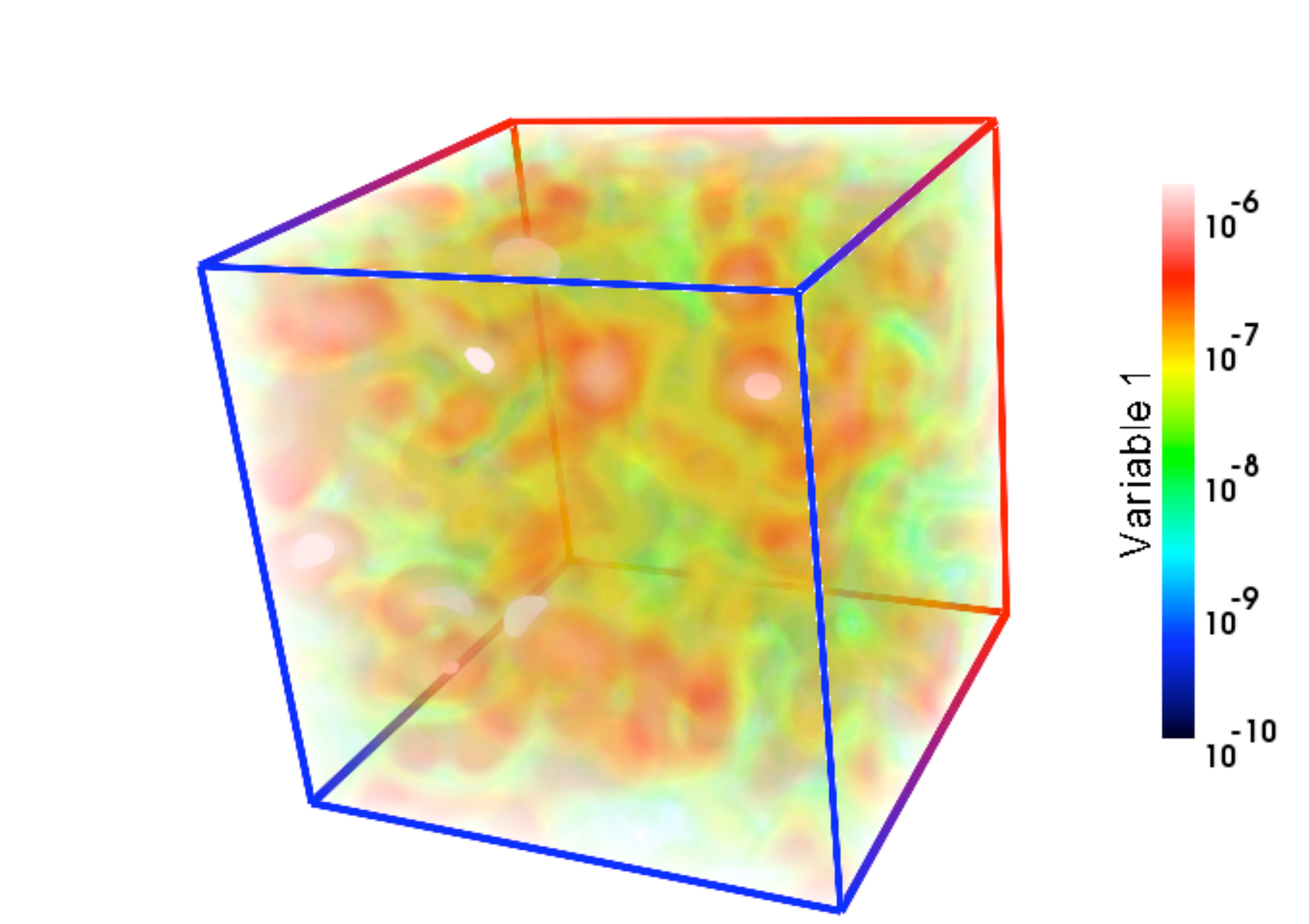}
\includegraphics[height=6.15cm]{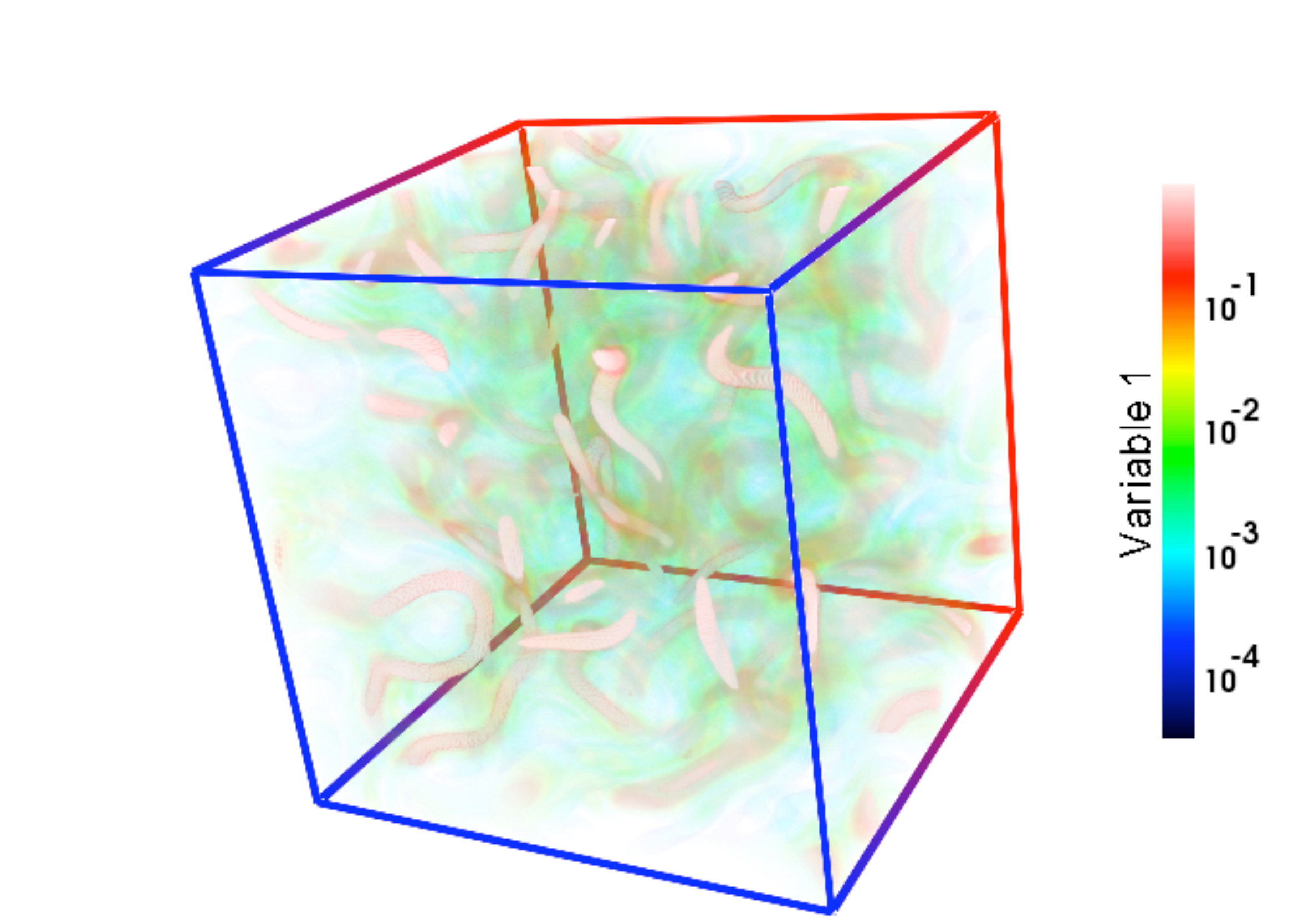}
\includegraphics[height=6.15cm]{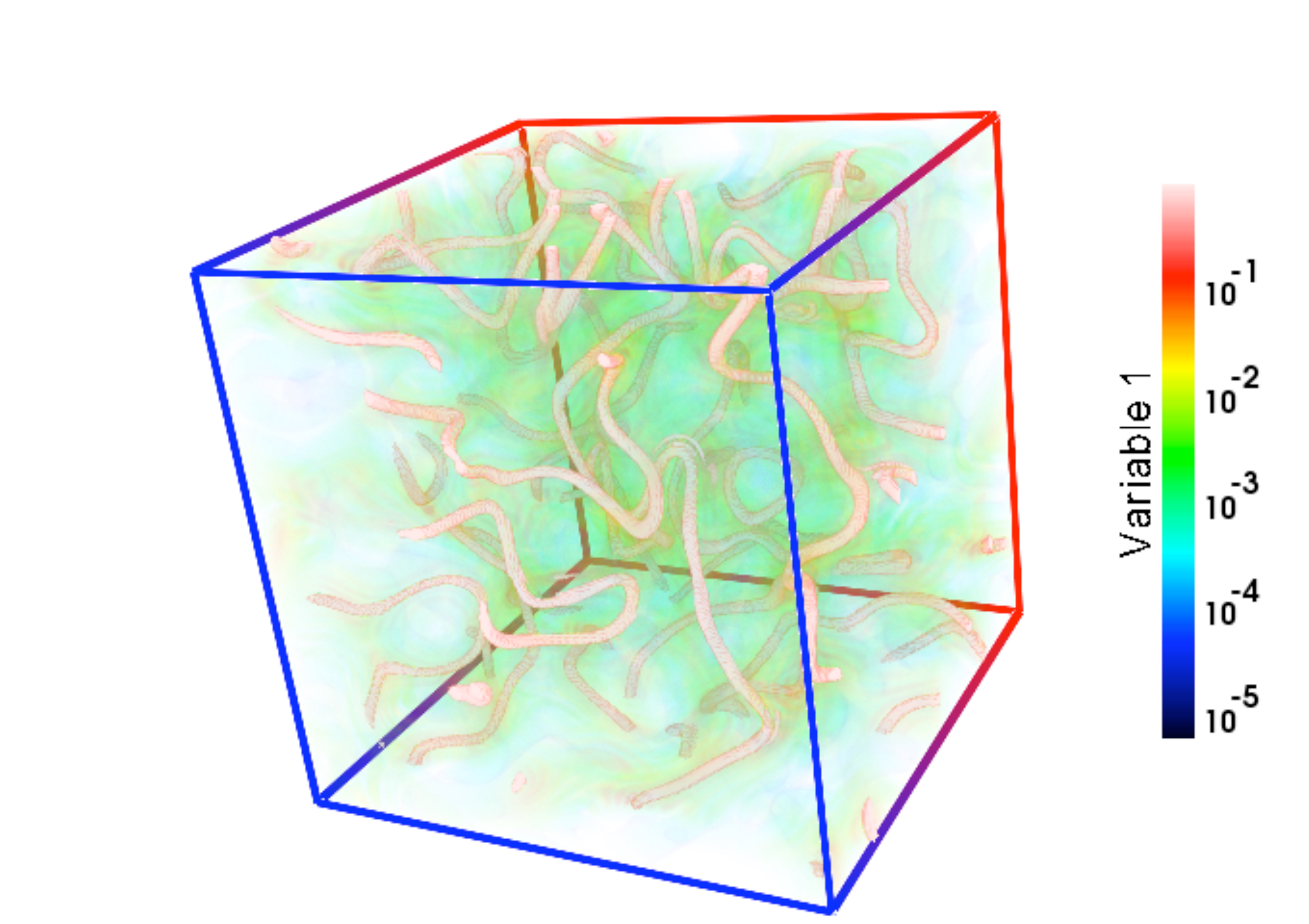}
\includegraphics[height=6.15cm]{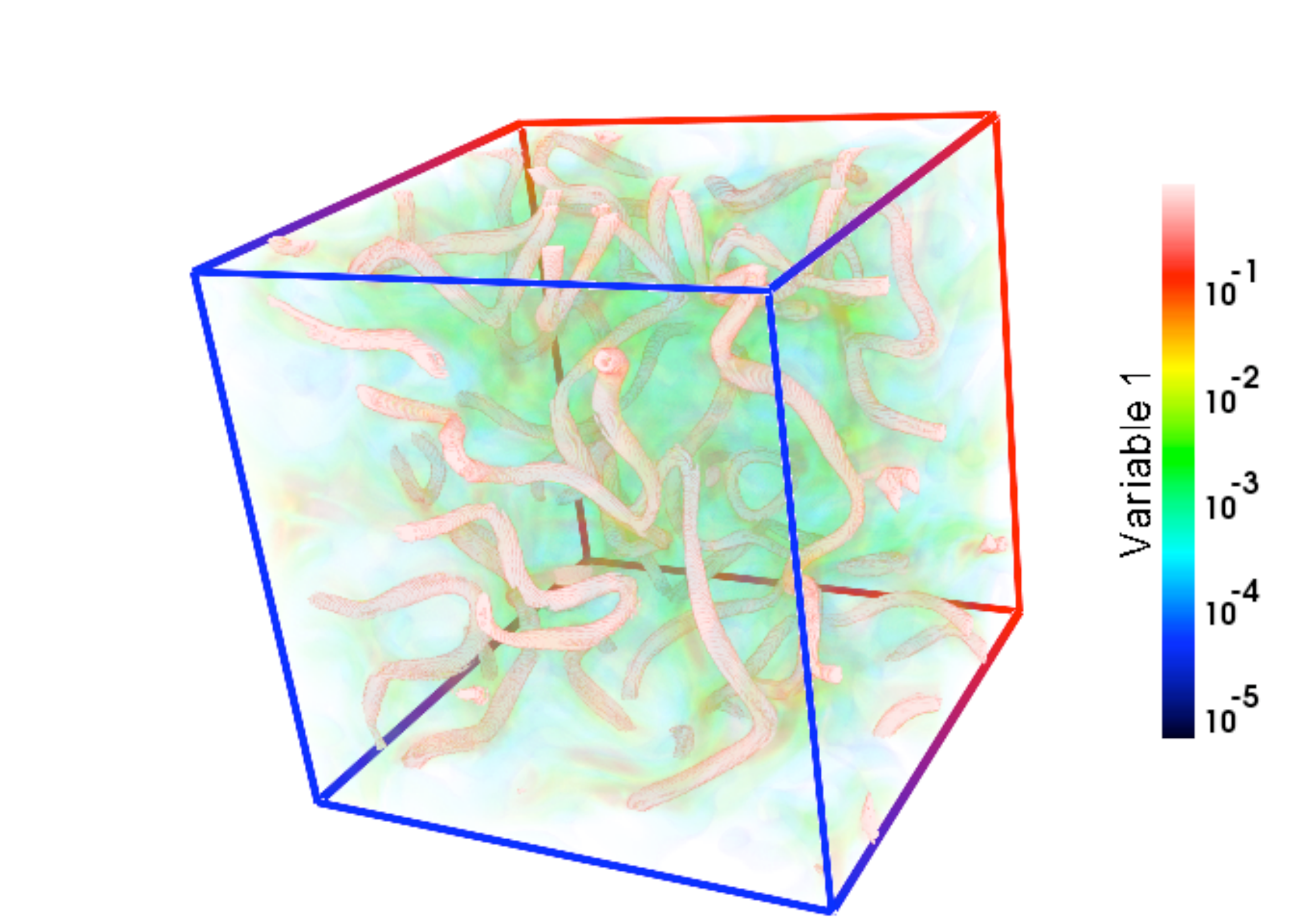}
\includegraphics[height=6.15cm]{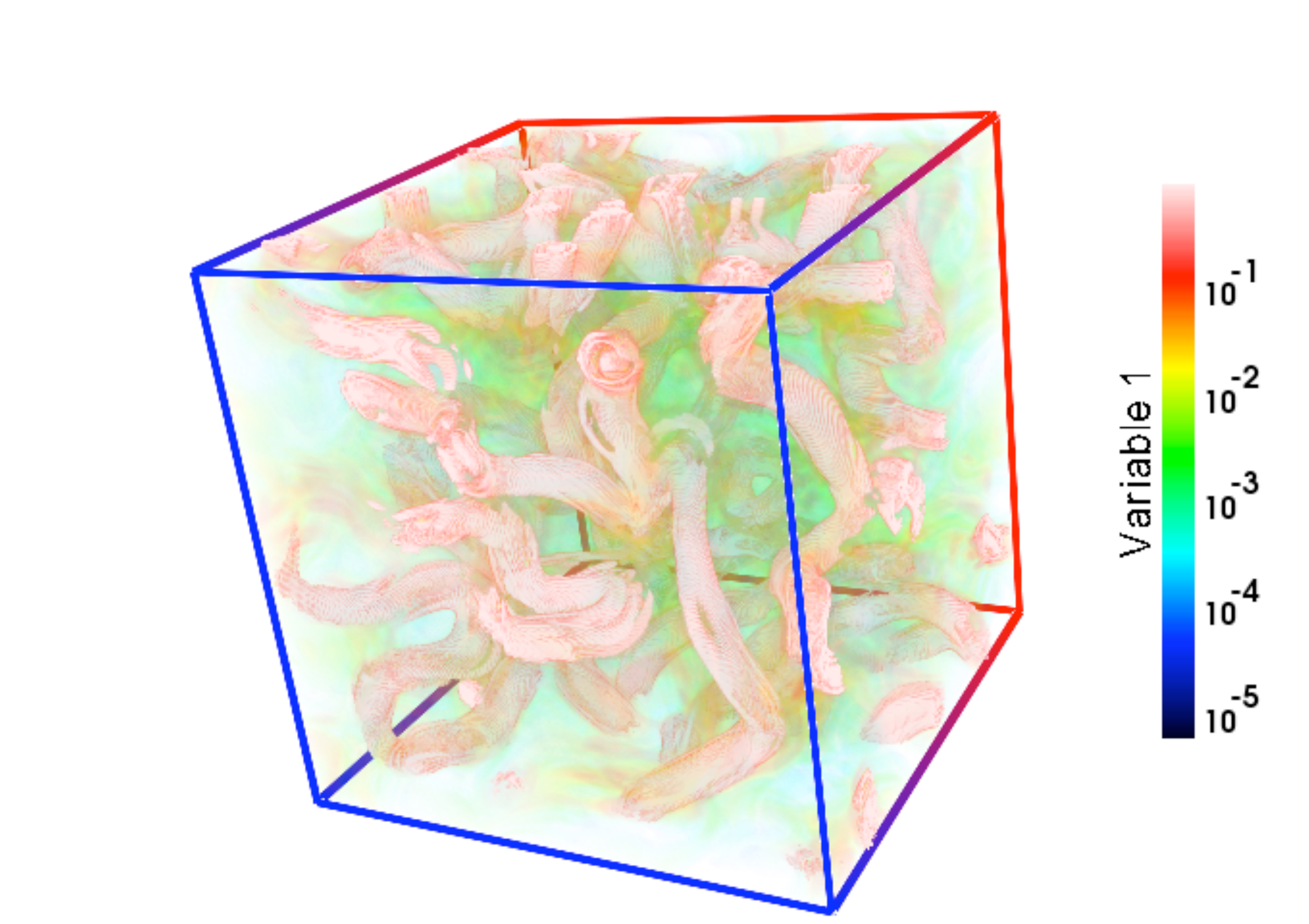}
\end{center}
\vspace*{-5mm}
\caption{Time evolution of the spatial distribution of the magnetic energy density $B^2$ (in units of $m^4$) along the process of the Higgs symmetry breaking. The images have been obtained with a N = 256 lattice simulation with an IR cut-off $k_{IR} = 0.1$ m, and parameters $g^2 = 2 \lambda = 0.25$, $V_c = 0.024$ and $e = 6\,\sqrt{\lambda}$. From left to right, top to bottom, the snapshots correspond to mt = 5.5, 11.0, 17.3, 19.0, 21.0 and 23.0. At early times, before the Higgs bubbles percolate, the magnetic field is still very small and has not acquired yet the distinctive shape of topological string configurations. At times $mt \sim 17-19$, the string-like spatial distributions of the magnetic energy density have finally developed, following the locus of points which corresponds to the intermediate regions between Higgs bubbles. The string-like distributions are most clearly seen at time $mt = 19$. Later, due to the time evolution of the gauge field's mass, the string segments fatten and start shedding away the magnetic field, see the main text. }
\label{fig:Mag3D}
\vspace*{-3mm}
\end{figure}

\subsection{Higgs bubbles}

In Fig~\ref{fig:figSpatialHiggs} we show a sequence of snapshots of the spatial distribution of the modulus of the Higgs field for a model with couplings $g^2 = 2 \lambda = 0.25$, initial inflaton velocity $V_c = 0.024$, and gauge coupling $e = 6\,\sqrt{\lambda}$. From left to right, top to bottom, the snapshots correspond to mt = 5.5, 11.0, 17.3 and 23.0, and capture the details of the symmetry breaking process towards the true VEV. We have choosen a simulation beginning at $mt = 5$, when the tachyonic modes of the Higgs are already well inside in the classical regime, as described by~(\ref{eq:InitialConditions}). Therefore, the distribution of the Higgs at $mt = 5.5$, i.e. sligthly after the initialization of the simulation, simply shows the bubble-like structures as developed in random positions, corresponding to the tail of high field values of the gaussian distribution~(\ref{eq:InitialConditions}) which describes the tachyonic modes. Since we are using a big lattice with N = 256 points per dimension, the number of bubbles we capture in a single box is statistically quite significant, and the resolution of their spatial profiles is also quite well captured, as can be clearly seen in the figures corresponding to mt = 5.5 and mt = 11. In the snapshot corresponding to mt = 11, the tachyonic instability of the excited modes has already led to the growth of the amplitude of the Higgs field towards the true vacuum, such that the Higgs field at the center of the bubbles at that time has reached already a 5\% amplitude of the true VEV. The higher the random value of the Higgs was in a given location at the initial time, the faster the amplitude of the Higgs has grown in such location and the neightboring region. However, at time $mt = 11$, for the parameters choosen, the system has already entered into a regime in which the non-linearities due to the self-coupling for the Higgs are not negligible anymore. In the non-linear regime, the initially IR tachyonic modes are transferring power into the higher momenta modes out of the initial tachyonic band. As a result, in configuration space one can see that the amplitude of the Higss has grown everywhere in space, although the bubble structures are yet preserved. At mt = 17.25 we see that the Higgs has already reached the true VEV. Indeed, the Higgs at the centers of the bubbles has overpassed the VEV and reached a slightly greater value (as allowed by energy conservation). Due to the non-linearities, the bubbles are growing in that moment and are about to percolate. Since the size of the bubbles grows in time, the effective volume of the regions of lower amplitude in between the bubbles is naturally shrinking. At mt = 
23, the bubbles have already percolated and one can clearly see in the intermediate regions between bubbles, that the Higgs amplitude is an 
order of magnitude lower than the true VEV. Those regions correspond to the locus of points in configuration space where a non-trivial 
winding has been developed, therefore leading to the formation of a spatial region where the Higgs amplitude is frustated to reach the 
true VEV. Of course, we are still far away from the stationary regime describing the evolution of topological defect networks. Rather, in the previous sequence of snapshots, we are looking at the dynamics of the symmetry breaking process itself, from the false to the true vaccuum, in a time scale in which the fields are still highly oscillatory and have not reached a scaling behaviour. 

The growth and collisions of the Higgs bubbles during symmetry breaking give rise to a significant anisotropic stress-tensor in the scalar fields, which drives the initial production of GW in this model. The specific correlation in configuration space between the locus of points where the gradients of the Higgs are maximum and the distribution of the GW energy density was already shown in \cite{GFS} (section V of that paper), see also \cite{DFKN}. We will not reproduce again here such correlations and rather we will focus on the correlations between the GW energy density distribution and the energy density of the new source of GW considered in this paper, the gauge fields.

Note that we will maintain the parameters choosen for Fig.~\ref{fig:figSpatialHiggs} fixed through the rest of this section, such that all the plots shown will represent some output from a simulation run with those values for the parameters.

\subsection{Magnetic string formation and evolution}

We already saw in momentum (Fourier) space how a new scale emerges in 
the spectrum of the gauge fields as the Higgs approches the true 
vaccuum, since then the gauge field aquires a mass through the Higgs 
mechanism. The covariant gradient energy of the Higgs is minimized 
for~\cite{chern}
$$A_\mu = \frac{1}{e} \Omega \partial_\mu \Omega\,,$$
where $\Omega= \phi/|\phi|$ is an element of U(1). This induces the 
magnetic field to concentrate its energy density within those regions 
in which the Higgs amplitude is smaller and phase gradients are 
larger. Thus, during the symmetry breaking of the Higgs, magnetic flux 
tubes will be concentrated in the regions between the Higgs bubbles. 
This can clearly be seen in the left pannel of Fig.~\ref{fig:3Dmag2DhiggsWinding}, where we plot a high value isosurface of the magnetic energy density, showing this way how the three dimensional configurations of the magnetic field forms flux tubes (strings). Together with the strings, we also show a two dimensional plane orthogonal to those long flux tubes, showing how the strings appear in those places where the Higgs is minimum. This correlation between Higgs' zeros and magnetic field strings is a universal feature of the Abelian-Higgs model, and is related to the Nielsen-Olesen vortices (and strings) predicted in the model many decades ago~\cite{strings}. 

Note that in the Abelian-Higgs model, the magnetic flux along the strings is quantized and is related to the Higgs winding number, 
$$\Phi_B = \int d^2{\bf x} \, \vec B\cdot \hat z = \oint d\vec x\cdot 
\vec A = \frac{2\pi n}{e}\,,$$
where $n$ is an integer called the \textit{winding} number. This topological number is conserved along the evolution unless there are over-the- 
barrier transitions during preheating, see Ref.~\cite{chern}. In the right pannel of Fig.~\ref{fig:3Dmag2DhiggsWinding}, we have shown the magnetic strings (the magnetic energy density) from a perspective in which one can see the (color coded) variation of the phase of the Higgs from 0 
to $2\pi$, as painted in a transverse plane to the string in the center of the box. Such plane correspond to a bidimensional cut of the three dimensional distribution of the values of the Higgs' phase between 0 and $2\pi$. Around the place where the central string segment touches the plane, one can clearly see that the Higgs phase winds non-trivially around the string. The plane is also crossed by another string segment close to one of the walls of the box, and again one can see the non-trivial structure of the winding around such string. Besides, the two-dimensional distribution of the Higgs phase in the choosen plane, also shows very clearly the locus of points (lines within the plane) where the Higgs phase jumps from $2\pi$ to 0. If we interpret the plane as a Riemann surface, the curve lines where the Higgs phase jumps would then be the edges of the Riemann surfaces.

\begin{figure}[htb]
\begin{center}
\includegraphics[height=6.85cm,angle=0]{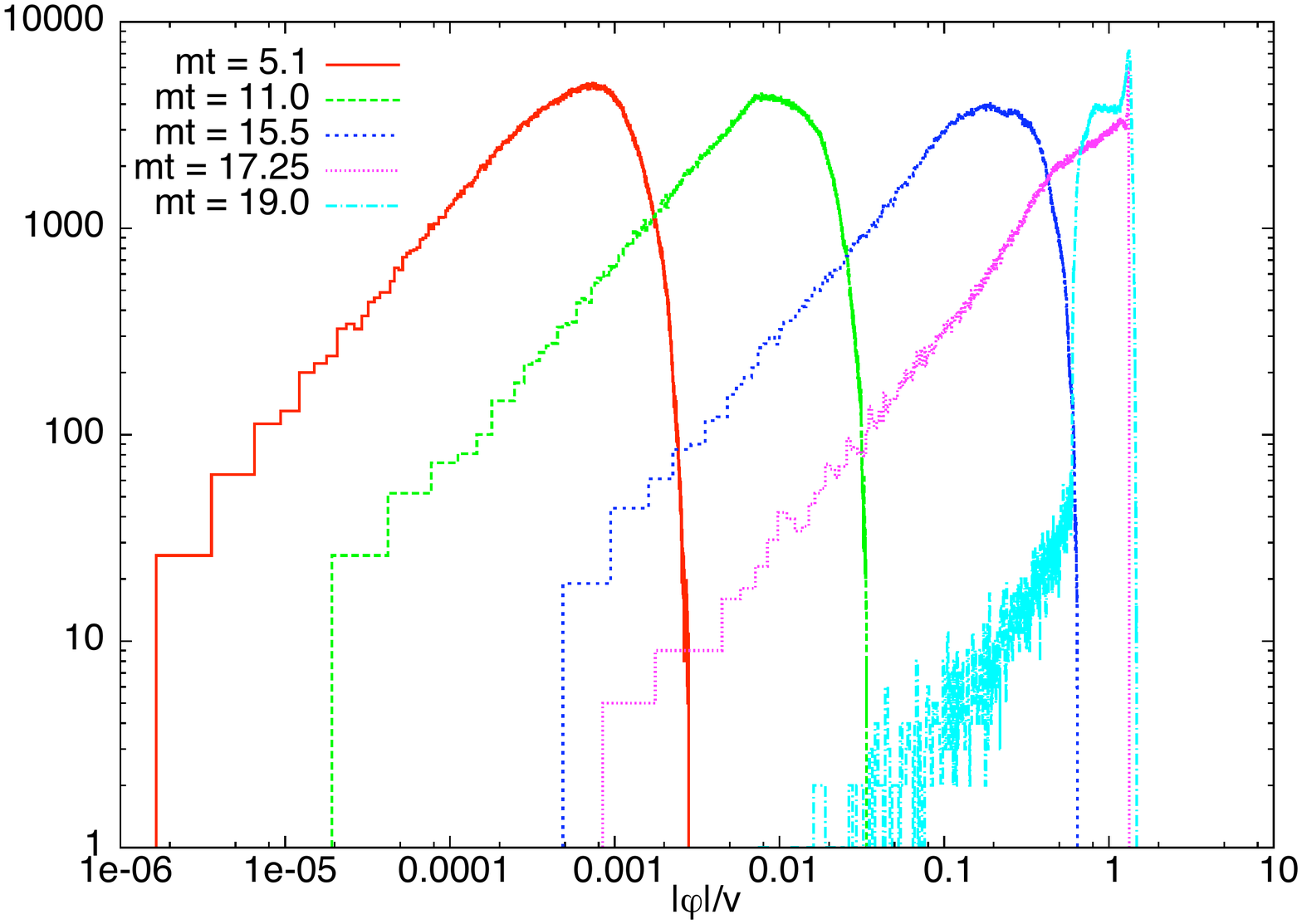}
\includegraphics[height=6.85cm,angle=0]{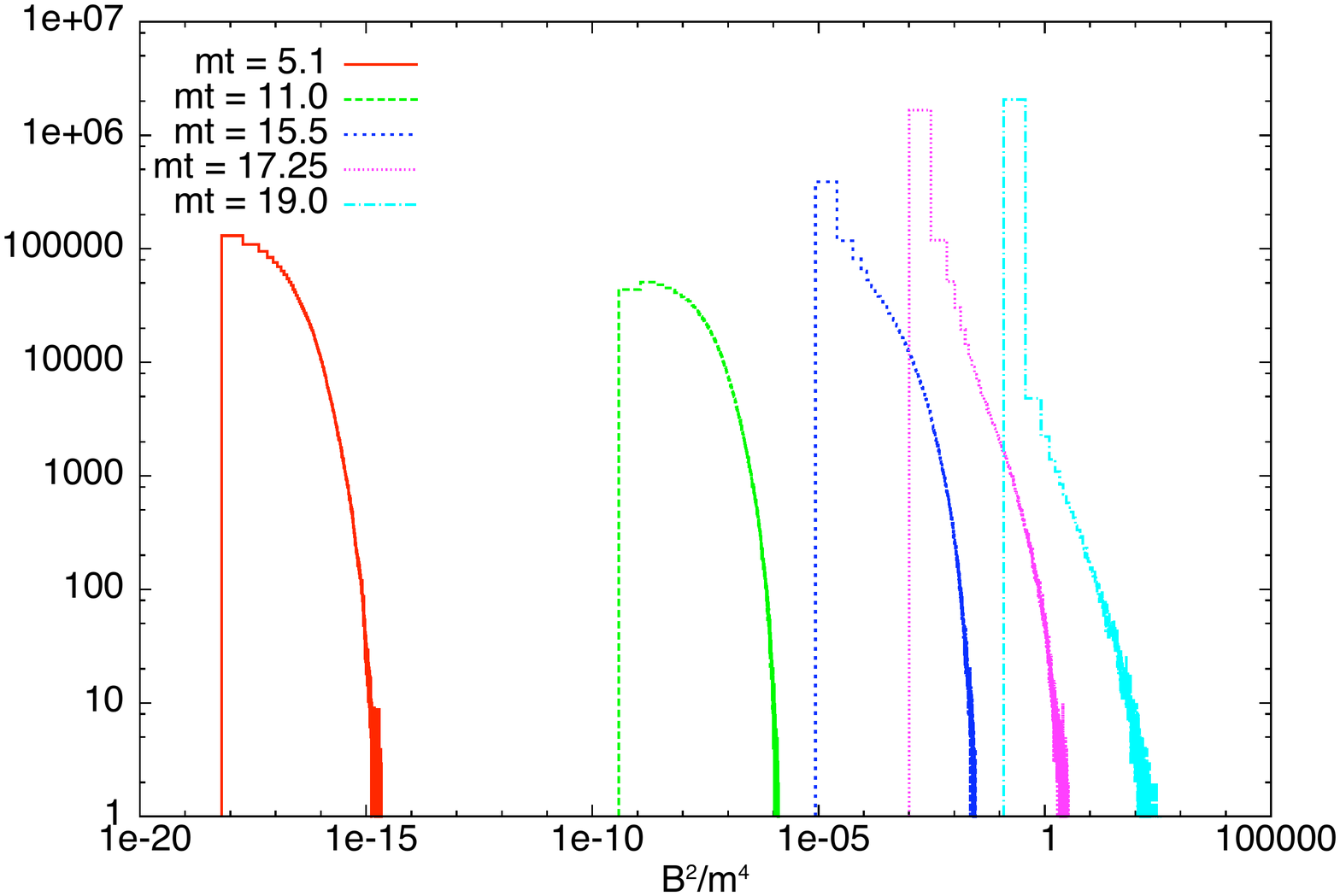}
\includegraphics[height=6.85cm,angle=0]{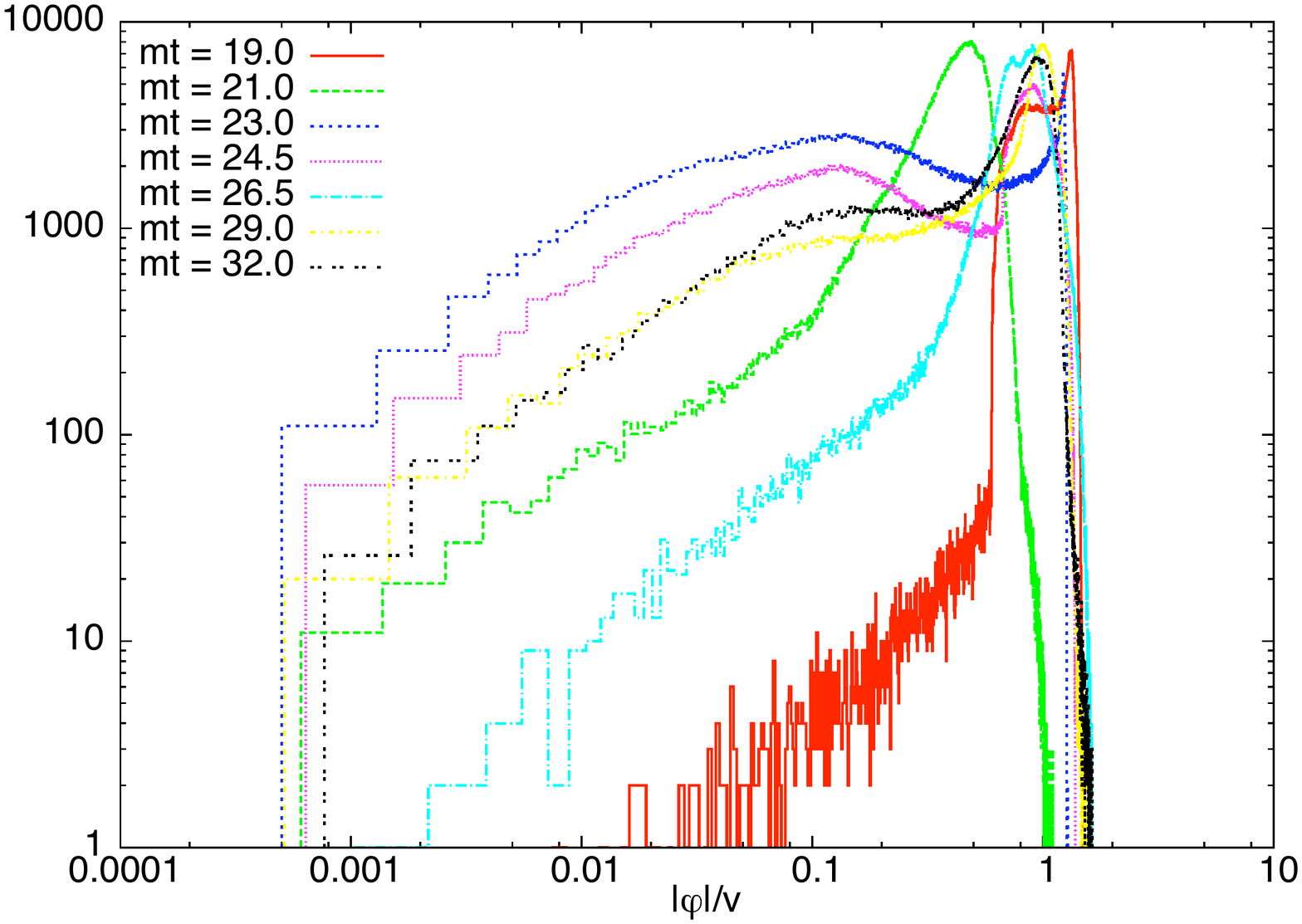}
\includegraphics[height=6.85cm,angle=0]{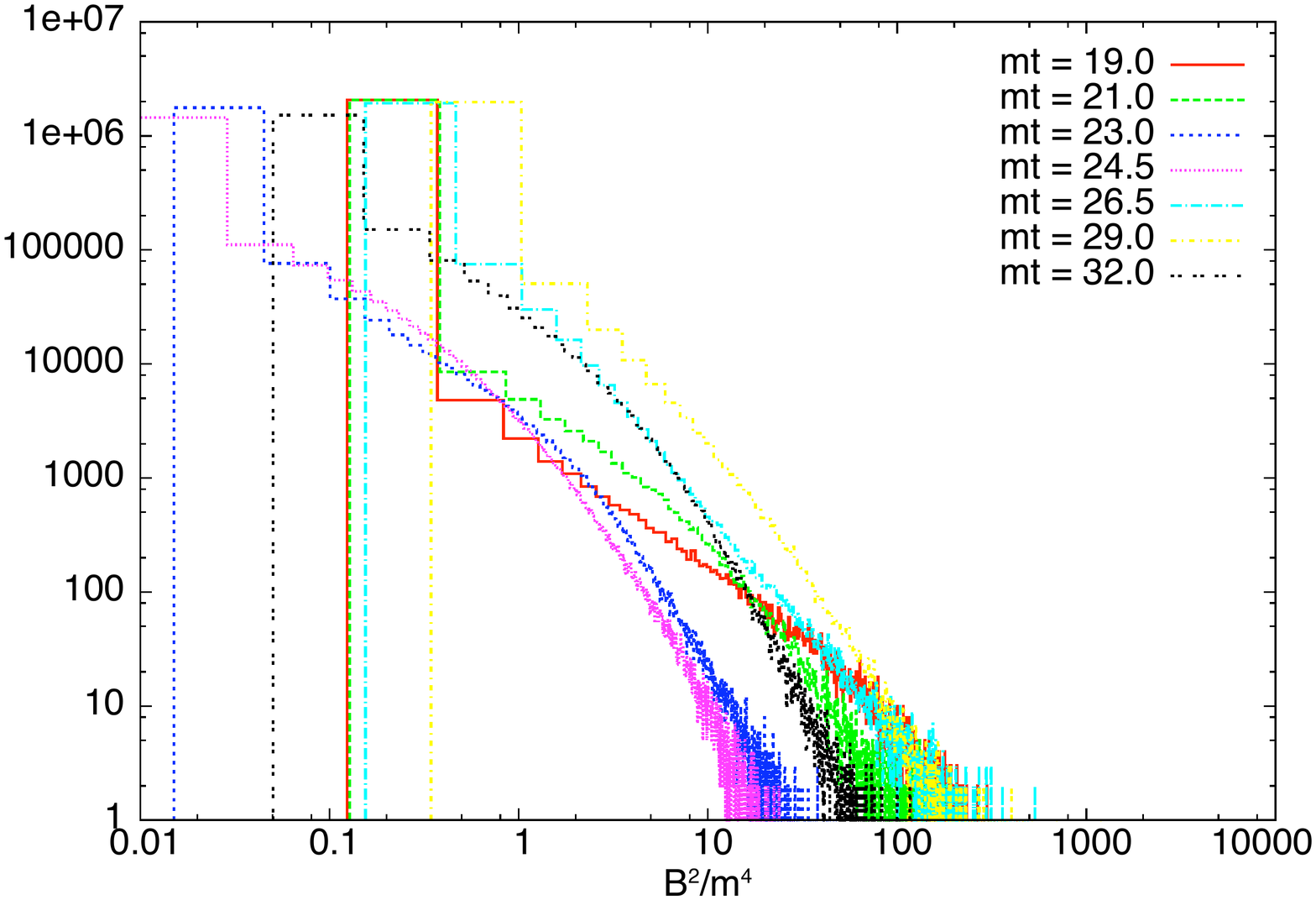}
\includegraphics[height=6.85cm,angle=0]{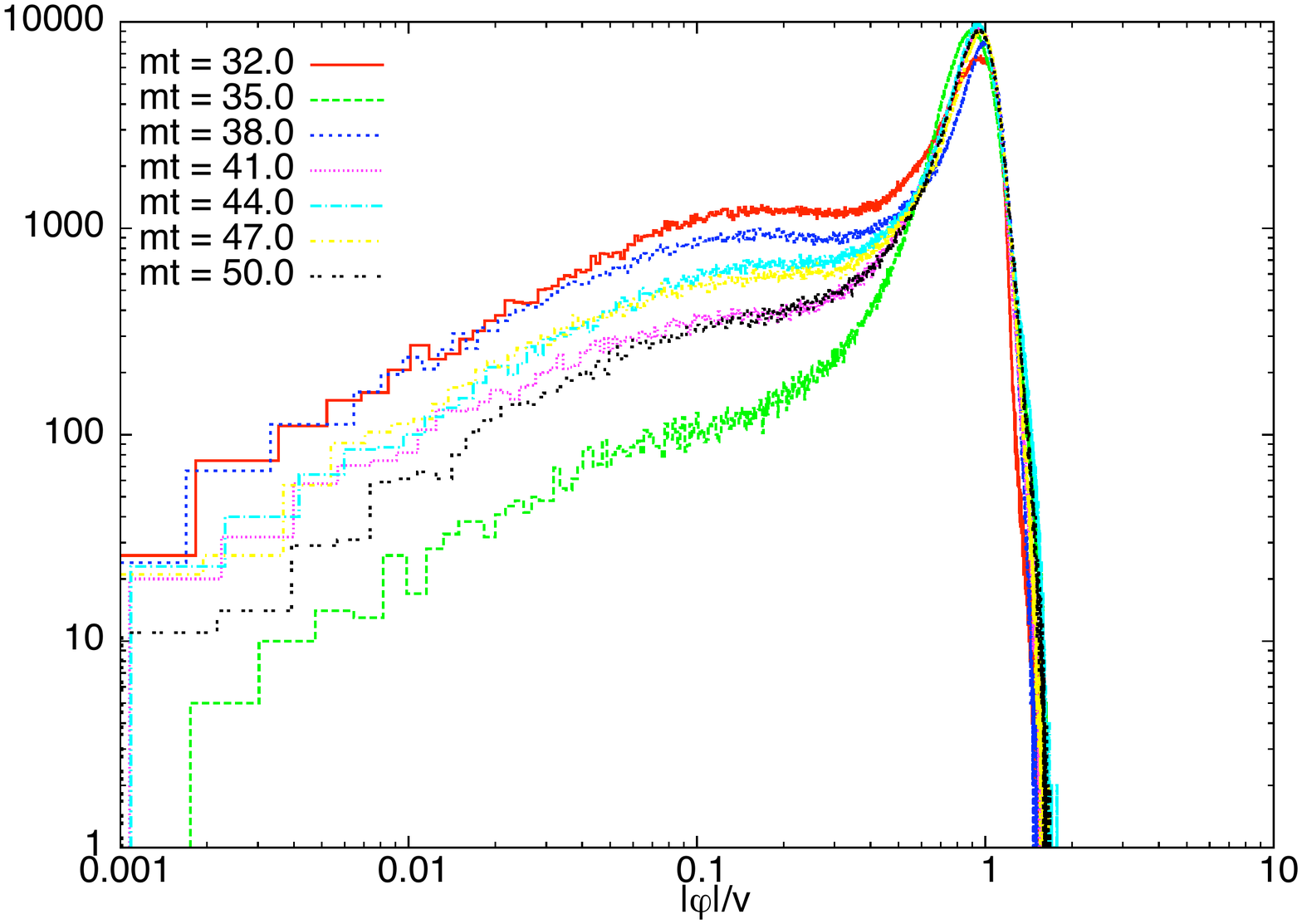}
\includegraphics[height=6.85cm,angle=0]{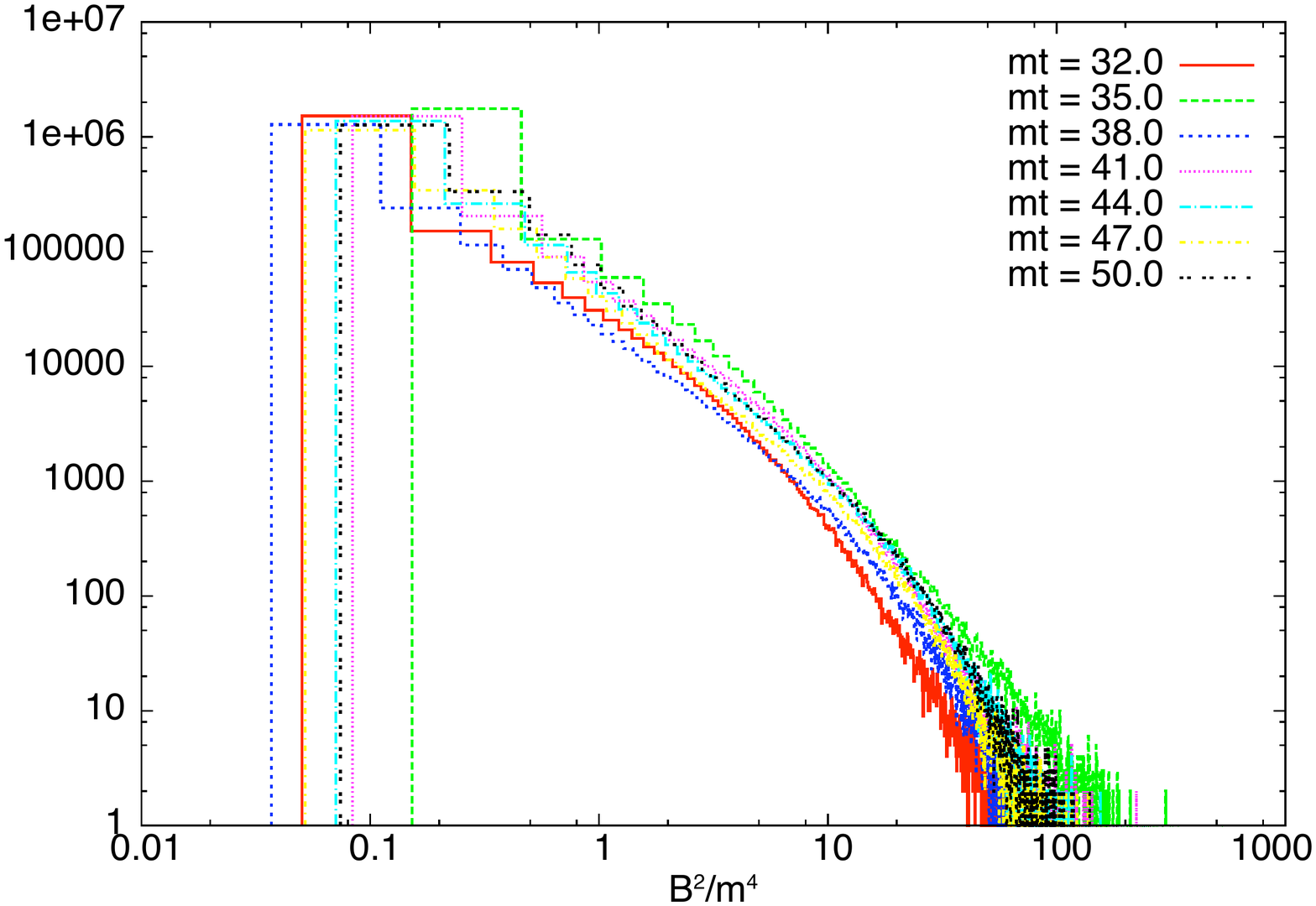}
\end{center}
\vspace*{-5mm}
\caption{Evolution in time of both the histograms of the Higgs field normalized to its VEV (left) and the magnetic energy $B^2$ normalized to $m^4$ (right). The first row corresponds to the initial times, from $mt = 5.05$ to $mt = 19$. The second row coresponds to times from $mt = 19$ to $mt = 32$. The third row corresponds to times from $mt = 32$ to $mt = 50$. It can be clearly distinguished that the Higgs moves very fast towards the true VEV in the intial stages of hybrid preheating and later oscillates close to VEV (see the left tails of the Higgs histograms). At any moment, even when the Higgs is oscillating in the broken phase with small amplitude compared to the VEV, there remains a significant fraction of points in the lattice where the amplitude of the Higgs is much smaller than the VEV, corresponding to the spatial regions located in between the Higgs bubbles.}
\label{fig:Histograms}
\vspace*{-3mm}
\end{figure}

\begin{figure}[htb]
\begin{center}
\includegraphics[height=6.15cm]{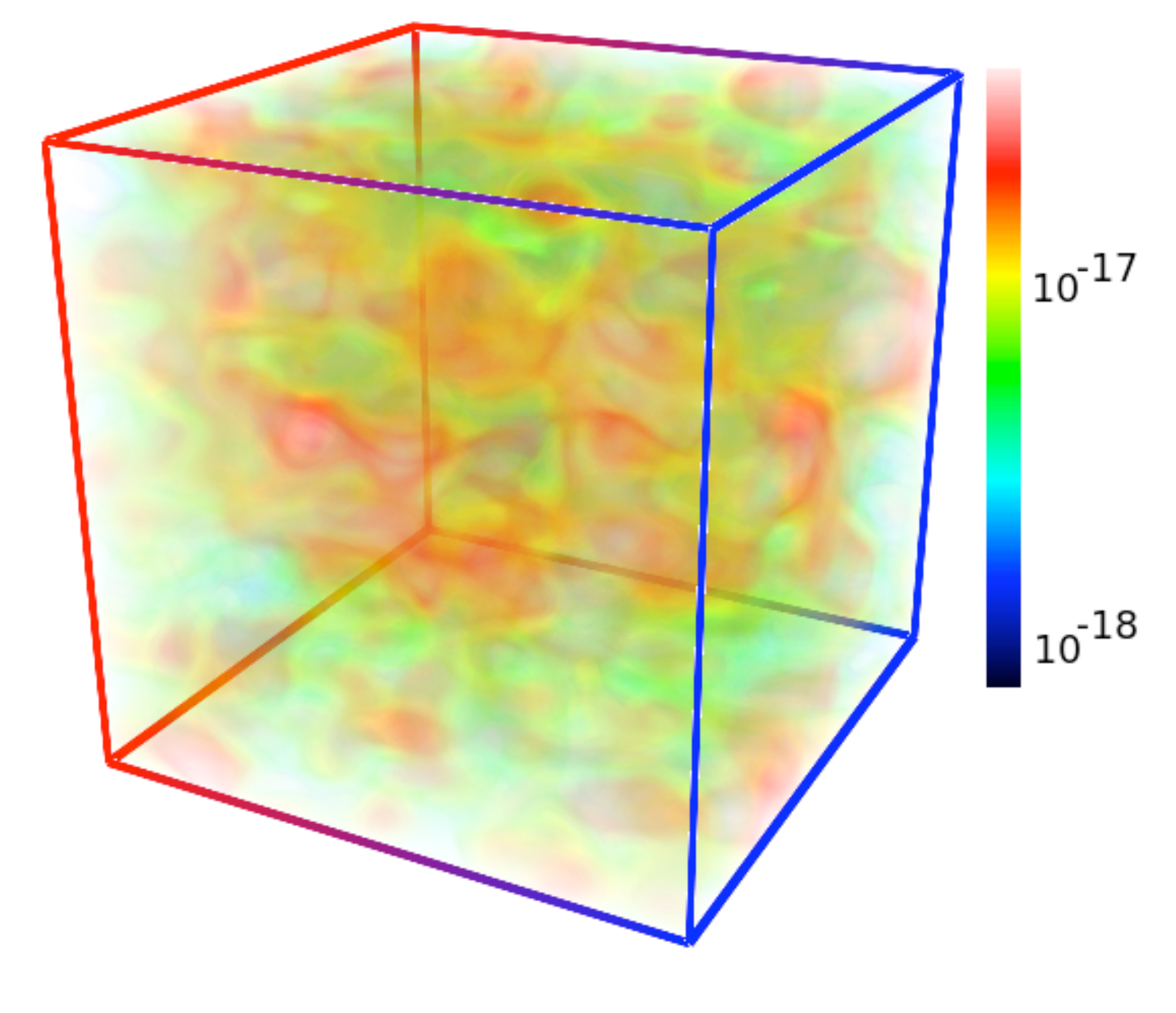}
\includegraphics[height=6.15cm]{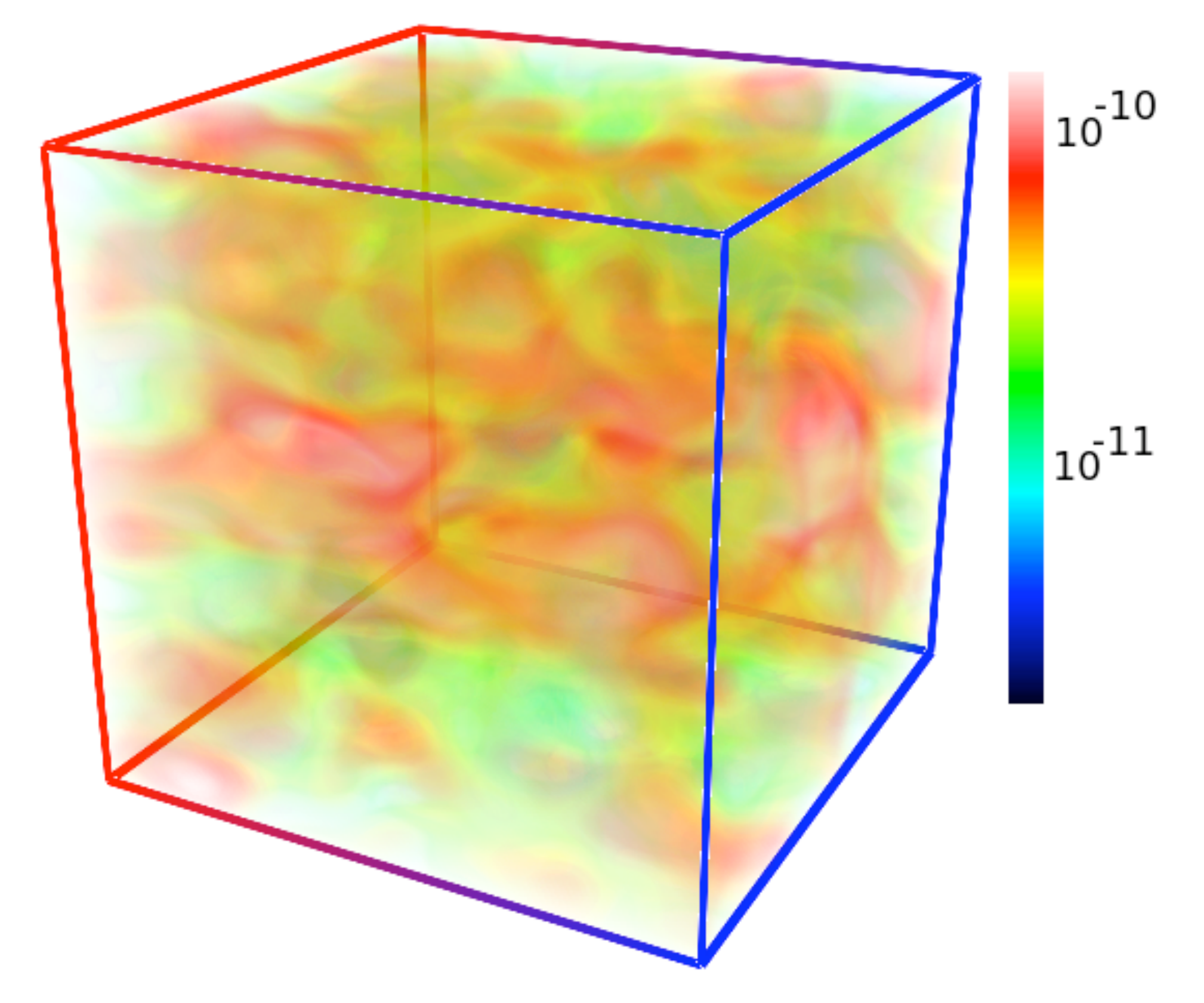}
\includegraphics[height=6.15cm]{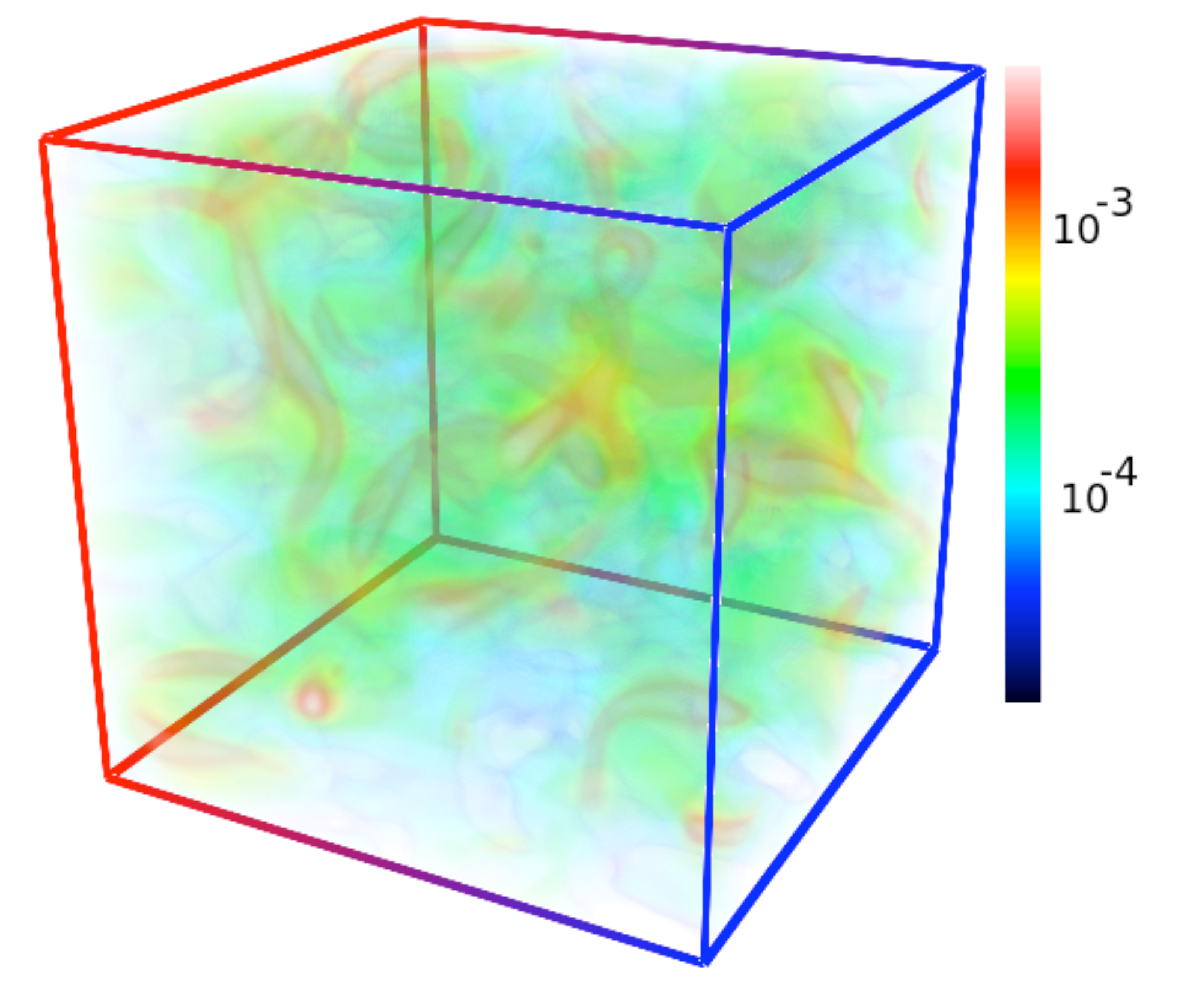}
\includegraphics[height=6.15cm]{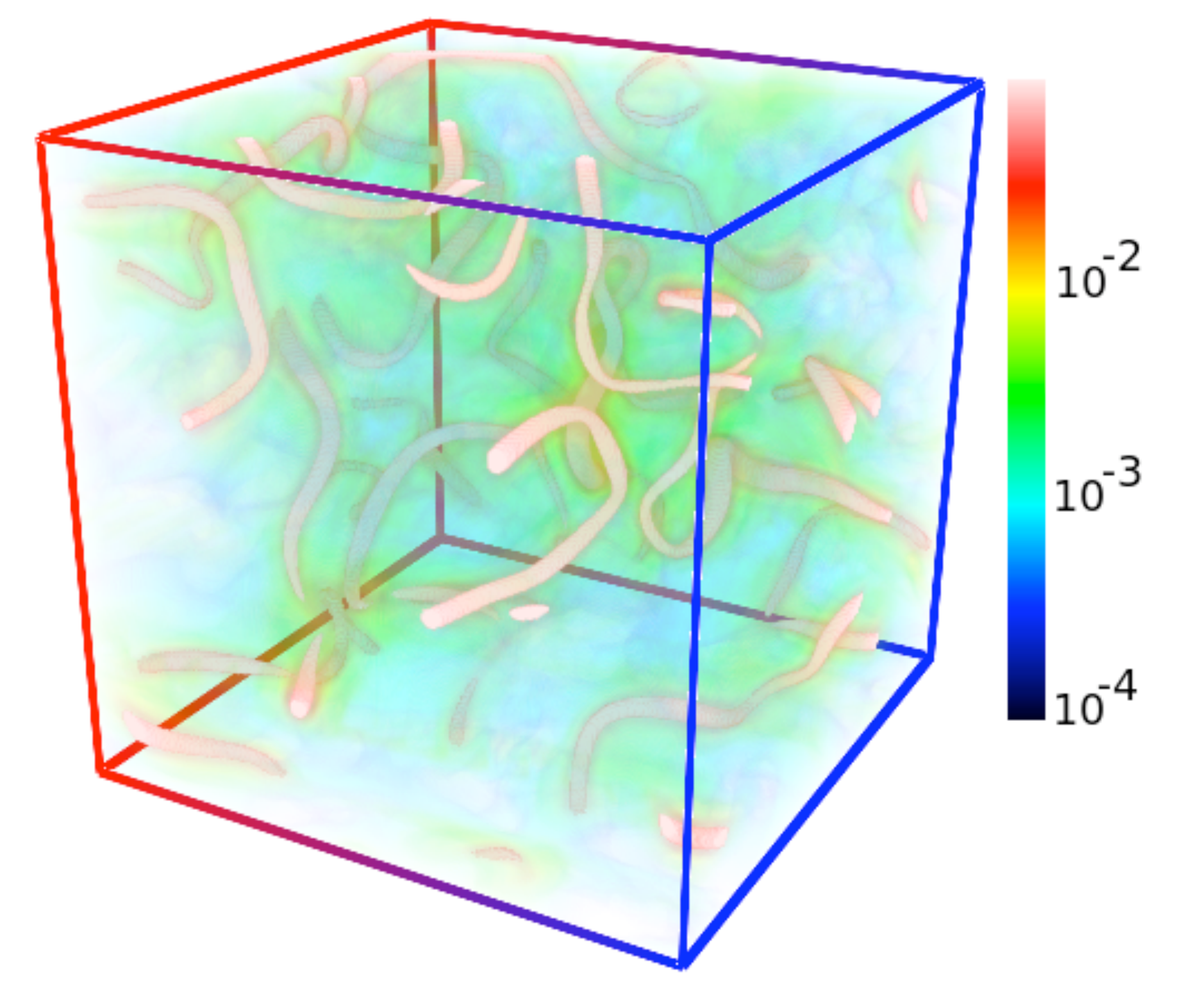}
\includegraphics[height=6.15cm]{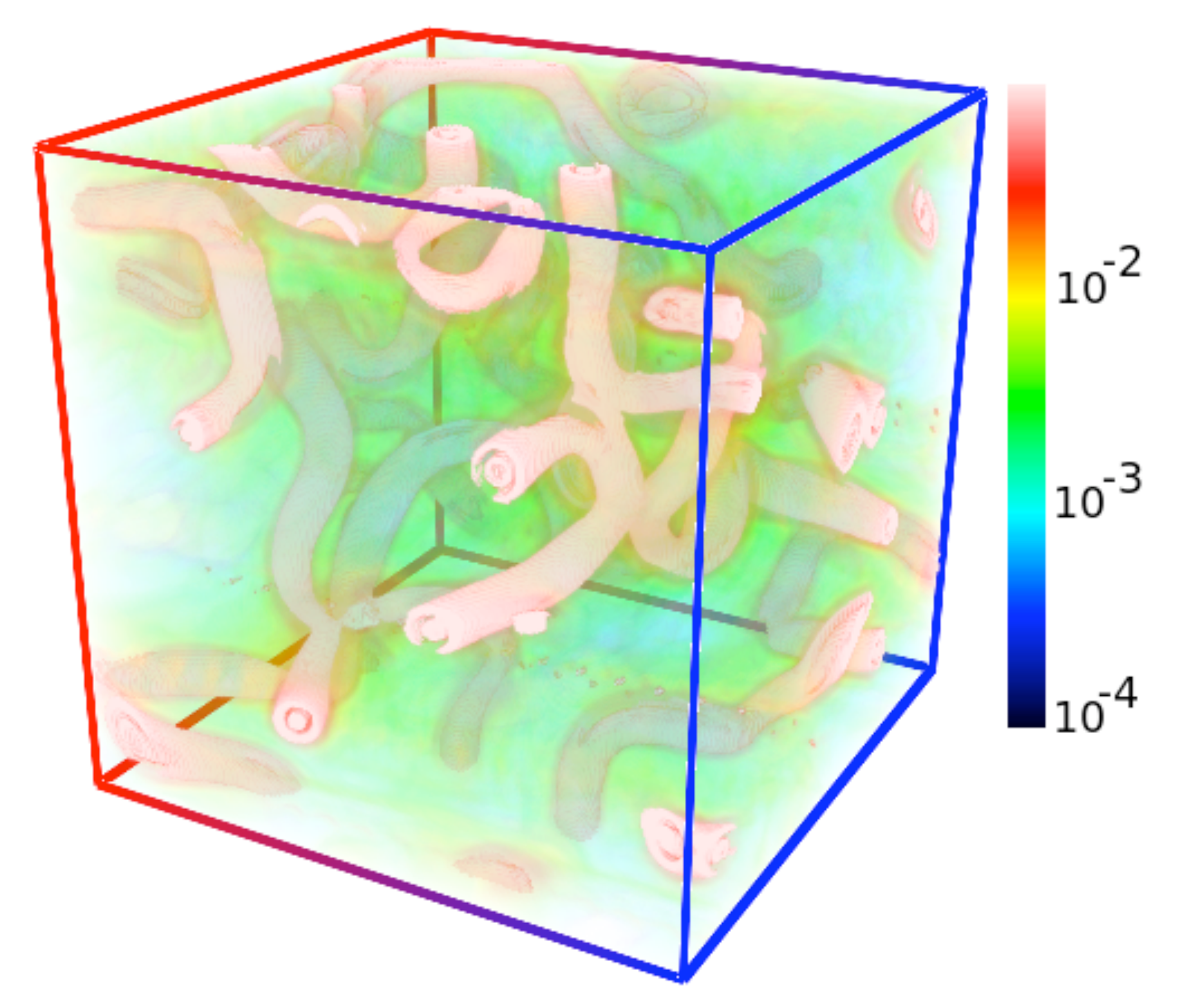}
\includegraphics[height=6.15cm]{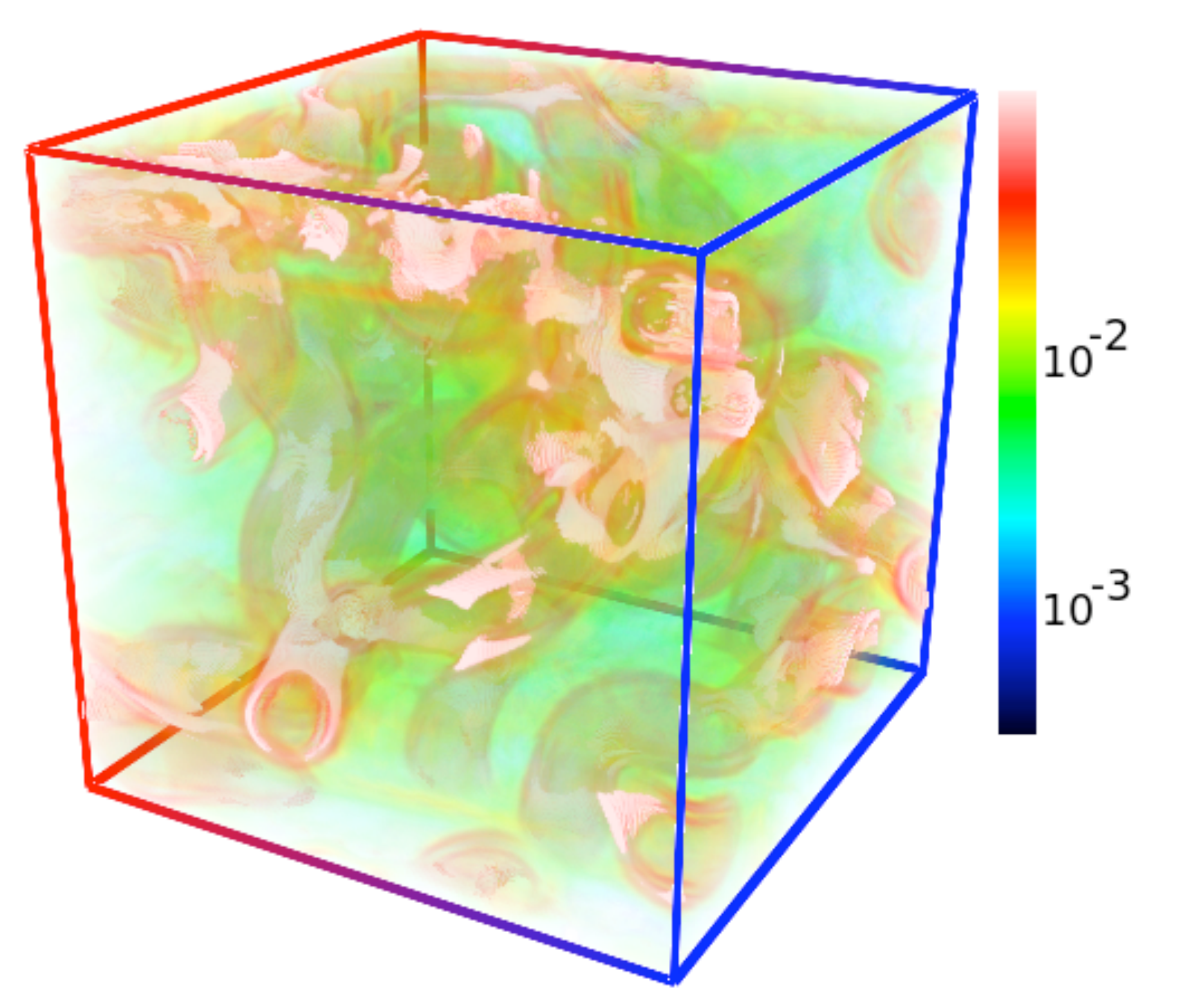}
\end{center}
\vspace*{-5mm}
\caption{Time evolution of the spatial distribution of GW $|\dot{h}_{ij}({\bf x},t)|^2$ (in arbitrary units) along the process of the symmetry breaking of the Higgs. Note that we use the same arbitrary units for all the plots, so the relative amplitude between one snapshot and another tells about the physical growth of the GW energy density. The images have been obtained with a $N = 256$ lattice simulation with IR cut-off $k_{IR} = 0.1 m$ and parameters $g^2 = 2 \lambda = 0.25$, 
$V_c = 0.024$ and $e = 6\,\sqrt{\lambda}$. From left to right, top to bottom, the snapshots correspond to mt = 5.5, 11.0, 17.3, 19.0, 21.0 and 23.0. At early times, before the Higgs bubbles percolate, the GW energy density is still very small and distributed in lumps over the lattice, with maximum values in the regions where the gradients of the Higgs are maximum. At times $mt \sim 17-19$, however, the string-like configurations of the gauge and Higgs fields induce similar string-like distributions of GW. The tubes of highest energy in GW can be seen most clearly at times $mt = 17-19$. Later, due to the time evolution of the strings, i.e. their fattening and shedding away of small-scale structures, the distribution of GW seems also to follow a similar pattern. Particularly noticeable here is the figure corresponding to time $mt = 21$, where inside one of the concentric tubes one can see another string-like configuraion in the core of the tube, very similar to what we observed in the magnetic energy density. Finally note that the orientation of the box have been choosen such that one can clearly see as best as possible some of the features developed by the spatial distribution of GW. Consequently, there is no specific correlation between the particular magnetic strings shown in figure~(\ref{fig:Mag3D}) and the ones shown here, since the boxes are being observed from very different orientations.}
\label{fig:GW3D}
\vspace*{-3mm}
\end{figure}

\begin{figure}[htb]
\begin{center}
\includegraphics[height=6.15cm]{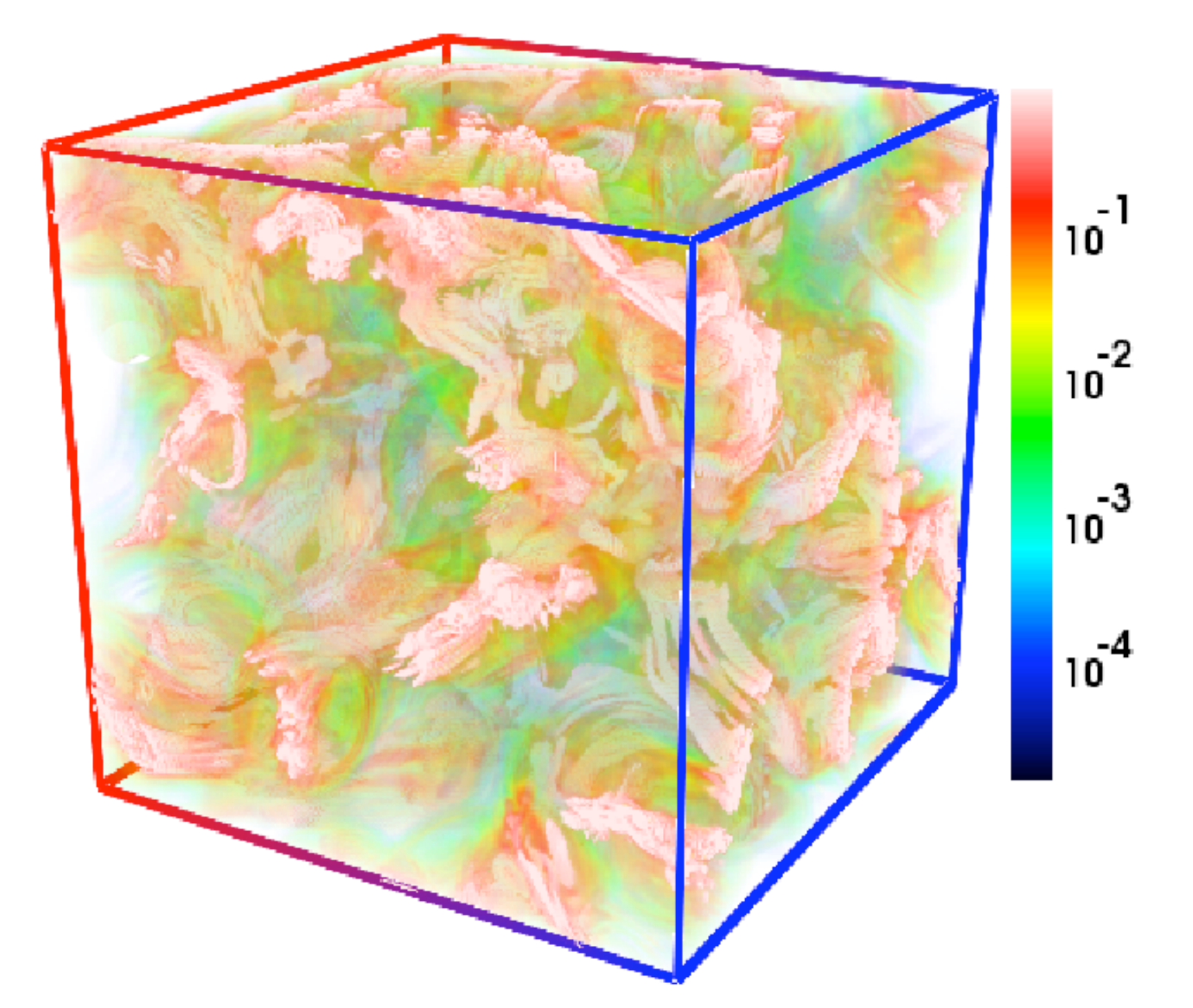}
\includegraphics[height=6.15cm]{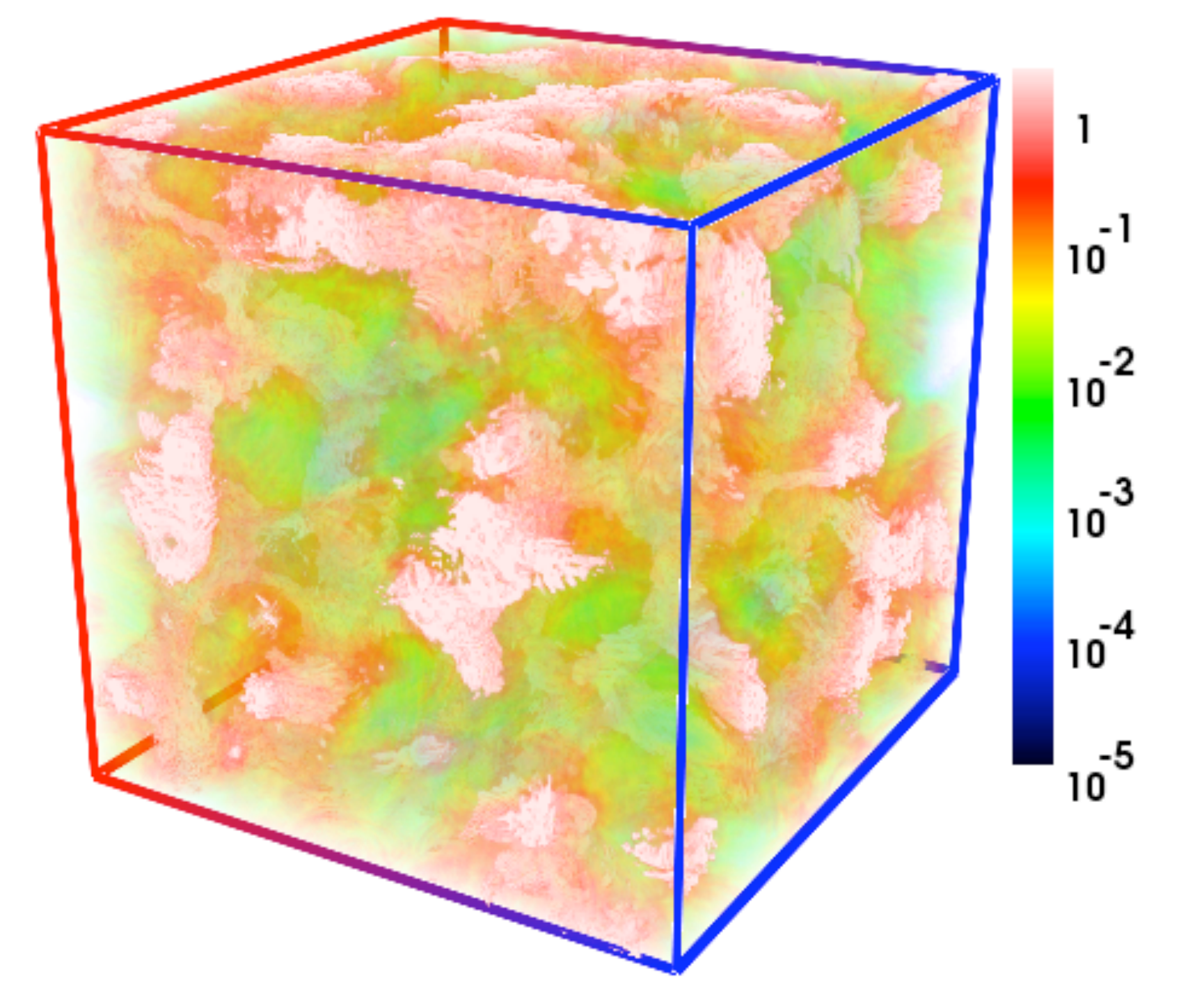}
\includegraphics[height=6.15cm]{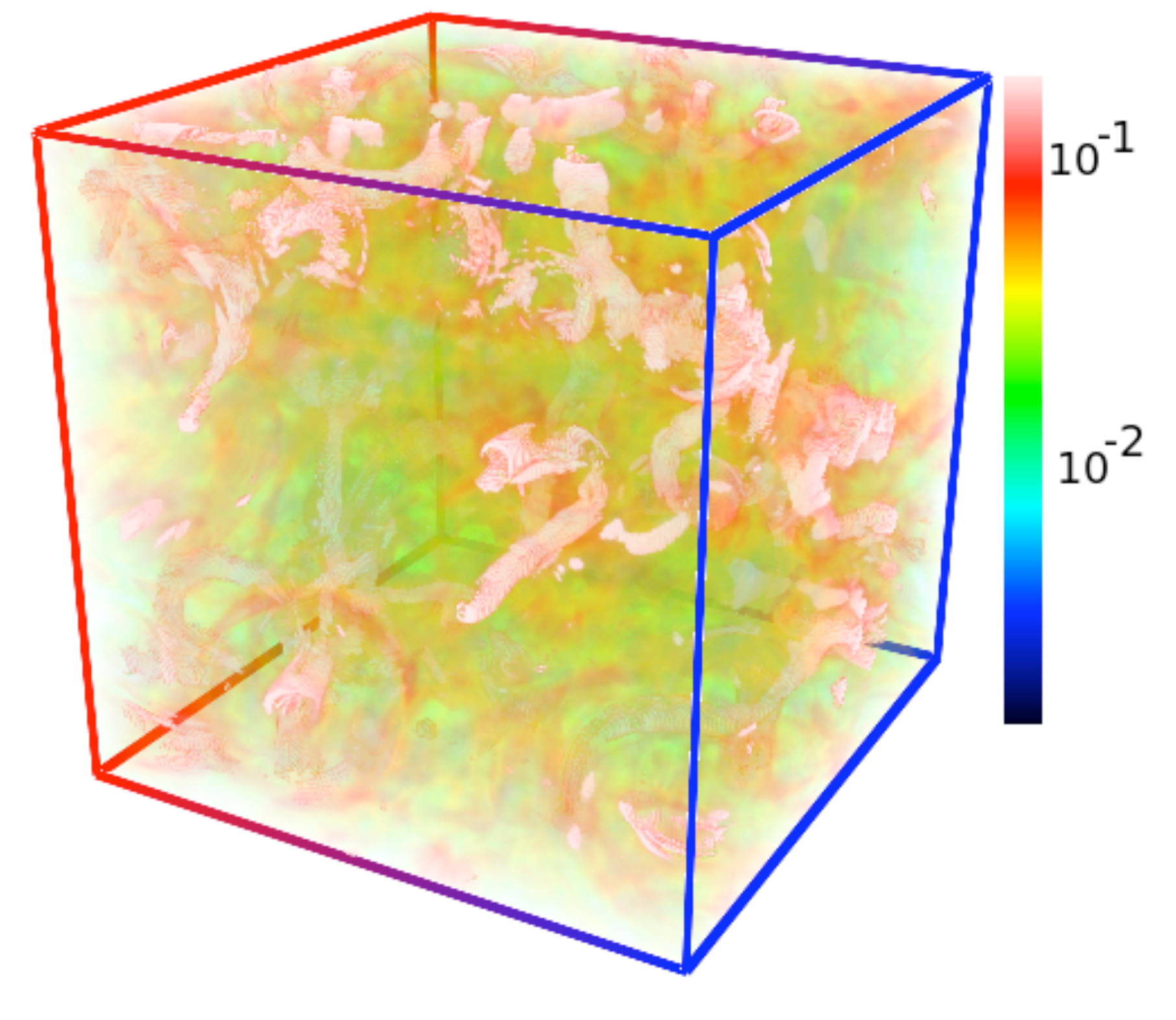}
\includegraphics[height=6.15cm]{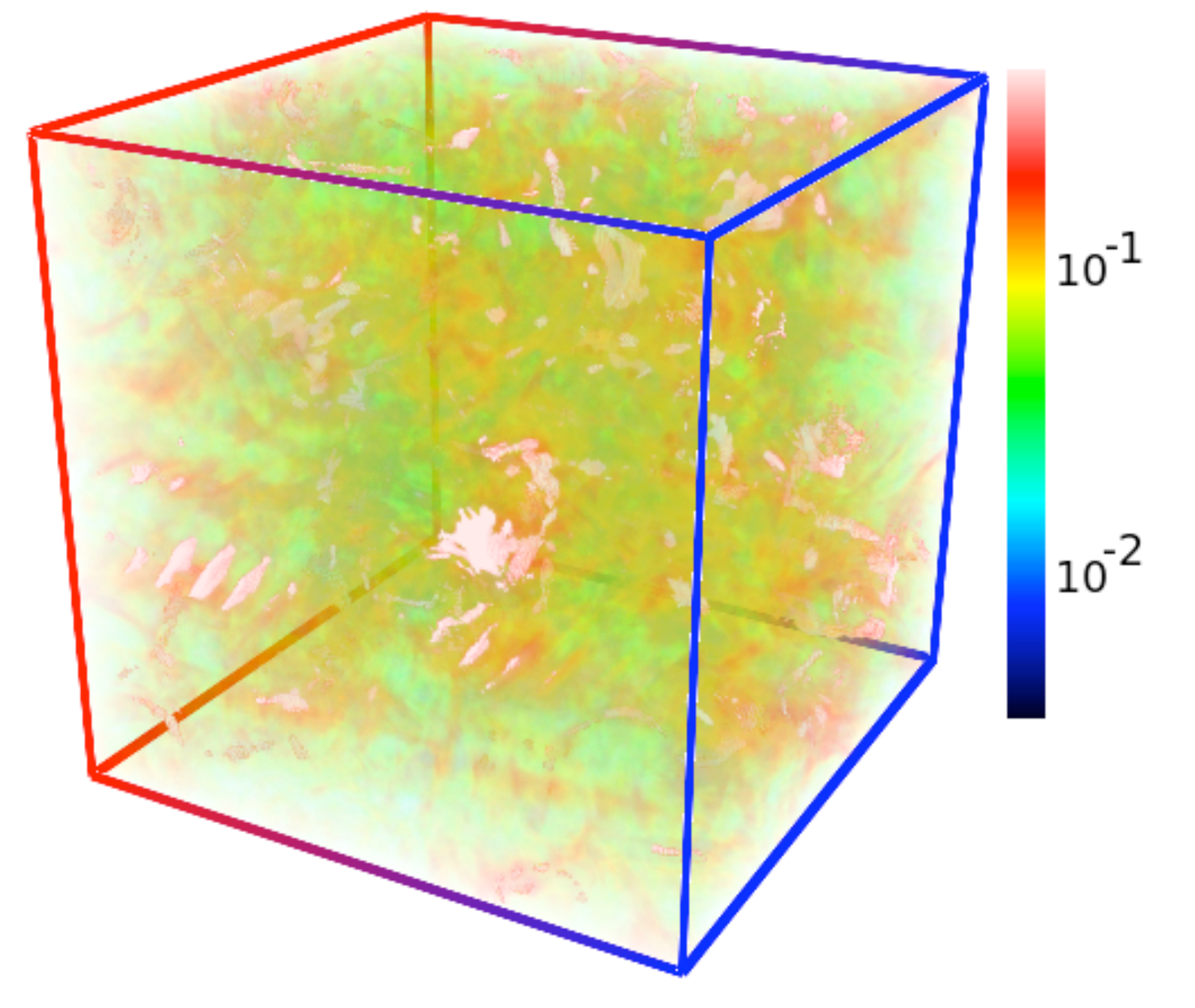}
\end{center}
\vspace*{-5mm}
\caption{Here we show the magnetic field energy density $B^2$ in units of $m^4$ (top) and the GW spatial distribution $\dot h_{ij}^2$ in arbitrary units (bottom), for times $mt = 29$ (left) and $mt = 35$ (right). Note that there are in fact three components in the magnetic field spatial distribution, the very core of the strings (barely seen here), as well as lumps of a small-scale structure which has been shedded away from the initial string-like configuration, plus a diffuse background (in green in the figures) which we interpret as radiation. The spatial distribution of GW follows a similar pattern. }
\label{fig:GW_and_Mag_3D_late_times}
\vspace*{-3mm}
\end{figure}

In Figure~\ref{fig:Mag3D} we show the time evolution of the spatial distribution of the magnetic energy density, from time $mt = 5.5$ till $mt = 23$. As explained above, the magnetic fluxes will tend to concentrate in those regions in which the Higgs amplitude (gradients) will be minimum (maximum). Thus, in the first two plots of Fig.~\ref{fig:Mag3D}, we see that during the initial times of symmetry breaking (when the Higgs bubbles have not yet percolated), the distribution of the magnetic field forms inhomogeneous lumps where the magnetic energy density is maximum in the spatial regions in between the initial nucleated Higss bubbles. However, the amplitude of the magnetic field is still very small to compete with the gradients of the Higgs, so the GW production is driven initially only by the Higgs inhomogeneities. When the Higgs bubbles percolate, the magnetic field is finally excited significatively and its amplitude grows by several orders of magnitude. At the same time, the regions between the percolating bubbles shrink, forming elongated tubes, i.e. topological defects, which correspondingly induce new spatial configurations in the distribution of the magnetic field. In particular, at times $mt= 17$ and $mt = 19$, see the third and fourth plots of Fig.~\ref{fig:Mag3D}, the magnetic field is compressed into thin tubes, located precisely in the locus of points where the Higss was prevented to reach the true VEV due to the development of a topological winding around the tube. Since the effective mass of the gauge field oscillates dynamically according to the amplitude of the Higgs around the VEV, the string-configurations of the gauge field do not have a constant width, but rather a time-dependent one, see fifth plot of Fig.~\ref{fig:Mag3D}. Indeed, the magnetic field string-configurations get thicker and eventually break into concentric layers which are shedded away, see the sixth plot of Fig.\ref{fig:Mag3D}. Nevertheless, the Higgs winding remains, following the Higgs' zeros at the cores of the strings. At time $mt=23$ we can even see in the figure a thin string at the core of a thicker one. Later, we will see that these features are also inherited by the spatial distribution of the energy density of GW. Since the magnetic field energy is pushed away from the core of the strings, one expects that, at late times, there will be three components of the magnetic field, one associated with the core of the strings, another one which has ``evaporated" from the core and now occupies the whole box in the form of very small-scale structures, plus a diffuse radiation component~\cite{magnetic}. This is indeed what we see in the first two plots of Fig.\ref{fig:GW_and_Mag_3D_late_times}. Again these features will also be inherited by the GW distribution of energy, see the last two plots of Fig.~\ref{fig:GW3D}.

\subsection{Histograms of the Higgs and magnetic field energy}

Histograms give yet another perspective on the dynamics of the Abelian- 
Higgs model at preheating. Power spectra gave us an idea of the 
energy distribution as a function of scale, and allowed us to pick up 
certain typical scales of the problem. Then we could look at the 
spatial distributions/configurations to search for specific features 
and find those given scales, like the string width and length. However, neither the power spectra nor the spatial distributions, tell us how common those features are. For that we have to look at the histograms of the values 
of the fields, and see how these distributions change with time. For 
instance, since magnetic strings seem to the localized around the 
zeros of the Higgs, if we can follow the time evolution of the Higgs 
histograms, we can see how often the Higgs field has a significant 
fraction of lattice points with its vev close to zero. At those points 
one expects new windings to appear, and indeed this is what it seems 
to occur, see left pannel in Fig.~\ref{fig:Histograms}.

One can also correlate the time evolution (oscillations) of the Higgs histograms with the production of magnetic field energy, associated 
with both the strings and the diffuse component coming from their evaporation. We show the time evolution of these histograms in the right pannel of Fig.~\ref{fig:Histograms}.

\subsection{Spatial distributions of GW}

Finally, let us consider the spatial configurations of GW. The expression for the GW energy density was given in 
(\ref{defrhogw}) and here we will look at the spatial distribution of $\dot{h}_{ij} \dot{h}_{ij}$. We will show in a sequence of snapthots the time evolution of $|\dot h_{ij}({\bf x},t)|^2$ in space, similarly to what we did with the 
Higgs and magnetic fields previously. Note that we will plot $|\dot h_{ij}({\bf x},t)|^2$ in arbitrary units, since we 
only want to hint the spatial features developed in the distribution of GW over the lattice, to show that these features follow precisely the string-like configurations of the source. As expected, the spatial distribution of GW will follow 
that of the dominant source term at each time. 

In Fig.~\ref{fig:GW3D}, a series of six snapshots of $|\dot{h}_{ij}({\bf x},t)|^2$ are shown sequentially from $mt = 5.5$ to $mt = 23$. In the first two figures, corresponding to times $mt = 5.5$ and $mt = 11$, the distribution of GW follows the spatial configurations of the Higgs bubbles during the initial tachyonic stage. The correlation between the Higgs features and the GW distribution was studied in detail in~\cite{GFS}, here we simply want to emphasize that initially the spatial configurations of GW are distributed as lumps over the lattice. However, in the middle figures of Fig.~\ref{fig:GW3D}, corresponding to times $mt = 17$ and $mt = 19$, the GW distribution has begun to concentrate within elongated regions which coincide precisely with those positions in space in which the Higgs and gauge fields have formed string-like configurations. The spatial distribution of GW is clearly concentrated around these strings. In the last two snapshots of the sequence of Fig.~\ref{fig:GW3D}, corresponding to times $mt = 21$ and $mt = 23$, we see that the string-like configurations of GW also fatten and break into small structures shedded away in the lattice, in full analogy with the behaviour of the magnetic energy density that we observed before. At later times, we show in Fig.~\ref{fig:GW_and_Mag_3D_late_times} how both the magnetic field and the GW are distributed all over the lattice, in the form of very small-scale structures. This behavior of the spatial distribution of GW, closely tracking the behavior of the abelian-Higgs strings as they are formed, evolve and later fragment into small-scale structures, is in perfect agreement with the successive appearence of the different peaks in the GW spectrum that we observed in Section V.


\section{Discussion and Perspectives}
\label{SecConclu}

Gravitational waves are a robust prediction of general relativity. There is indirect evidence of their existence
from inspiraling binary pulsars, although no single direct detection has been claimed. A stochastic background
of GW may soon be discovered, either directly with laser interferometer antennas or indirectly through the pattern
of polarization anisotropies they induce in the cosmic microwave background. Such a detection would open
a completely new and unexplored window into the Early Universe, possibly as rich as that which has been
recently revealed in the CMB. There are many sources of GW that can generate a stochastic backgrounds
and thus it is necessary to characterize those backgrounds with as much detail as possible. Apart from
known astrophysical point-like sources beyond the confusion limit (where we cannot resolve them), there
are also predictions for GWB from cosmic defects and hypothetical strongly first order phase transitions
in the Early Universe. Moreover, Cosmological Inflation makes a robust prediction of a stochastic GWB
produced during the quasi-exponential expansion of the Universe, with very specific spectral signatures:
a Gaussian, almost scale-invariant spectrum with an amplitude directly related to the energy scale of inflation.
If the scale is high enough (close to the Grand Unification Theories' scale) then these GW will also leave
an imprint in the (curl) polarization anisotropies of the CMB. Unfortunately, this GWB is still too weak to be
discovered with the near-future GW interferometric antennas, although it does cover a sufficiently broad
frequency range to be detectable by future GW Observatories (GWO) like BBO or DECIGO.

Furthermore, a robust prediction of inflation is that it must have ended, converting the vacuum energy
responsible for the tremendous expansion into the matter and radiation we observe today. Such a process,
known as reheating, is typically very violent and very inhomogeneous, with large density waves moving
at relativistic speeds and colliding among each other, thus converting a large fraction of their gradient energy
into gravitational waves. In some cases, the conversion is so sudden and violent that a significant fraction of
the total energy that goes into radiation ends in a stochastic background of GW, which could be detected in
the future. The energy spectrum of such a GWB is very non-thermal and far from scale invariant, but actually
peaked at a frequency which is related to the typical mass scale responsible for the end of inflation (either the
mass of the inflaton or that of the field that triggers the end of inflation, like in hybrid models), which could
be orders of magnitude smaller than the Hubble scale at the end of inflation. However, if the energy scale
of inflation is large (of order the GUT scale) then this stochastic GWB will be peaked at GHz frequencies,
far from the present sensitivity of GW interferometers. Nowadays, our only chance of detecting the GWB from
reheating is to consider the low-scale models of inflation - like hybrid models - with the appropriate parameters
to convert a large fraction of the initial vacuum energy into GW. The analyses done so far have
considered only scalar fields whose gradient energies source the anisotropic stresses needed for GW
production. However, vector fields (gauge or not gauge) are expected to be an even better source of GW, 
due to their anisotropic curl components, so that preheating scenarios with gauge fields may have a larger 
contribution to the GWB than scalar models. In fact, previous studies of gauge fields at preheating, 
in the context of Electroweak Baryogenesis and in the generation of Primordial Magnetic Fields, have identified
the formation of long-wave semiclassical gauge field configurations like sphalerons and helical magnetic
flux tubes which evolve with time very anisotropically and could contribute significantly to the production of GW.

In this paper, we have developed a formalism to calculate the production of GW by coupled systems of scalar 
and gauge fields on the lattice. The numerical method that we have constructed can be applied to different 
sources of GW where out-of-equilibrium gauge fields play an important role, such as thermal phase transitions, 
cosmological networks of local defects or non-perturbative decays of scalar condensates into gauge fields. 
We have studied in detail the dynamics and the production of GW during preheating after hybrid inflation, 
in the context of abelian-Higgs models that go through dynamical symmetry breaking triggered by the
expectation value of the inflaton field. As the inflaton is driven slowly (as opposed to a quench)
below the critical value, the mass squared of the abelian Higgs field becomes negative and drives the
spinodal growth of long-wave modes of the Higgs. Since the Higgs is charged, its rapid growth induces
a corresponding growth of gauge field configurations. At the end of inflation there are no temperature
fluctuations that can induce over-the-barrier transitions. However, long-wave quantum fluctuations become
semi-classical and act as a stochastic force that allow transitions over the false vacuum and thus induce
(locally) the generation of a topological winding number of the Higgs field. After symmetry breaking,
there is not enough energy to unwind the Higgs phase, leaving behind a Nielsen-Olesen string.
Such cosmic string configurations can be seen explicitly in our spatial distributions of both Higgs and
gauge fields. They play a crucial role in the production of GW at preheating and we observe that the spatial 
distribution of GW is indeed concentrated around the strings. Those strings will eventually decay (we see that 
they become wider and disperse away their energy density in the form of small-scale structures of the fields,
although the winding phase around the core of the string remains, since it is topologically stable), which
eventually shuts-off the GW production.

The complicated dynamics occurring at preheating in this abelian Higgs-inflaton model has been studied
using both power spectra analyses, as well as the fields' distributions in configuration space, together with
histograms of the fields' values, as a function of time, in order to correlate the different features observed
and their evolution. The picture that arises is the following. At the end of inflation the tachyonic growth of
the Higgs Gaussian random field creates an inhomogeneous distribution of fields characterized by
``bubbles" of Higgs energy density that expand and collide. The gauge field concentrates at the valleys
between the bubbles, where the Higgs has low values, forming long flux tubes of magnetic energy density.
The dynamics of the bubles when they expand and collide leads to regions in space where the Higgs field 
reaches the false vacuum and there are over-the-barrier transitions, with topological windings associated 
with them. These Nielsen-Olesen vortices are connected with each other in a cosmic string which runs along 
the core of the magnetic flux tubes. There are strings that encompass the whole simulation box and even beyond, 
thanks to periodic boundary conditions. We observe this process both in configuration space and with the histograms 
of Higgs vevs.

We have followed the dynamics of the strings during and after the symmetry breaking, although still on time 
scales shorter than the Hubble time and on length scales smaller than the Hubble radius. Once the strings are 
formed, they evolve by increasing their size and shedding away layers of magnetic energy density. At the cores 
of the strings there always remain a thin magnetic flux line but the energy seems to pour away from the strings 
in the form of waves concentric with the string. Nevertheless, we observe (in a transverse plane to the string) 
that at the core of the string there remains a conserved winding number of the Higgs. We have followed this winding 
number up to long times and we confirm that it is still there, in spite of the fact that the magnetic flux tube is 
so dilute that we cannot see it coherently: it seems to have ``evaporated". What remains is a diffuse background of 
small-scale structures of the Higgs and gauge fields permeating the whole box, together with the remnants of the 
strings.

The formation, evolution and fragmentation of the strings are accompanied by a significant production of gravitational 
waves which inherit specific features from the string dynamics. In position space, we observe how the distribution of 
GW follows very closely the evolution of the strings, being first concentrated around the straight segments of strings, 
then fattening as the strings become wider and finally being dispersed over the lattice as the strings emit small-scale 
structures of the fields. In Fourier space, this dynamics is encoded into the successive appearance of 
very distinct peaks in the GW spectra. The position of each peak is directly related to the physical scales
in the problem: the Higgs mass, which governs the width and interactions of Higgs field's strings, the gauge field mass,
which governs the width and interactions of gauge field's strings, and the typical momentum amplified by tachyonic 
preheating, which determines the characteristic size of the bubbles when they collide and the correlation length of the 
straight segments of strings. The former two determine the peaks in the high momentum (UV) range of the spectrum, while 
the latter corresponds to the long-wave (IR) peak. The IR peak appears first, when the bubbles collide and the strings 
are formed, while the UV peaks are formed later on, when the strings evolve and decay into small-scale structures of the 
Higgs and gauge fields. When the different scales are close to each other, the different peaks are superimposed and 
the amplitude in GW increases. When the gauge coupling constant is significantly smaller than the Higgs' self-coupling, 
the results reduce to the GW spectra produced without gauge field, characterized by a single peak. 

\begin{figure}[htb]
\begin{center}
\includegraphics[width=17cm]{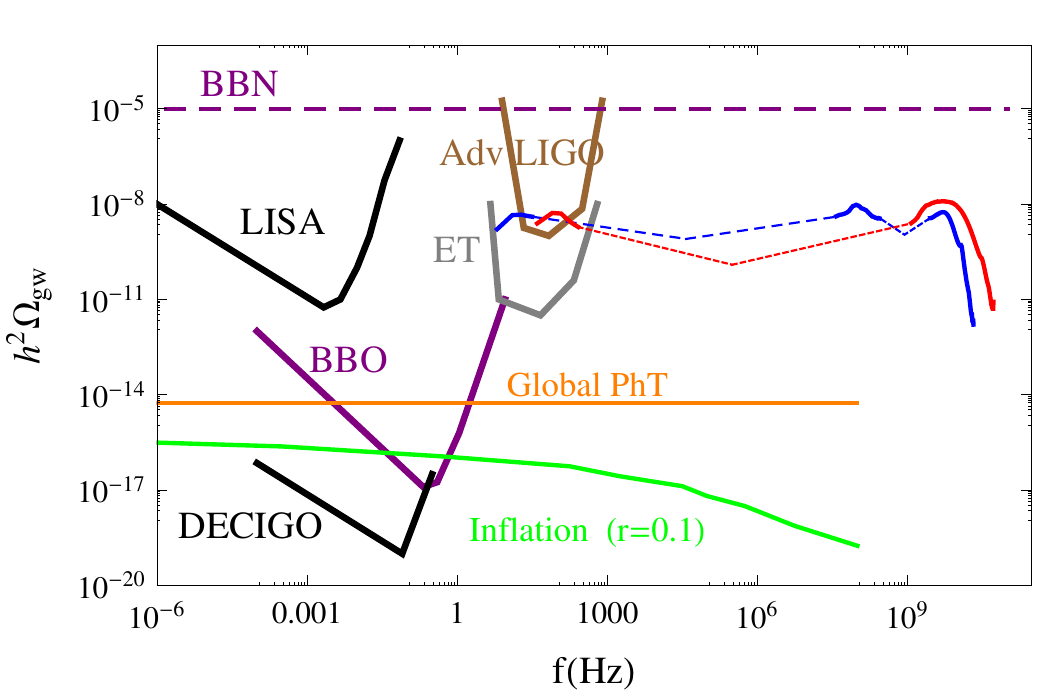}
\end{center}
\vspace*{-5mm}
\caption{The predicted stochastic background of GW from preheating in the abelian-Higgs model for two different sets of parameters, 
$\lambda\sim10^{-3}, g\sim10^{-8}, e\sim0.1, v\sim10^{11}$ GeV (red curve) and $g \sim \sqrt{\lambda} \sim 10^{-6}, e \sim 10^{-4}, v \sim 10^{13}$ GeV, and negligible initial velocity (blue curve), together with the expected sensitivity from future GWO like Advanced LIGO/VIRGO, LISA, ET, BBO and DECIGO. Note that in order to reach GWO sensitivity we had to extrapolate the position of the IR peaks using the expressions in Eq. (\ref{IRpeak}), and make an educated guess for the shape of the spectra between the peaks (dashed lines). Also plotted are the expected GWB from
a global phase transition and that from an inflationary model with tensor to scalar ratio $r=0.1$.}
\label{sensi}
\vspace*{-3mm}
\end{figure}

We have calculated the GW spectra produced in this abelian Higgs model of preheating after hybrid inflation with state-of-the-art simulations, although still limited in spatial resolution and box sizes. In order to probe reliably the different scales in the problem in each simulation, we developed a lattice calculation of GW production with gauge fields that is accurate up to second order in the lattice spacing. Our numerical results for the GW spectra today 
are well described by Eqs.~(\ref{f123}-\ref{omegagw*}). The present-day frequency and amplitude of these GW are very sensitive to the model parameters and the frequency of the different peaks may differ by many orders of magnitude, as illustrated in Fig.~\ref{sensi}. As in the same model with only scalar fields, very small coupling constants are still neccessary for these GW to fall into a frequency range that is accessible by interferometric experiments. Whether this is natural or not depends on the underlying theory for inflation and particle-physics models of hybrid inflation with such small coupling constants have indeed been already proposed in the literature, see~\cite{DFKN} and references therein.  We also observed that the frequency of the IR peak in the GW spectrum can be smaller than the peak frequency produced in the same model with only scalar fields, so the gauge field may enlarge the regions of the parameter space that may lead to an observable signal. More generally, there are many other models of inflation and preheating where gauge fields may play an important role and wich may lead to GW that could be observed in the future.    

After preheating the system enters into a turbulent regime where, at least in the abelian Higgs model that 
we considered, gravitational waves are no longer produced and the GW energy density saturates. We expect this 
result to be rather generic in abelian scalar gauge theories, because in that case the gauge fields that are 
produced at preheating acquire a mass, either directly through the Higgs mechanism or due to their interactions 
with scalar fields' fluctuations. It would be interesting to study models with other gauge groups, like 
$SU(2) \times U(1)$, where gauge fields remain effectively massless after symmetry breaking. This may happen for instance during 
preheating after hybrid inflation close to the electroweak scale (possibly a secondary stage of inflation, 
not necessarily related to the CMB anisotropies, only responsible for reheating the universe),    
where the photon spectra may exhibit inverse cascade during the turbulent evolution towards thermal 
equilibrium~\cite{magnetic}. This could significantly lower the typical frequency of the resulting GW today and 
relax the conditions on the parameters for these GW to be observable. The details of the GW spectra produced from preheating should also be rather sensitive to the particular gauge group under consideration because this determines the nature of the defects that can be formed. The defects do not have to be stable since these GW are produced when they are being formed. 

An intriguing possibility is the following. Given that preheating is so extremely inhomogeneous, and since
these inhomogeneities get imprinted in the gravitational wave background, which immediately decouples
from the plasma, one may envision a stage of technological development in the not so far future in which
GWO with sufficient angular resolution may resolve the structures that gave rise to the GWB right at the
moment when the Universe reheated. In the usual preheating scenario at high energy scales with only
scalar fields, the physical structures will have today a size that is completely undetectable when projected
over the sky, and thus the GWB will look essentially homogeneous from Earth. However, if gauge fields with
long string-like configurations (of horizon size and possibly even with superhorizon correlations) were
behind the generation of the GWB, then one could expect to see inhomogeneities in the angular distribution
of those gravitational waves. In particular, an array of GWO could detect the string-like anisotropies in the
GWB across the sky. At the moment, the angular resolution of LIGO is not better than a degree projected
in the sky. However, in the future one could resolve much finer structures in the GWB thanks to a dense
network of ground-based laser interferometers, as proposed e.g. in Ref.~\cite{GWBanisotropies}. Thinking 
ahead of our times, it may not be unrealistic to imagine that in the not so far future the GWB will
be mastered with sufficient detail to resolve the anisotropies in this elusive background and thus recover
vital information about the physics responsible for the violent conversion of energy from inflation to a
radiation and matter dominated epoch (the Big Bang of the Old Theory). No other probe can give us so
much information, since GW decouple immediately upon production and thus retain the spatial and
energy distributions of the sources that produced them. We can compare with the Cosmic Microwave
Background, which gives us detailed information about the epoch of photon decoupling thanks to the
exquisite measurements of the angular correlation of both temperature and polarization anisotropies.
The CMB provides a snapshot of how the Universe was like 380,000 years after the Big Bang. On the other
hand, the GWB would open a window into the physics of the Big Bang itself, allowing us to infer from its
detailed features whether it was as violent and inhomogeneous as we predict, to determine what kind of
fields were present and whether the rich phenomenology that we associate with preheating (topological
defects, baryogenesis and/or leptogenesis, primordial magnetic seed creation, non-thermal production
of dark matter, etc.) was actually realized in nature.


\section*{Acknowledgments}

It is a pleasure to thank Thibault Damour, Andres D\'iaz-Gil, Gary Felder, Margarita Garc\'ia Perez, Lev Kofman, 
Patrick Peter, Misha Shaposhnikov, Julien Serreau, Dani\`ele Steer and Jean-Philippe Uzan for very enlightening and helpful 
discussions. We acknowledge the use of the Beowulf cluster of the Instituto de F\'isica Te\'orica. DGF acknowledges 
support from a Marie Curie Early Stage Research Training Fellowship associated with the EU RTN ``UniverseNet" 
during his stay at CERN TH-Division. JGB thanks the Institute de Physique 
Th\'eorique de l'Universit\'e de Gen\`eve for their generous hospitality during his sabbatical in Geneva. 
This work is supported by the Spanish MICINN under project AYA2009-13936-C06-06 and by the EU FP6 Marie Curie 
Research and Training Network ``UniverseNet" (MRTN-CT-2006-035863).


\appendix

\section{No Massless Gauge Fields in Abelian Scalar Gauge Theories During Preheating}
\label{AppNogo}

In this Appendix, we extend to an arbitrary number of fields the argument of Section~\ref{SecTheory} that massless gauge fields are not produced during preheating in abelian scalar gauge theories.  

Consider an abelian theory $U(1)^K$, with $K$ gauge fields $A^{(k)}_\mu$ coupled to $N$ complex scalars $\varphi_n$
\be
-\mathcal{L} = \frac{1}{4}\,\sum_{k=1}^{K} F^{(k)}_{\mu\nu} \, F^{(k)\,\mu\nu} + \sum_{n=1}^{N} \left(D_\mu \varphi_n\right)^*\,D^\mu \varphi_n 
\ee
with
\be
D_\mu \varphi_n = \partial_\mu \varphi_n - i\,\sum_{k=1}^{K} e_{kn} \, A^{(k)}_\mu \, \varphi_n
\ee
and $F^{(k)}_{\mu\nu} = \partial_\mu A^{(k)}_\nu - \partial_\nu A^{(k)}_\mu$. The equations of motion are
\be
\partial^\nu F^{(k)}_{\mu\nu} + 2\,\sum_{n=1}^{N} |\varphi_n|^2\,e_{kn}\,\sum_{l=1}^{K} e_{ln}\,A^{(l)}_{\mu} 
= 2 \,\sum_{n=1}^{N} e_{kn}\,\mathrm{Im}\left[\varphi_n^* \partial_{\mu} \varphi_n\right] \hspace*{0.5cm} \mbox{ for } k = 1, ..., K \ .
\ee
Denoting by $v_n$ the VEV of $\varphi_n$, the $K \times K$ mass matrix of the gauge fields is
\be
\mathcal{M}_{kl} = 2\,\sum_{n=1}^{N} v_n^2 \, e_{kn} \, e_{ln} \hspace*{0.5cm} \mbox{ for } k, l = 1, ..., K \ .
\ee
The matrix $\mathcal{M}$ is real and symmetric, so the system can be diagonalized with an orthogonal matrix $U$: 
$U^T\,\mathcal{M}\,U$ is diagonal with $U^{-1} = U^T$. For the system to admit a massless gauge field, at least 
one eigenvalue of the mass matrix should vanish. Suppose that the $j^{\mathrm{th}}$ eigenvalue $\lambda_j$ vanishes, 
so that
\be
\label{lambda0}
\lambda_j = \left(U^T\,\mathcal{M}\,U\right)_{jj} = 2\,\sum_{n=1}^{N} v_{n}^2\,\tilde{e}_{jn}^2 = 0
\ee
where we have defined
\be
\tilde{e}_{kn} = \sum_{l=1}^{K} e_{ln}\,U_{l j} \hspace*{0.5cm} \mbox{ for } k = 1, ..., K \mbox{ and } n = 1, ..., N \ .
\ee
In the new basis
\be
\label{newbasis}
\tilde{A}^{(k)}_{\mu} = \sum_{l=1}^{K} U_{lk}\,A^{(l)}_\mu \hspace*{0.5cm} \mbox{ for } k = 1, ..., K \ ,
\ee
$\tilde{A}^{(j)}_\mu$ is a massless candidate. It satisfies the equation
\be
\label{eomj}
\partial^\nu \tilde{F}^{(j)}_{\mu\nu} + 2\,\sum_{n=1}^{N} |\varphi_n|^2\,\tilde{e}_{jn}\,\sum_{l=1}^{K} \tilde{e}_{ln}\,\tilde{A}^{(l)}_{\mu} 
= 2 \,\sum_{n=1}^{N} \tilde{e}_{jn}\,\mathrm{Im}\left[\varphi_n^* \partial_{\mu} \varphi_n\right] \hspace*{0.5cm} \mbox{ for } k = 1, ..., K \ ,
\ee
where $\tilde{F}^{(j)}_{\mu\nu}$ is its gauge field strength. There are two cases to consider. If all the scalar fields have a non-zero VEV, 
$v_n \neq 0$ $\forall n = 1, ..., N$, then the condition (\ref{lambda0}) implies that 
$\tilde{e}_{jn} = 0$ $\forall n = 1, ..., N$. In this case, we see from (\ref{eomj}) that the massless gauge 
field $\tilde{A}^{(j)}_\mu$ decouples from all the scalars, $\partial^\nu \tilde{F}^{(j)}_{\mu\nu} = 0$. On 
the other hand, for this field to remain coupled we need $\tilde{e}_{jn} \neq 0$ for at least one value of $n$. 
The condition (\ref{lambda0}) then implies that $\varphi_n$ has a zero VEV. However, since $\tilde{e}_{jn} \neq 0$, we 
see from (\ref{eomj}) that $\tilde{A}^{(j)}_\mu$ acquires an effective mass proportional to 
$\tilde{e}_{jn}\,\langle |\delta \varphi_n|^2 \rangle$ due to its interaction with the fluctuations of $\varphi_n$. 
Thus the $U(1)$ gauge fields are either effectively massive or decoupled from the scalars and the other $U(1)$ gauge fields.


\section{Lattice Formulation with $\O(dx^2)$ Accuracy}
\label{AppLattice}

In this appendix we describe the discretized equations of motion that we evolve on the lattice. 
As is well known, special care has to be maid when discretizing a gauge theory in order to preserve 
gauge invariance on the lattice. For instance, this is necessary for constraint equations to follow 
from the dynamical equations that are evolved. The basic formalism that we use is standard (see e.g. 
\cite{ambjorn}) but, in the presence of gravity waves, we have to generalize it in order to reproduce 
the continuum theory up to $\O(dx^2)$ and $\O(dt^2)$ accuracy in the lattice spacing $dx$ and timestep $dt$. 

In the lattice formulation, the scalar fields $\tilde{\chi}$ and $\tilde{\varphi}$ are defined at spacetime lattice points $x$ 
while the gauge field $\tilde{A}_\mu$ is defined in the middle of the segments between lattice points, at $x + \hat{\mu} / 2$ where $\hat{\mu}$ 
is a vector of length $dx^\mu$ in the $\mu$ direction, $dx^i = dx$ is the lattice spacing and $dx^0 = dt$ is the 
timestep. In this appendix, we denote the lattice fields with a tilde to distinguish them from their continuum analogs. 
One then introduces link variables, also defined in the segments between lattice points, and related to the gauge field according to
\be
\tilde{U}_\mu\left(x + \frac{\hat{\mu}}{2}\right) = e^{-i e dx^\mu \tilde{A}_\mu(x + \frac{\hat{\mu}}{2})} \ .
\ee
In this equation the repeated index $\mu$ is \emph{not} summed.
In the case of abelian symmetry, we can choose to treat either the gauge field or the links as the fundamental 
objects that we evolve numerically. We found that the first option led to faster simulations, so in the following 
we will express everything in terms of the gauge field. The discussion is easily generalized to the case where the 
links are considered as the fundamental variables. 

The starting point is the lattice action (\ref{Sdis}) which is a discretized version of the continuum action 
(\ref{eq:lagrangianGauge}). The lattice expressions for the forward partial derivative, the forward gauge covariant derivative
and the gauge field strength are given respectively by
\bea
\label{forpar}
\partial^+_\mu \tilde{\chi} = \frac{1}{dx^\mu}\,\left[\tilde{\chi}(x + \hat{\mu}) - \tilde{\chi}(x)\right] \ , \\
\label{forcov}
D^+_\mu \tilde{\varphi} = \frac{1}{dx^\mu}\,\left[\tilde{U}_\mu\left(x + \frac{\hat{\mu}}{2}\right)\,\tilde{\varphi}(x + \hat{\mu}) - \tilde{\varphi}(x)\right] \ , \\
\label{Ftilde}
\tilde{F}_{\mu\nu} = \partial^+_\mu \tilde{A}_\nu - \partial^+_\nu \tilde{A}_\mu \ .
\eea 
Again, repeated indices are \emph{not} summed in Eqs.(\ref{forpar}, \ref{forcov}). In the limit $dx^\mu \rightarrow 0$, 
these expressions reduce to the continuum partial derivative, gauge covariant derivative and gauge field strength, respectively. 
The definitions (\ref{forpar}-\ref{Ftilde}) imply that the action (\ref{Sdis}) is invariant under the lattice gauge 
transformation (\ref{gaugedis}), which is a discretized version of a continuum gauge transformation. 

The equations of motion following from (\ref{Sdis}) are obtained by the lattice equivalent of functional differentiation. 
They read
\bea
\label{latphi}
\partial^{- \mu} \partial^+_\mu \tilde{\chi} &=& \frac{\partial V}{\partial \tilde{\chi}}\\
\label{latX}
D^{- \mu} D^+_\mu \tilde{\varphi} &=& \frac{\partial V}{\partial \tilde{\varphi}^*}\\
\label{latA}
\partial^{- \mu} \tilde{F}_{\mu \nu} &=& -2e\,\mathrm{Im}\left[\tilde{\varphi}^*\,D^+_\nu \tilde{\varphi}\right] 
\eea
where we have defined the backward partial derivative
\be
\label{backpar}
\partial^-_\mu \tilde{\chi} = \frac{1}{dx^\mu}\,\left[\tilde{\chi}(x) - \tilde{\chi}(x - \hat{\mu})\right] 
\ee
and the backward gauge covariant derivative
\be
\label{backcov}
D^-_\mu \tilde{\varphi} = \frac{1}{dx^\mu}\,\left[\tilde{\varphi}(x) - \tilde{U}_\mu\left(x - \frac{\hat{\mu}}{2}\right)\,\tilde{\varphi}(x - \hat{\mu})\right]
\ee
where again repeated indices are \emph{not} summed. 

We evolve these equations in the temporal gauge ($\tilde{A}_0 = 0$, $\tilde{U}_0 = 1$) with the staggered leapfrog method, 
where the fields and their time derivatives are evaluated at times that differ by half a timestep. Explicitely, we have
\bea
\label{phidis}
\tilde{\dot{\chi}}\left(x + \frac{dt}{2}\right) &=& \tilde{\dot{\chi}}\left(x - \frac{dt}{2}\right) + 
dt\,\left[\partial^-_i \partial^+_i \tilde{\chi} - \frac{\partial V}{\partial \tilde{\chi}} \right]_{(x)} \\
\label{Xdis}
\tilde{\dot{\varphi}}\left(x + \frac{dt}{2}\right) &=& \tilde{\dot{\varphi}}\left(x - \frac{dt}{2}\right) + 
dt\,\left[D^-_i D^+_i \tilde{\varphi} - \frac{\partial V}{\partial \tilde{\varphi}^*} \right]_{(x)} \\
\label{Adis}
\tilde{E}_i\left(x + \frac{\hat{i}}{2} + \frac{dt}{2}\right) &=& \tilde{E}_i\left(x + \frac{\hat{i}}{2} - \frac{dt}{2}\right) + 
dt\,\left[\partial^-_j \tilde{F}_{j i} + 2e\,\mathrm{Im}\left[\tilde{\varphi}^*\,D^+_i \tilde{\varphi}\right]\right]_{(x + \frac{\hat{i}}{2})}
\eea
together with $\tilde{f}(x + dt) = \tilde{f}(x) + dt\,\tilde{\dot{f}}\left(x + \frac{dt}{2}\right)$ and 
$\tilde{A}_i\left(x + \frac{\hat{i}}{2} + dt\right) = \tilde{A}_i\left(x + \frac{\hat{i}}{2}\right) + 
dt\,\tilde{E}_i\left(x + \frac{\hat{i}}{2} + \frac{dt}{2}\right)$. 

Finally, the discretized version of Gauss constraint (\ref{Gcont}) reads
\be
\label{Gdis}
\partial^-_i \tilde{E}_i = 2e\,\mathrm{Im}\left[\tilde{\varphi}^*\,\tilde{\dot{\varphi}}\right] \ .
\ee
It follows from the dynamical equations (\ref{Xdis}, \ref{Adis}) if it is satisfied initially, because these equations are
derived from the same, gauge-invariant lattice action (\ref{Sdis}). Gauss constraint was satisfied down to machine 
precision in all our runs.

In the limit $dt \rightarrow 0$, one can show that the equations of motion (\ref{phidis}-\ref{Adis}) implies that the total energy density
\be
\rho = \left\langle \, \frac{1}{2}\,\tilde{\dot{\chi}}^2 + |\tilde{\dot{\varphi}}|^2 + \frac{1}{2}\,\tilde{E}_i \, \tilde{E}_i +
\frac{1}{2}\,\partial_i^{+} \tilde{\chi}\,\partial_i^{+} \tilde{\chi} + D_i^{+} \tilde{\varphi}\,\left(D_i^{+} \tilde{\varphi}\right)^* + 
\frac{1}{4}\,\tilde{F}_{ij}\,\tilde{F}_{ij} + V \,\right\rangle 
\ee
is conserved, $d\rho / dt = 0$. Here $\langle ... \rangle$ denotes the average over all the lattice points. 
Energy was conserved up to $\sim 0.1 \%$ in all our runs.

It is important to note that these discretized equations of motion reproduce the continuum ones up to $\O(dt^2)$ and 
$\O(dx^2)$ when $dt, dx \rightarrow 0$. This can be checked explicitely, remembering that the gauge field and thus its 
equations of motion are defined in the segments between lattice points. In particular, we have
\bea
D^-_i D^+_i \tilde{\varphi} &\simeq& D_i D_i \varphi (x) + \O(dx^2)\\
\label{Fcont}
\tilde{F}_{ij} &\simeq& F_{i j}\left(x + \frac{\hat{i}}{2} + \frac{\hat{j}}{2}\right) + \O(dx^2)\\
\partial^-_j \tilde{F}_{ji} &\simeq& \partial_j F_{j i} \left(x + \frac{\hat{i}}{2}\right) + \O(dx^2)\\
\mathrm{Im}\left[\tilde{\varphi}^*\,D^+_i \tilde{\varphi}\right] &\simeq& \mathrm{Im}\left[\varphi^*\,D_i \varphi\right]_{(x + \frac{\hat{i}}{2})} + \O(dx^2)
\eea
  
We now come to the discretized equations of motion for gravity waves $\tilde{h}_{ij}(x)$, that we define at the lattice points as the scalar fields. Fortunately, we don't have to start from a discretized action for them because GW are gauge invariant and no constraint equation is associated to them. However, in the presence of gauge fields, special care is 
to be paid in order to reproduce the continuum equation (\ref{hcont}) up to $\O(dx^2)$ and $\O(dt^2)$ accuracy. In order 
to do so, we cannot replace the terms (\ref{Tcont}) in $\Pi_{ij}^{\mathrm{TT}}$ by their lattice analogs defined above, because these reduce to the continuum values at lattice points up to $\O(dx)$ or $\O(dt)$ accuracy only. We thus have to construct new lattice expressions that reproduce the continuum up to second order and which lead to a 
gauge-invariant stress-energy tensor. For the partial derivative of the inflaton, this is just the symmetric derivative
\be
\partial^S_i \tilde{\chi} = \frac{1}{2 dx}\,\left[\tilde{\chi}(x + \hat{i}) - \tilde{\chi}(x - \hat{i})\right]
\ee
which indeed reduces to $\partial_i \chi(x) + \O(dx^2)$ for $dx \rightarrow 0$. We can achieve the same for the gauge covariant derivative of the Higgs, by defining
\be
D^{S}_i \tilde{\varphi} = \frac{1}{2}\,\left(D^+_i \tilde{\varphi} + D^-_i \tilde{\varphi}\right) = \frac{1}{2 dx}\,
\left[\tilde{U}_i\left(x + \frac{\hat{i}}{2}\right)\,\tilde{\varphi}(x + \hat{i}) - 
\tilde{U}_i\left(x - \frac{\hat{i}}{2}\right)\,\tilde{\varphi}(x - \hat{i})\right] \ .
\ee
Note also that, under the lattice gauge transformation (\ref{gaugedis}), this transforms as 
$D^{S}_i \tilde{\varphi} \rightarrow e^{i \tilde{\alpha}}\,D^{S}_i \tilde{\varphi}$, as it should. 

For the lattice gauge field strength $\tilde{F}_{i j}$, note that it reduces with $\O(dx^2)$ accuracy to the 
continuum $F_{ij}$ evaluated in between lattice points, see Eq.(\ref{Fcont}). To achieve the same at the lattice points themselves, we consider the clover average
\be
\tilde{F}^c_{i j}(x) = \frac{1}{4}\,\left[\tilde{F}_{ij}\left(x + \frac{\hat{i}}{2} + \frac{\hat{j}}{2}\right) + 
\tilde{F}_{ij}\left(x + \frac{\hat{i}}{2} - \frac{\hat{j}}{2}\right) + 
\tilde{F}_{ij}\left(x - \frac{\hat{i}}{2} + \frac{\hat{j}}{2}\right) +
\tilde{F}_{ij}\left(x - \frac{\hat{i}}{2} - \frac{\hat{j}}{2}\right) \right] \ .
\ee
Similarly, the lattice electric field $\tilde{E}_i$ reduces with $\O(dx^2)$ and $\O(dt^2)$ accuracy to its continuum conterpart evaluated at $x + \hat{i} / 2 + dt / 2$. We thus consider its clover average
\be
\tilde{E}^c_i(x) = \frac{1}{4}\,\left[\tilde{E}_i\left(x + \frac{\hat{i}}{2} + \frac{dt}{2} \right) + 
\tilde{E}_i\left(x + \frac{\hat{i}}{2} - \frac{dt}{2} \right) + \tilde{E}_i\left(x - \frac{\hat{i}}{2} + \frac{dt}{2} \right) 
+ \tilde{E}_i\left(x - \frac{\hat{i}}{2} - \frac{dt}{2} \right) \right]
\ee
which reduces to $E_i(x) + \O(dx^2) + \O(dt^2)$ for $dx, dt \rightarrow 0$. In terms of these, the lattice analog of 
(\ref{Tcont}) is given by
\be
\label{Tdis}
\tilde{\Pi}^{\mathrm{TT}}_{ij} = \left[ \partial^S_i \tilde{\chi} \, \partial^S_j \tilde{\chi} + 
2\,\mathrm{Re}\left[D^S_i \tilde{\varphi} \left(D^S_j \tilde{\varphi}\right)^*\right] + 
\tilde{F}^c_{ik} \tilde{F}^c_{jk} - \tilde{E}^c_i \tilde{E}^c_j\right]^{\mathrm{TT}} \ . 
\ee
It fulfills the conditions stated above, namely it is invariant under the lattice gauge transformation (\ref{gaugedis}) 
and reproduces the continuum with second order accuracy in $dx$ and $dt$. 

Contrary to $\tilde{E}_i$ which was displaced by half-timesteps, the clover average of the electric field $\tilde{E}^c_i$ 
is defined at the timesteps themselves, as the other terms in (\ref{Tdis}). This allows to preserve the leapfrog scheme 
for the gravity waves
\be
\tilde{\dot{h}}_{ij}\left(x + \frac{dt}{2}\right) = \tilde{\dot{h}}_{ij}\left(x - \frac{dt}{2}\right) + 
dt\,\left[\partial^-_k \partial^+_k \tilde{h}_{i j} + 16\pi G\,\tilde{\Pi}^{\mathrm{TT}}_{ij}\right]_{(x)} \ .
\ee 
This is easily implemented at each iteration, by first advancing $\tilde{E}_i$ by half a time step, then advancing the 
gravity waves by a full timestep and finally advancing $\tilde{E}_i$ by the remaining half timestep. 

As discussed in subsection~\ref{subsecLatt}, the $\mathcal{O}(dx^2)$ calculation of GW allows for a much better control 
on the UV part of the GW spectra.



\end{document}